\documentclass[12pt]{article}
\usepackage[utf8]{inputenc}
\usepackage[title]{appendix}
\usepackage{graphicx}
\usepackage{graphbox}
\usepackage[margin=1in]{geometry} 
\usepackage{xcolor}
\usepackage{amsmath}
\usepackage{amssymb}
\usepackage{amsthm}
\usepackage{pdfpages}
\usepackage[bookmarks=false]{hyperref}
\usepackage{titlesec}
\usepackage[flushleft]{threeparttable}

\usepackage{mathtools,nccmath,bbm}
\usepackage{booktabs}
\usepackage[inline]{enumitem}
\usepackage[backend=biber,authordate,noibid]{biblatex-chicago}
\usepackage[nameinlink,noabbrev,capitalize]{cleveref}
\usepackage[singlespacing]{setspace}
\usepackage[multiple]{footmisc}
\usepackage{standalone}
\newtheorem{prop}{Proposition}

\newtheorem{hypothesis}{Hypothesis}
\newtheorem{result}{Result}

\newtheorem{innercustomgeneric}{\customgenericname}
\providecommand{\customgenericname}{}
\newcommand{\newcustomtheorem}[2]{%
  \newenvironment{#1}[1]
  {%
   \renewcommand\customgenericname{#2}%
   \renewcommand\theinnercustomgeneric{##1}%
   \innercustomgeneric
  }
  {\endinnercustomgeneric}
}

\allowdisplaybreaks[1]

\newcustomtheorem{customre}{Result}

\usepackage[utf8]{inputenc}
\usepackage[TS1,T1]{fontenc}
\usepackage{array, booktabs}
\usepackage{graphicx}
\usepackage{caption}

\usepackage{lscape}

\usepackage{xcolor}

\definecolor{colora}{RGB}{148 196 193}
\definecolor{colorb}{RGB}{206 106 108}
\definecolor{colorc}{RGB}{237 173 163}

\hypersetup{
    colorlinks=false,
    citebordercolor=colora,
    filebordercolor=colorb,
    linkbordercolor=colorc,
    menubordercolor=colorc,
    urlbordercolor=colorb,
    runbordercolor=colorb,
}

\newcounter{savefootnote}
\newcounter{symfootnote}
\newcommand{\symfootnote}[1]{%
  \setcounter{savefootnote}{\value{footnote}}%
  \setcounter{footnote}{\value{symfootnote}}%
  \ifnum\value{footnote}>8\setcounter{footnote}{0}\fi%
  \let\oldthefootnote=\thefootnote%
  \renewcommand{\thefootnote}{\fnsymbol{footnote}}%
  \footnote{#1}%
  \let\thefootnote=\oldthefootnote%
  \setcounter{symfootnote}{\value{footnote}}%
  \setcounter{footnote}{\value{savefootnote}}%
}

\usepackage{ragged2e}

\usepackage{dcolumn}
\newcolumntype{d}[1]{D{.}{.}{#1}}

\begin{document}
\let\origaddcontentsline\addcontentsline
\renewcommand{\addcontentsline}[3]{}
\begin{center}
\phantom{.}

\vspace{4\baselineskip}

\textbf{\Large
On Prior Confidence and Belief Updating\symfootnote{We thank Erik Eyster, Liang Yucheng, Daniel Martin, Ryan Oprea, and Sevgi Yuksel for their helpful comments. The University of Alberta provided generous funding for this project.}}

\vspace{2\baselineskip}

Kenneth Chan\footnote{National University of Singapore; \href{mailto:ken_chan@ucsb.edu.sg}{kenneth\_chan@ucsb.edu}.}, Gary Charness\footnote{University of California, Santa Barbara.}, Chetan Dave\footnote{University of Alberta; \href{mailto:cdave@ualberta.ca }{cdave@ualberta.ca}.}, J.~Lucas Reddinger\footnote{Purdue University; \href{mailto:reddinger@ucsb.edu}{reddinger@ucsb.edu}.}\\

\vspace{2\baselineskip}

May 13, 2025

\vspace{\baselineskip}

\end{center}

\thispagestyle{empty}
\paragraph{Abstract}

We experimentally investigate how confidence over multiple priors affects belief updating. Theory predicts that the average Bayesian posterior is unaffected by confidence over multiple priors if average priors are the same. We manipulate confidence by varying the time subjects view a black-and-white grid, the proportion representing the prior in a Bernoulli distribution. We find that when subjects view the grid for a longer duration, they have more confidence, under-update more, and place more (less) weight on priors (signals). Overall, confidence over multiple priors matters when it should not, while confidence in prior beliefs does not matter when it should.

{\RaggedRight
}

\clearpage

\linespread{1.5}\selectfont

\section{Introduction}\label{sec:intro}
\setcounter{footnote}{0} 
\setcounter{page}{1}

Bayes' rule is the objectively correct belief updating process with microeconomic foundations in decision-making under uncertainty \parencite{ortoleva2024}. It is thus the standard for updating prior beliefs upon receiving new information in classical economic models. However, the behavioral and experimental economics literature provides, perhaps unsurprisingly, substantial evidence that individuals do not behave in accordance with the Bayesian paradigm \parencite{Benjamin2019}.  Examining how people incorporate new information into their existing beliefs, relative to Bayes' law, is critical to the formalization of new belief updating models to explain individual information synthesis and subsequent decision-making.\footnote{For example, in finance and macroeconomics, \gentextcite['s][][]{GennaioliShleifer2010} formalization of \gentextcite['s][][]{KahnemanTversky1972b} representativeness heuristic allows \textcite{BordaloGennaioliShleifer2018} to develop the notion of diagnostic expectations and explain credit cycles. That broad notion of diagnostic expectations allows \textcite{BianchiIlutSaijo2024} to better reconcile macroeconomic models with data than with use of the rational expectations standard.}

Our objective here is to consider how an individual's \textit{confidence} relates to the belief-updating process and the resultant updated belief. We consider two notions of confidence. First, we consider confidence in the prior belief distribution as being related to the dispersion of the belief distribution; specifically, for some distributions, greater confidence implies less dispersion in the belief distribution.\footnote{Variance (or standard deviation) is a useful measure of dispersion for a model of Bayesian updating.} This notion of confidence offers an intuitive result in which a Bayesian agent updates more when less confident in a prior belief.\footnote{Specifically, the difference between the mean updated belief and the mean prior belief is increasing in the variance of the prior belief distribution.} The most well-known example is the updating of a Gaussian prior belief distribution with a signal drawn from a Gaussian distribution. When updating, an agent places more weight on the prior belief when there is less variance in the prior distribution.\footnote{Other well-known cases include (i) a beta prior belief distribution with signals drawn from a binomial distribution, and (ii) a Dirichlet prior distribution and a signal drawn from a multinomial distribution.} Our experiment examines a Bernoulli distribution, for which the greatest updating occurs given a 50\% prior belief (maximal uncertainty regarding a binary state).

Our second notion of confidence is that over multiple prior distributions, corresponding  to uncertainty or second-order beliefs over multiple priors. In this setting, an agent has a set of prior beliefs considered plausible and assigns probability weights to these priors.\footnote{This set of weights can be interpreted as second-order prior beliefs.} If one is only interested in average updated beliefs, a counter-intuitive result obtains: confidence over multiple priors does not affect how a Bayesian agent updates average beliefs as long as the average prior belief remains unchanged. To update beliefs, a Bayesian agent updates each individual prior and the beliefs over the set of priors. But if one is only interested in the average Bayesian posterior belief, a Bayesian agent simply updates beliefs using the average prior belief.  Our focus is on this second notion of confidence over multiple priors. Henceforth, when referring to confidence, we mean confidence over multiple priors unless stated otherwise. Moreover, as described in our experimental environment below, multiple priors arise for subjects as they see a stimulus for a very short period of time in a treatment, which induces a prior. Thus although there is an objective prior known to the experimenter, subjects' uncertainty about it can cause the set of beliefs about the stimuli they see to dilate.

Confidence over multiple priors is relevant to settings where there is a lack of precise information or ambiguity \parencite{Ellsberg1961}. In situations where there is a lack of precise information about the prior belief (\emph{i.e.}, how the states are determined), multiple subjective prior beliefs can exist. Even when the prior is objective, uncertainty about the signal's reliability can cause the set of beliefs to dilate \parencite{ShishkinOrtoleva2023}, resulting in multiple posterior beliefs.\footnote{There will be a posterior belief for each possible signal reliability that the agent considers.} Consequently, there are now multiple priors for subsequent steps of updating. Situations in which individuals hold multiple priors could reflect realistic conditions in many domains, such as financial markets, medical decision-making, and political forecasting. Studying how individuals update their beliefs in the presence of multiple priors provides valuable insight into belief-updating models as well as behavior that is guided by beliefs.

Our experimental treatment induces exogenous variation in confidence over multiple priors. We then elicit confidence over multiple priors with our incentive-compatible mechanism, a process that can be employed to measure confidence in any stated probability by a subject in an experimental design. The experimental design thus allows us to explore the relationship between confidence across multiple priors and the belief-updating process.

Our belief updating task is a re-framed version of the seminal taxi-cab problem discussed in \textcite{KahnemanTversky1972a}. Instead of being informed of a numerical prior probability, a subject views a $10 \times 10$ grid consisting of black and white squares; the proportion of white squares represents the actual probability of success (prior to any signal). In one treatment the grid is flashed for 0.25 seconds, giving the subject only a rough sense of the proportion.\footnote{\textcite{EspondaOpreaYuksel2023} use a 0.25 second viewing time to provide a noisy \emph{signal} for updating, while we use it to induce a noisy \textit{prior}.} This feature induces multiple priors as subjects are unsure of the actual proportion of white squares. In the other treatment, the viewing time is 30 seconds, giving subjects sufficient time to count the number of squares exactly. After seeing the grid, we inform subjects that a square from the grid is randomly selected with uniform probability and we provide each subject with a signal on the color of their realized square (with a fixed, symmetric accuracy of 60\% or 80\%).\footnote{A \emph{symmetric accuracy} rate $r$ implies $P(\text{black signal}\,|\,\text{black square})=P(\text{white signal}\,|\,\text{white square})=r$.} We then elicit each subject's updated belief regarding the color of the selected square before and after seeing the signal. We incentivize the elicitation of beliefs by paying subjects a fixed amount if their stated value is within three percentage points of the actual prior or the true Bayesian posterior.

As stated above, our main goal is to study how one's \textit{confidence} in prior beliefs affects one's belief-updating process. To measure confidence, we develop an incentive-compatible elicitation method as follows.  Following the elicitation of a subject's point-estimate $\pi$ of the probability of an event, we offer the subject two options: a fixed payment that is obtained when $\pi$ is within three percentage points of the true parameter (a \emph{subjective gamble}), and a simple gamble that obtains the same fixed payment with probability $q$ (an \emph{objective gamble}). Each subject ultimately provides a switching point $q^*$, above which the subject prefers the objective gamble and below which the subject prefers the subjective gamble. This switching point provides a measure of the subject's confidence in their prior.\footnote{We elicit confidence on both the prior belief \textit{and} the updated belief. Inference on the subject's confidence depends on how beliefs are incentivized; we discuss this in greater detail in \cref{sec:design-confidence}.}

To analyze our data, we construct two different measures to quantify the degree of over- and under-updating relative to the Bayesian posterior. We then examine the relationship between one's confidence in one's prior belief and any over- or under-updating relative to the Bayesian benchmark. We find that higher confidence is associated with more conservative updating.  Our estimates of the \textcite{Grether1980} model further validate this result. We find that a subject places less weight on their prior and more weight on the signals when they view the grid for 0.25 seconds. We also find that our subjects are insensitive to confidence in a single prior distribution. Contrary to the Bayesian prediction, we find that our subjects respond to confidence as represented by multiple priors but do not respond to confidence as represented by (less) dispersion in a belief distribution.

We contribute to the existing literature in three main ways. First, we establish a relationship between confidence over multiple priors and belief updating. While results for prior mixture models are well-established, these results have not been experimentally tested in the context of individual belief updating. Our results show that \textit{confidence} over multiple priors matters even when it should not, and people update more when less confident. We find a systematic departure from Bayesian updating that gives insight regarding theoretical models that aim to formalize belief-updating rules in environments with second-order beliefs. Second, we propose and implement a simple incentive-compatible mechanism to elicit confidence in one's belief. Finally, we present empirical evidence that subjects are surprisingly overconfident in their ability to perform Bayesian updating.

The paper is structured as follows: \cref{sec:literature} reviews relevant literature, \cref{sec:design} describes our design in detail, \cref{sec:results} presents our results, and \cref{sec:discussion} concludes.

\section{Related Literature} \label{sec:literature}

The survey in \textcite{Benjamin2019} summarizes well the belief updating literature, including studies of mistakes in probabilistic reasoning. Laboratory experiments have documented several non-Bayesian updating patterns. Some of the more prominent biases include base-rate neglect \parencite{KahnemanTversky1972b, EspondaVespaYuksel2024}, conservatism bias \parencite{PhillipsEdwards1966, AugenblickLazarusThaler2024}, correlation neglect \parencite{EysterWeizsacker2016, EnkeZimmerman2019}, confirmation bias \parencite{Nickerson1998, CharnessDave2017}, and motivated beliefs \parencite{EilRao2011, Couts2019, Thaler2021, CharnessOpreaYuksel2021, MobiusNiederleNiehausRosenblat2022}.  \textcite{AgranovReshidi2023} also show that in a sequential updating problem with multiple signals, the order in which the signals are presented to the subjects affects the final belief profile.\footnote{This is a violation of the divisibility property in belief updating \parencite{Cripps2018}.} Beyond the lab, a growing literature documents non-Bayesian belief updating in finance \parencite{BordaloGennaioliShleifer2018, AugenblickLazarus2023}, job search \parencite{ConlonPilossophWiswallZafar2018, BrownChan2024}, and sports-betting markets \parencite{AugenblickRabin2021, AugenblickLazarusThaler2024}.

To accommodate these belief-updating biases, behavioral economists have constructed numerous non-Bayesian updating models \parencite{Grether1980, RabinShrag1999, EpsteinNoorSandrioni2010, Cripps2018, Woodford2020, LevyInesRazin2022}. Among these, the most popular model for empirical analyses is that of \textcite{Grether1980}, which accommodates several updating biases and can be estimated with linear regression. We apply this model to our environment in \cref{sec:model-grether} and present the results in \cref{sec:results-grether}.

When there is a set of prior beliefs, the ambiguity literature has proposed a few non-Bayesian updating models. One is termed \textit{Full Bayesian Updating} \parencite{pires2002, gilboamarinacci2016} in which an agent ignores the second-order belief over the priors and simply updates each prior with Bayes' rule. Another is \textit{Maximum Likelihood Updating} \parencite{Dempster1967,gilboaschmeidler1993}, in which an agent selects the most likely prior after observing the signal. In contrast to these extreme models of belief updating, \textcite{kovach2024ambiguity} introduces \textit{Partial Bayesian updating}, a generalized version of \textit{Full Bayesian Updating} and \textit{Maximum Likelihood Updating}, in which an agent considers a subset of priors deemed plausible, and updates these priors with Bayes' rule. \textcite{klibanoffhannay2007} introduces a model of updating that retains dynamic consistency in preferences, and \textcite{Ortoleva2012} introduces the \textit{Hypothesis Testing model} in which an agent may switch between prior beliefs upon receiving a surprising signal.

While deviations from Bayes' rule comprise numerous systematic patterns, we are primarily interested in examining over- and under-updating relative to the Bayesian benchmark. \textcite{Benjamin2019} notes that empirical biases in belief updating generally take two forms: base-rate neglect, a phenomenon where people underweight the prior belief, and conservatism, a tendency to under-react to new information. Base-rate neglect results in over-updating relative to the Bayesian benchmark, while conservatism results in under-updating relative to the Bayesian benchmark. The dominating bias then gives the direction of the overall effect.

Our work is closely related to two recent studies that examine the causes of over- and under-updating. \textcite{AugenblickLazarusThaler2024} show that individuals tend to under-infer from signals when test reliability is high (greater than 60\% accuracy) and over-infer when reliability is low (below 60\% accuracy).  \textcite{BaBohrenImas2023} show that when the number of possible outcomes or states increases, subjects tend to over-update when there are more than two states, while subjects under-update when there are two states. The aforementioned papers employ a noisy cognition model \parencite{Woodford2020, EnkeGraeber2023} to explain over- and under-updating. In particular, \textcite{EnkeGraeber2023} argue that when humans are cognitively uncertain they tend toward a cognitive default. Our results cannot be explained by the standard model of noisy cognition alone. In contrast to these studies, we focus on prior confidence as an explanation for over- or under-updating. We also diverge from these papers with our use of \gentextcite['s][][]{KahnemanTversky1972a} famous taxi-cab problem, while the aforementioned studies both use some variant of the balls-and-urn (or book-bag-and-poker-chip) design \parencite{PhillipsEdwards1966} commonly used in belief updating studies \parencite{Benjamin2019}.\footnote{\gentextcite['s][][]{KahnemanTversky1972a} task is also recently used by \textcite{EspondaVespaYuksel2024} and \textcite{AgranovReshidi2023}.} We note that while the experiments of these two papers were conducted online, our study was conducted in-person with a university subject pool.\footnote{\textcite{CharnessCoxEckelHoltJabarian2023} discuss the merits of in-person laboratory experiments.}

\section{Experimental Design and Hypotheses} \label{sec:design}

The main goal of our experiment is to investigate whether confidence over multiple priors influences how one updates beliefs in tasks with two components: a belief-updating exercise and a confidence-in-stated-belief elicitation.  In each task, subjects provide four probability reports in the following order: (1) their prior belief, (2) their confidence in their prior belief, (3) their updated belief after receiving a signal, and (4) their confidence in their updated belief. In this section we first describe the belief-updating exercise, describe the elicitation method, and conclude with our hypotheses.

\subsection{Belief Updating} \label{sec:design-updating}

\begin{figure}[h!]
    \centering
    \includegraphics[width=3in]{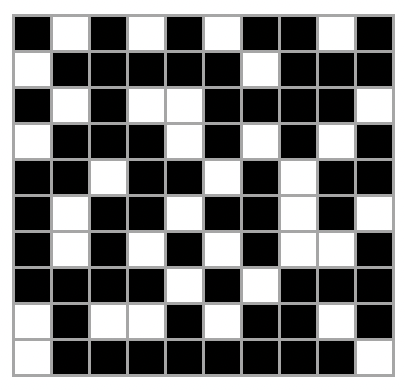}
    \caption{An example grid displayed in an experimental task}
    \label{fig:grid}
\end{figure}

Our belief updating task is a version of \gentextcite['s][][]{KahnemanTversky1972a} taxi-cab problem. To make the task less abstract, we follow the framing of \textcite{EspondaVespaYuksel2024} in which subjects are asked to play the role of a manager who is evaluating whether a project is a success or failure.  Subjects are told that each square in a $10 \times 10$ grid represents a project, with white representing a success and black a failure (see \cref{fig:grid}).  We induce prior beliefs by showing a grid realization to the subject. The proportion of successes to failures varies across tasks (grids): the number of successful projects is one of 0, 20, 40, 50, 70, or 90. We included a task with 0 successful projects, so correct responses are straightforward, to evaluate a subject's comprehension of the environment.

We use a within-subject design to study how an individual updates beliefs with different levels of confidence. To achieve this, we have two treatment conditions that subjects complete in order. Subjects first complete tasks in a \textit{Low Confidence} treatment that flashes each grid for 0.25 seconds. Subjects then complete tasks in a \textit{High Confidence} treatment that displays each grid for 30 seconds (subjects are allowed to proceed after five seconds).\footnote{We chose 30 seconds because it allows a subject to tally a grid twice. In our experiment, most subjects proceeded without using all 30 seconds to view the grid.} The difference in the duration of the display of the grid induces variation in one's confidence, hence the treatment names.\footnote{Another possible way to induce different degrees of confidence in the prior belief is to inform the subject that the prior is either 75\% or 25\% \parencite{Liang2024}. We induce the multiple prior in this manner because it avoids describing the prior in a compound manner, which is known to pose challenges for subjects in the context of risk \parencite{Halevy2007}.}  We postulate that confidence is lower in the Low Confidence treatment relative to the High Confidence treatment, the latter giving subjects sufficient time to tally squares. All subjects complete the Low Confidence treatment before the High Confidence treatment to minimize the formation of beliefs regarding our experimental parameters.

After we show a subject a realized grid, we tell the subject that a project (\textit{i.e.}, a square from the grid) has been randomly selected with equal probability. The subject is then asked to state the probability that the selected project is a success. We refer to this reported probability as the subject's \emph{prior} belief.  We pay \$3 if this stated probability is within three percentage points of the actual probability and zero otherwise \parencite{DufwenbergGneezy2000, CharnessDufwenberg2006}.  We choose this simple incentive-compatible mechanism to simplify the confidence-in-beliefs elicitation method, which we discuss in \cref{sec:design-confidence}. Moreover, studies have shown that complex belief-elicitation methods like the binarized scoring rule \parencite{HossainOkui2013} can bias subjects' reported beliefs \parencite{DanzVesterlundWilson2022}, despite being incentive compatible in theory.

Next, to permit belief updating, we tell subjects that they will be shown a computer test result to help with their evaluation of the selected project. Subjects are assigned randomly to a treatment with either a 60\% or an 80\% test reliability.\footnote{We held signal reliability constant for each subject to minimize potential confusion.} The test reliability indicates the proportion of times when the computer test correctly predicts whether the project is a success or a failure with symmetric false positive and false negative rates. For example, when the reliability is 80\%, if the selected project is a success (failure) the test result will be positive (negative) with 80\% chance and negative (positive) with 20\% chance.

After informing subjects of the result of the test (positive or negative), we ask the subject to report the probability of the selected project being a success, which we refer to as the subject's \emph{update}. We pay each subject \$3 for reporting a probability within three percentage points of the Bayesian posterior. It is important to note that when the agent is uncertain about their prior or updated beliefs, our belief elicitation method remains incentive compatible by encouraging the agent to report the belief they are most confident in---that is, the point with the highest probability mass within a three-percentage-point interval. 

In our experiment, when the prior about the success of the project is non-degenerate, we elicit the subject’s updated belief for both positive and negative test results \parencite[that is, we use a version of the strategy method; see][]{BrandtsCharness2011}.\footnote{The strategy method has also been used in other belief updating experiments, such as \textcite{CiprianiGuarino2009, Toussaert2017, CharnessOpreaYuksel2021, EspondaVespaYuksel2024}.}. We have one task where the prior is degenerate at 0; in this case, we only elicit the update when the test result is positive using an incomplete strategy method.\footnote{The purpose of this task was to gauge the subject's understanding of the task. The positive signal realization of this task allows us to determine if the subject understood the updating task. If this task is selected for payment and the negative signal is realized, we draw again.} Each subject completes eleven tasks in each of the two treatments. Subjects receive no feedback for their decision, and we draw the parameters of the task, realized project type, and signals at the end of the experiment to determine which of the subject's responses determine their payoff. As a result of this, it is not equally likely for all responses to be selected for payment.

\subsection{Confidence Elicitation} \label{sec:design-confidence}

We elicit confidence after each belief elicitation---that is, after a subject states their prior belief and also following the updated belief.\footnote{This captures confidence over multiple priors.} To elicit a confidence-in-stated-belief, we offer the subject a choice between either (i) a subjective gamble, which obtains $\$3$ if their stated belief is within three percentage points of the true value, and (ii) an objective gamble that obtains $\$3$ with probability $x$ and nothing otherwise. This method is similar to \gentextcite['s][][]{HeathTversky1991} experiments that study ambiguity preferences. The key difference is that their subjects self-report their confidence in their responses, and subsequently they offer subjects the choice of betting on their initial responses or a lottery that corresponds to their level of confidence. This was done to study subjects' preferences over ambiguity \parencite{Ellsberg1961}.

We seek to identify the point $x^*$ at which a subject is indifferent between the subjective and objective gambles by varying $x$. We then interpret $x^*$ as the confidence level for an expected utility maximizer. For example, if the subject is 70\% confident that their stated prior belief is within three percentage points of the actual prior (or, following the signal, the Bayesian posterior), the subject expects to be paid 70\% of the time.  Thus the subject should only accept lotteries that pay \$3 at least 70\% of the time. By reporting 70\% in the confidence elicitation, a subject ensures that they will only receive lotteries that obtain \$3 at least 70\% of the time.

We elicit the switching point $x^*$ using a BDM mechanism \parencite{BeckerDeGrootMarschak1964} following \gentextcite['s][][]{Healy2020} instructions; we explain the BDM with a multiple price list \parencite{HoltLaury2002} to improve comprehension of the mechanism.  Because subject comprehension of the BDM mechanism may be a concern \parencite[see][]{CasonPlott2014}, we include a simple question with no successful projects where the theoretically optimal response is obvious.  In this question, the grid that is shown contains all black squares, and subjects should have maximal confidence in their belief.  Our data for this degenerate prior case demonstrate that our subjects have an excellent understanding of the mechanism and the tasks.

With our initial belief elicitation method, it is easy for subjects to recognize that the switching point is their level of confidence that their stated belief is within three percentage points of the actual value. A more complex scoring rule (such as the binarized scoring rule) would have created a challenge in constructing the equivalent lottery and for subjects to identify the optimal decision. Subjects struggle with the binarized scoring rule on its own in a setting with an objective probability \textcite{DanzVesterlundWilson2022}. For the binarized scoring rule, reporting the beliefs truthfully only maximizes the probability of winning the prize. Reporting the actual confidence does not maximize the subject's payoff.\footnote{For the quadratic scoring rule, there are multiple payoffs depending on how close the subject's response is to the actual value. Construction of an equivalent lottery in the second-stage is difficult. It is also not robust to risk preferences.}

Our confidence elicitation is incentive-compatible for standard expected utility maximizers but also for some non-expected utility maximizers. Let $U(m, q)$ be the utility of a simple gamble that obtains monetary payoff $m$ with probability $q$.\footnote{In the case of expected utility theory, $U(m, q) = q \cdot u(m) + (1-q) \cdot u(0)$ for some function $u$ that expresses the utility of each respective payoff.} For our confidence elicitation method to be incentive-compatible, we require the three following assumptions. First, the agent only has preferences over monetary payoffs and the probabilities of winning. For example, we assume that the agent does not have a preference to earn money for the sake of merit-based self-signaling \parencite[\textit{e.g.},][]{BenabouTirole2006}, wherein the agent will report higher confidence. Second, we assume utility is strictly increasing in $q$. This assumption allows us to interpret $q^*$ as the confidence, and the strict condition ensures that the agent is not indifferent between two different values of $q^*$.  Our third and most substantial assumption is that the agent is ambiguity neutral. The subjective gamble over one’s stated confidence in the belief elicitation task relies on subjective probabilities, whereas the alternative gamble is based on an objective probability of winning \$3.  When an agent is ambiguity averse (seeking), the agent will report a lower (higher) $q^*$ than the agent's true confidence.  In \cref{app:incentive}, we show that appending the confidence elicitation to the first-stage belief elicitation does not distort the incentive for subjects to report the belief they are most confident in during the first stage.

In summary, our confidence elicitation method consists of two stages that can be used for belief elicitation in any setting with a verifiable truth.  In the first stage, each subject states their belief (\textit{i.e.}, a probability report) or makes a guess that is incentivized.  In the second stage, each subject is offered a choice of sticking to their reported probability or a payoff-equivalent simple gamble that obtains $q$ of the time.  Instead of using alternatives such as a quadratic or a binary scoring rule \parencite[see][]{Brier1950, HossainOkui2013}, we choose our particular incentive-compatible mechanism for its intuitive appeal and ease of comprehension.  We simply propose paying a subject if their guess is within a reasonably small interval of a correct value because a subject can readily recognize that truthful revelation of their confidence is in their best interest \parencite{DanzVesterlundWilson2022}. 

\subsection{Session Summary and Implementation Details} \label{sec:design-session}

We recruited 118 subjects for sessions held in April 2024 at the University of California, Santa Barbara.\footnote{The UCSB Human Subjects Committee exempted our protocol 69-23-0349. All subjects gave informed consent. We implemented the experiment in Qualtrics and recruited subjects with ORSEE \parencite{Greiner2015}.} We paid each subject a \$7 show-up fee. The session duration was one hour, with $\$17.70$ in average earnings. Of the $118$ subjects, $70$ ($59.3\%$) identified as female and 21 years was the average age of all subjects.  At the beginning of each session, we presented detailed instructions using slides, subjects were encouraged to ask questions as the instructions were being read, and they retained hard-copy instructions for reference.

After we read the instructions to subjects, they respond to four comprehension check questions and finally three practice tasks (with Low Confidence treatment) to familiarize themselves with the interface and the task. Later, when subjects begin the High Confidence treatment, they complete one practice question before the paid tasks.

For the main experiment, each subject completes twenty-two tasks; the eleven in each treatment are presented in a random order. We present the tasks in random order to ensure independence; eliciting a subject's belief for both signal realizations on the same page or consecutively could affect the independence of the subject's responses. Each of these tasks involves four probability reports: a prior belief, an updated belief, and a confidence elicitation for each of these prior and updated beliefs. Subjects simply type an integer between 0 and 100 in a text box to represent a percentage.  For each probability report, we randomly select one of the twenty-two tasks for payment. We pay each subject for four of their responses, one for each type of probability report.\footnote{More precisely, we first randomly select the prior and treatment of the task. Based on the prior and signal accuracy of task, we draw the project type and a signal realization. We then use the subject's response for the drawn prior and signal realization to determine the payoff. We follow this process four times, one for each of the four types of probability reports.} Because these four draws are independent, the probability reports that we pay generally correspond to different tasks. We did this to avoid any possible hedging behavior within each task. Following the updating tasks, subjects complete a demographic questionnaire.

\subsection{Theory: Confidence and Bayesian Updating} \label{sec:theory}

We now provide a model that corresponds to our experiment. Consider a binary state space $\Omega \coloneqq \{S, F\}$ that corresponds to the randomly-selected project being either a success or a failure. 

In our experiment the subject views a grid in which the number of white squares correspond to the prior that a randomly selected project is a success. Our subjects should believe that the number of successful project on the grid is an integer value between 0 and 100, and the subject would have a belief over the number of successful projects shown in the grid. More formally, we have an agent who is uncertain about their prior belief regarding the proportion of successful projects, $\pi_0 \coloneqq P(S) \in[0,1]$. Let us assume that the agent considers a finite set $\Pi_0 \subset [0,1]$ of $N$ possible priors $\pi_{0,i}$. The agent has a belief distribution over $\Pi_0$ (second-order belief), assigning subjective probability mass $k_{0,i}$ to prior $\pi_{0,i}$ (first-order belief) for all $i \in \{1,\hdots,N\}$. Ultimately the agent's prior belief $\pi_0$ about the project being a success is a weighted average of prior beliefs
\begin{equation*}
    \pi_0 \coloneqq \sum^N_{i=1} k_{0,i} \pi_{0,i}.
\end{equation*}

The agent receives an informative signal $\sigma \in \{\sigma_P, \sigma_N\}$. Define the conditional probability of observing signal $s$ conditioned on the project being a success or a failure as $P(\sigma|S)$ and $P(\sigma|F)$. The posterior belief is also a weighted average of posteriors. To obtain this posterior, the agent must update each prior $\pi_{0,i}$ to a posterior $\pi_{1,i}$ \emph{as well as} update each respective prior weight $k_{0,i}$ to a posterior weight $k_{1,i}$. Intuitively, the signal provides information about both the project type and the relative likelihood of the priors. However, for a Bayesian agent, updating with the average prior belief is equivalent to the aforementioned procedure.\footnote{This result is well established in statistics. We provide a more general proof in \cref{proofs} with finitely many states.}

This result is slightly counter-intuitive because the updating process is comprised of two countervailing effects. First is the convexity or concavity of the Bayesian updating function of each individual prior. In our task, when a positive (negative) signal is observed, the function is concave (convex). If the weights over the prior beliefs are held fixed, less updating occurs compared to updating given an average prior belief. Second, after a signal is observed, the Bayesian agent will place more weight on those prior beliefs which have been revealed to be more likely,  in turn resulting in a greater degree of updating. These two countervailing effects cancel each other out exactly, providing the result that confidence over multiple prior does not affect the agent's average posterior belief.

This property is not exclusive to Bayesian updating. In \cref{proofs}, we identify a class of non-Bayesian updating rules that share this property. This class of non-Bayesian updating rules has two features: First, the updating of the first-order belief allows for any distortion on the conditional probability; this distorted conditional probability may then be updated in a Bayesian manner.\footnote{An example of an updating rule that satisfies this condition is the \textcite{Grether1980} model with only power distortions on the conditional probability. In contrast, we allow for more types of distortion on the conditional probability than just the power function. Note that the \textcite{Grether1980} model with power distortion on the prior does not satisfy this condition.} Second, the updating of the second-order beliefs can be Bayesian; the agent uses the distorted perceived probability of the realization of the signal. This distortion is caused by the distortion on the conditional probability in updating the first-order belief. If we reject this property in our experiment, we are rejecting this class of non-Bayesian updating rules as well.

\subsection{A Measure of Over-updating}

Our study's primary outcome of interest is over- or under-updating.  We follow \gentextcite['s][][]{BaBohrenImas2023} approach and define two measures of excess updating relative to the Bayesian benchmark. \textcite{BaBohrenImas2023} refers to these as \emph{over-} and \emph{under-reaction}. We prefer \emph{over-} and \emph{under-update} and reserve \emph{reaction} for inference or consequent action.  Over-updating can arise from under-weighting of the prior or over-reacting to a signal; these two updating biases are modeled differently under the \textcite{Grether1980} model, each having a different implication. 

We let $\pi_0$ be a subject's reported prior beliefs, $\pi_1$ be the subject's reported updated beliefs, and $\pi^\mathrm{Bayes}_1$ be the subject's corresponding Bayesian posterior belief given the prior they previously reported. Our subject may not have an accurate perception of the actual prior, especially in the Low Confidence treatment, this makes it necessary to compare their updated beliefs against the Bayesian benchmark of their reported prior. 

Recall that our belief elicitation method incentivizes subjects to report the point estimate that has the largest probability mass within three percentage points interval, and not the average prior belief. In order to estimate $\pi^\mathrm{Bayes}_1$ from the subject's prior belief in the experiment, we have to assume that the prior that subjects are reporting is approximately the average prior belief. We attempt to verify this assumption by looking at the distribution of the reported prior in the Low Confidence treatment. 

\begin{table}[h!]
\centering
\begin{tabular}{cccc}
\toprule
Actual Prior & \multicolumn{2}{c}{Mean Prior} & Theoretical Prediction \\
\cmidrule{2-3}
& Full Sample & Restricted Sample & \\
\midrule
0   & 2.5  & 0.05 & 0, 1, 2, 3  \\
20  & 21.9 & 16.2 & 17 \\
40  & 41.1 & 39.3 & 42 \\
50  & 51.1 & 50.5 & 47 \\
70  & 73.6 & 74.9 & 77 \\
90  & 86.0 & 91.2 & 92 \\
\bottomrule
\end{tabular}
\caption{Mean Reported Prior From Experiment and Theoretical Predictions \\ \footnotesize \textit{Notes:} The restricted sample drops responses that are more than twenty percentage points from the actual prior. The theoretical prediction is the point that has the most density within a three-percentage-point interval; this will be the point reported if the second-order belief follows this distribution. The theoretical prediction is not unique for an actual prior of zero.}
\label{tab:prior_comparison}
\end{table}

\Cref{tab:prior_comparison} presents the average reported prior in our experimental data and the theoretical prediction of the reported prior, given our elicitation method.\footnote{Section S2.1 of the Supplementary Material offers a plot of the kernel density estimate of the reported priors in the Low Confidence treatment.} We can see that the average prior belief and the point that has the largest probability mass in our sample's reported prior is no more than six percentage points different from the theoretical prediction. When we drop responses that are off by twenty percentage points, the average prior is even closer to the theoretical prediction. If the subject's second-order belief is similar to the sample's distribution of reported prior, their reported belief would be approximately close to the average belief for us to compute the average Bayesian posterior based on subject's reported belief. In the High Confidence treatment, most subjects' belief are extremely close to the actual prior.

Our first measure captures the magnitude of the over-update in percentage points:
\begin{equation}
  \textrm{over-update} \coloneqq
    \begin{cases} 
      \pi_1 - \pi_1^\mathrm{Bayes} \quad & \text{if signal is positive, and} \\
      \pi_1^\mathrm{Bayes} - \pi_1 \quad & \text{if signal is negative.} 
    \end{cases}
\end{equation}
When a positive signal is observed, we expect the subject to update upwards, that is, the agent over-updates if the subject's updated belief $\pi_1$ is greater than the subject's Bayesian belief $\pi_1^\text{Bayes}$, resulting in $\textrm{over-update}>0$. Similarly, in the event of a negative signal, if the subject's updated belief $\pi_1$ is less than the subject's Bayesian belief $\pi_1^\text{Bayes}$, this implies $\textrm{over-update}>0$.  Thus over-update is positive when the subject's belief moves too far in the correct direction and negative when it moves too little in the correct direction. This measure is also negative if a subject updates in the wrong direction. We also construct a metric to capture the magnitude of an over-update relative to the magnitude of a Bayesian update: \footnote{Notice that an over-update of ten percentage points has different implications if the magnitude of the Bayesian update is five percentage points compared to forty percentage points. The former represents a relatively large updating error, while the latter represents a relatively small error.}
\begin{equation}
    \textrm{over-update-ratio} \coloneqq \frac{\textrm{over-update}}{\left|\pi_0 - \pi_1^\mathrm{Bayes}\right|} . 
\end{equation}
It is important to note that updating in the wrong direction results in $\textrm{over-update}<0$ and $\textrm{over-update-ratio}<-1$.  We address this empirically with a robustness check that excludes observations with updates in the wrong direction (about $27\%$ of our observations).  We find similar results, so we present results from the full data.

\subsection{Estimating the Grether (1980) Model} \label{sec:model-grether}

The \textcite{Grether1980} model is a generalized version of Bayes' rule that accommodates several updating biases. \textcite{chan2025} presents an axiomatic characterization of the model and conducts an experiment that empirically validates this model. In our setting,
\begin{equation*}
\pi_1(\sigma) = \frac{P(\sigma|S)^\alpha \pi_0^\beta}{P(\sigma|S)^\alpha \pi_0^\beta + \mathrm{P}(\sigma|F)^\alpha (1-\pi_0)^\beta},
\end{equation*}
where $\alpha \geq 0$ is the weight that the agent places on the signal and $\beta \geq 0$ is the weight on the prior. Thus $\alpha<1$ represents an under-reaction to a signal and $\alpha>1$ an over-reaction.  Meanwhile $\beta<1$ represents base-rate neglect and $\beta>1$ over-weighting the prior.

We can rewrite this model as an odds ratio,
\begin{equation*}
\frac{\pi_1(\sigma)}{1-\pi_1(\sigma)} = \left(\frac{P(\sigma|S)}{P(\sigma|F)} \right)^\alpha  \left( \frac{\pi_0}{1-\pi_0} \right)^\beta.
\end{equation*}
This formulation is widely used in the analysis of experimental data because in logarithms one obtains a model whose parameters can be estimated with a linear regression:
\begin{equation*}
\ln \frac{\pi_1(\sigma)}{1-\pi_1(\sigma)} = \alpha \, \ln \frac{P(\sigma|S)}{P(\sigma|F)} + \beta \ln \frac{\pi_0}{1-\pi_0} .
\end{equation*}
Thus, the model permits estimation of the weights that subjects place on prior information and on information conveyed by the signal.

In our experiment, we are interested in the weights in the High and Low treatments. To estimate the difference we specify the following regression:

\begin{equation}
\ln \frac{\pi_1(\sigma)}{1-\pi_1(\sigma)} = \left(\gamma_0 + \gamma_2\mathbbm{1}_H\right) \, \ln \frac{P(\sigma|S)}{P(\sigma|F)} + \left(\gamma_1 + \gamma_3\mathbbm{1}_H\right) \ln \frac{\pi_0}{1-\pi_0} + \varepsilon \label{eq:grether-reg-spec}
\end{equation}
where $\mathbbm{1}_H$ is an indicator variable for the High Confidence treatment. We use this specification to estimate $\alpha_T$ and $\beta_T$ for each treatment $T \in \{L,H\}$:
\begin{equation}
\alpha_L = \gamma_0, \qquad \beta_L = \gamma_1, \qquad \alpha_H = \gamma_0 + \gamma_1, \quad \text{ and } \quad \beta_H = \gamma_3 + \gamma_4. \label{eq:grether-reg-recovery}
\end{equation}

While previous studies provide subjects with explicit numerical priors, our subjects must report their prior beliefs after seeing the grid, thus measurement error may be a concern. To address this, we use the actual prior as an instrument for the prior log-odds ratio.\footnote{The instrumental variable regression was also used by \textcite{MobiusNiederleNiehausRosenblat2022}. Their subjects faced cognitive tasks of varying difficulty, regarding which subjects then reported beliefs about their performance. The authors use the task difficulty as an instrumental variable for their log ratio of prior beliefs.}

\subsection{Hypotheses}

Our primary research question asks how confidence over multiple priors affects how one updates their beliefs. While standard economic theory predicts that confidence over multiple priors should not matter, we hypothesize otherwise. 

\begin{hypothesis}
The over-update and over-update-ratio measures are larger for the Low Confidence treatment relative to the High Confidence treatment. \label{hypothesis1}
\end{hypothesis}

\begin{hypothesis}
Greater reported confidence (over multiple priors) is associated with lower over-update and over-update-ratio measures. \label{hypothesis2}
\end{hypothesis}

\begin{hypothesis}
Greater weight is placed on prior beliefs in the High Confidence treatment relative to the Low Confidence treatment. Specifically, in our \textcite{Grether1980} regression \cref{eq:grether-reg-spec,eq:grether-reg-recovery}, $\beta_H>\beta_L$. \label{hypothesis3}
\end{hypothesis}

\section{Results} \label{sec:results}

\subsection{Experimental Design Validation}

We first verify that the subjects understood the experiment and that our method of displaying the grid for a short duration achieved its intended manipulation.

\subsubsection{Comprehension of Tasks}

Ensuring subject comprehension of our environment is important given our confidence-in-stated beliefs elicitation and general concerns about the comprehension of the BDM \parencite{CasonPlott2014}. To test subjects' understanding of our confidence elicitation procedure, we include a task with a degenerate prior (\textit{i.e.}, with zero successful projects) as represented by a grid with only black squares.  One can readily discern that all squares are black, even when the grid is only displayed for a quarter of a second, as in the Low Confidence treatment.  In this task, 115 subjects (97.5\%) reported the prior correctly, suggesting that they understood the task of reporting the prior.\footnote{That is, three out of 118 total subjects did not report 3\% or less when asked about the probability of a successful project selected from the grid. Two subjects erred only in the Low Confidence treatment, while one subject erred in both treatments.} Next, if a subject understands the confidence elicitation method, they should report confidence of 100\% when the grid contains all black squares, assuming the subject is indeed confident. Fourteen subjects (11.9\%) failed to provide such reports.\footnote{None of these fourteen subjects provided an incorrect prior belief. Four erred only in the Low Confidence treatment, seven only in High Confidence, and three subjects erred in both treatments. Among the eleven subjects who erred in only one treatment, seven were 99\% confident and ten were at least 95\% confident.}

In total, each subject completed four comprehension checks, reporting a prior and confidence for both Low and High treatments. Thirteen subjects made one error in total, four subjects made two, while the remaining 101 subjects (85.6\%) made no errors.  Our subjects thus generally demonstrate a good understanding of the confidence elicitation task. Section S2.2 of the Supplementary Material offers qualitatively similar results for the subsample of subjects who passed these comprehension checks. We proceed to analyze and present results using the full sample.

\begin{figure}
    \centering
    \includegraphics[trim=0.1in 0.1in 0.1in 0.1in]{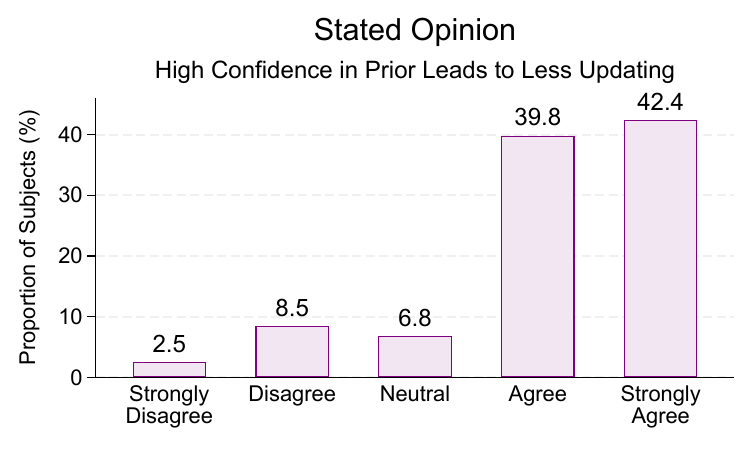}
    \caption{Self-reported relationship between confidence in prior and updating}
    \label{fig:opinion}
\end{figure}

Upon completion of the tasks, we ask subjects to state their agreement (on a Likert scale) with the following: ``The more confident that I am in my initial belief about the proportion of successful projects, the less I should respond to the outcome of the computer test result.''\footnote{This is in a similar spirit to \textcite{DellavignaPope2018}, who solicited predictions from academic experts on their experiment.}  Our subjects largely agreed with our overarching hypothesis: 97 of 118 (82.2\%) subjects agree that when they have higher confidence, they should update less (see \cref{fig:opinion}).

\subsubsection{Accuracy of Priors}

Our design induces a prior belief by showing subjects a grid for a limited amount of time. As a result, subjects may have incorrect prior beliefs. We now present results showing the accuracy of subjects' prior beliefs across the two treatments.

\begin{figure}
    \centering
    \includegraphics{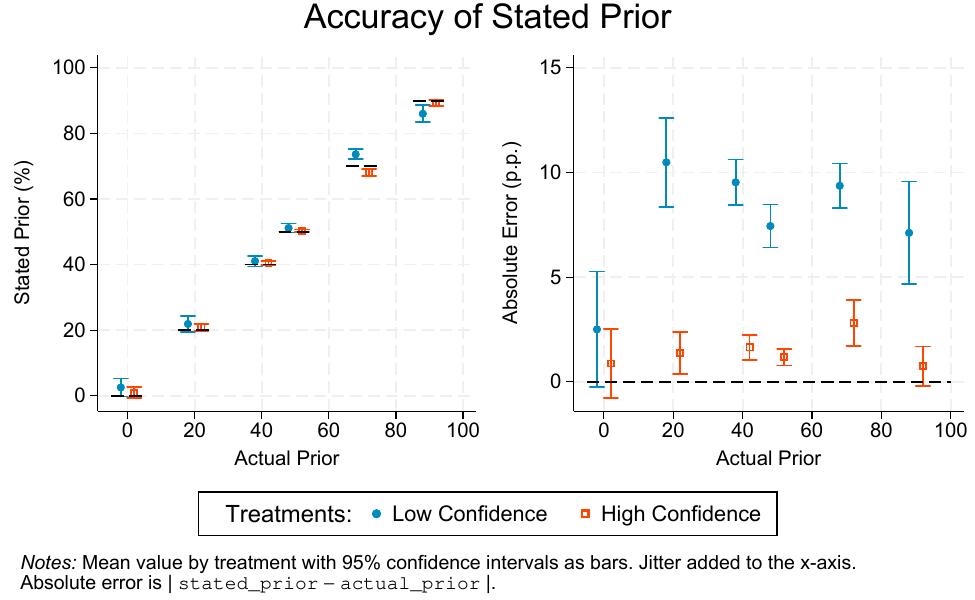}
    \caption{Stated prior accuracy by treatment}
    \label{fig:errors}
\end{figure}

As shown in the left panel of \cref{fig:errors}, at the aggregate level subjects are surprisingly accurate in stating the actual prior even in the Low Confidence treatment. The average reported prior from our subjects is close to the actual prior from the experiment. In the right panel, we plot the absolute deviation in subjects’ stated priors from the actual prior. As expected, in the Low Confidence treatment, we see that at the individual level, subjects have highly inaccurate prior beliefs. In the High Confidence treatment, errors may result from the miscounting of squares or from the incentives not requiring perfect precision in the stated prior belief. Regardless these errors should be smaller in the High Confidence treatment, which is exactly what the data show. Next, displaying the grid for only 0.25 seconds in the Low treatment gives subjects on average a noisier perception of the actual prior relative to the High treatment; we see wider confidence intervals for the Low Confidence treatment in the data, as shown in the right panel of \cref{fig:errors}.

\subsubsection{Confidence over Multiple Priors Across Treatments}

\begin{figure}
    \centering
    \includegraphics{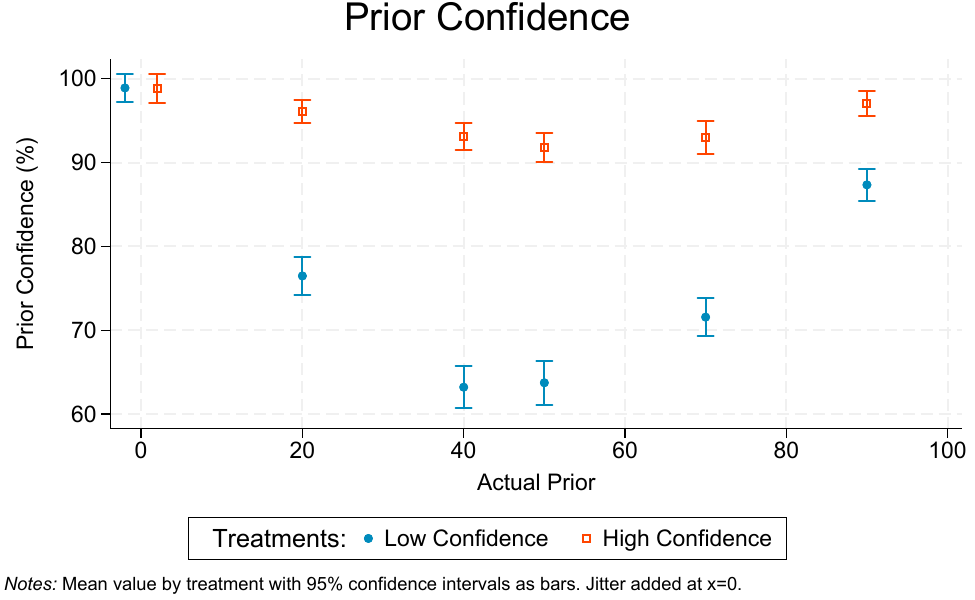}
    \caption{Confidence over Multiple Priors by treatment}
    \label{fig:confidence}
\end{figure}

Our design varies the display time of grids to induce different confidence levels in a subject's prior. We find that subjects’ confidence in their perceived (and hence stated) priors vary in an expected direction by treatment, shown in \cref{fig:confidence}, which demonstrates that subjects are less confident in the Low Confidence treatment compared to the High Confidence treatment.

\begin{customre}{0}
    Subjects in the Low Confidence treatment expressed lower confidence in their prior compared to the High Confidence treatment.
\end{customre}

\subsection{Over-updating}

\begin{figure}
    \centering
    \includegraphics{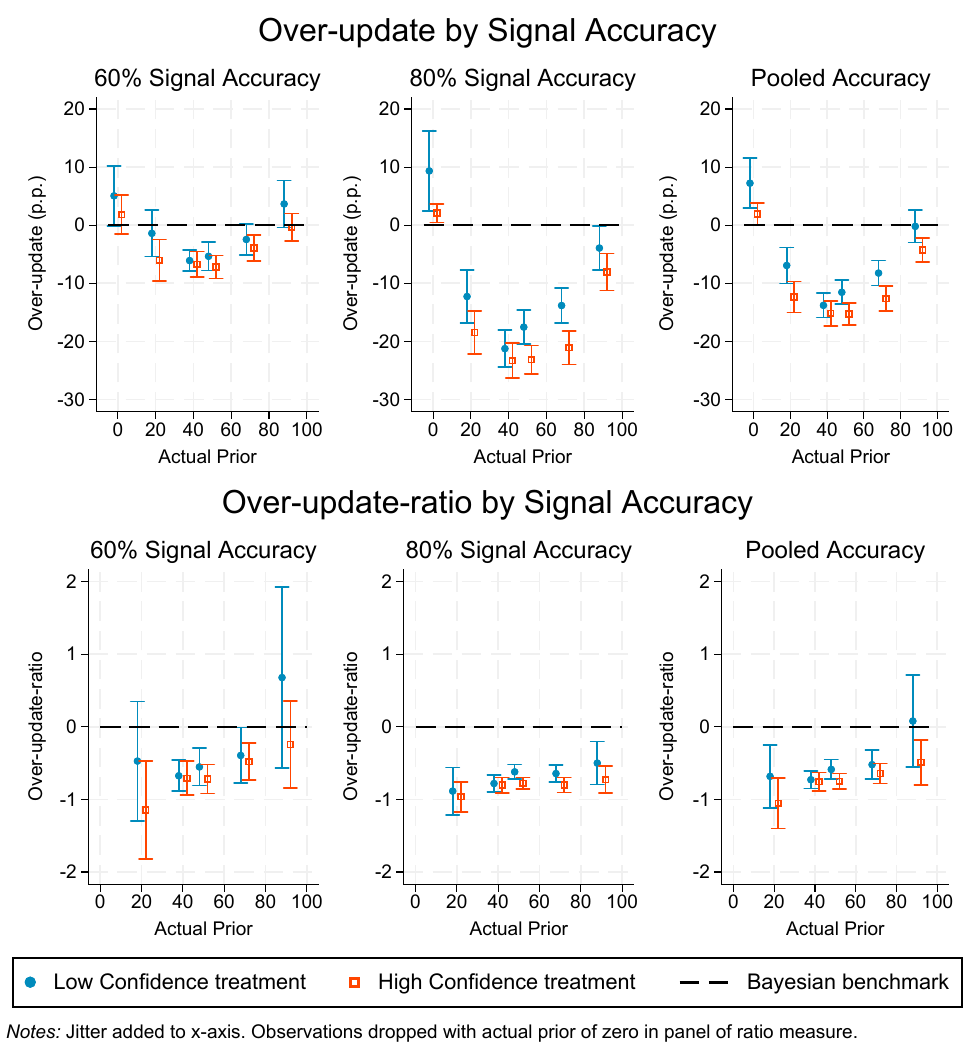}
    \caption{Mean over-update and over-update-ratio by signal accuracy and treatment}
    \label{fig:over-update}
\end{figure}

We plot both of our over-updating measures (over-update and over-update-ratio) in \cref{fig:over-update}.\footnote{We drop the task in which the prior is degenerate because it involves no updating and our over-update-ratio is not well-defined.} The first result is that subjects under-update relative to the Bayesian benchmark. This is consistent with the results from \textcite{BaBohrenImas2023}, who found under-updating relative to the Bayesian benchmark in updating tasks with only two states.\footnote{\textcite{BaBohrenImas2023} experiment was a ``balls and urn'' framing, while ours is the \textcite{KahnemanTversky1972a} belief updating problem. This validates their finding across a different framing of belief-updating tasks.}

\begin{result}
Subjects under-update relative to the Bayesian benchmark in both treatments. 
\end{result}

This result may seem inconsistent with base-rate neglect, which predicts over-updating in our experiment and is a main finding in this framing of the belief-updating problem \parencite{KahnemanTversky1972a, EspondaVespaYuksel2024}. Our experiment differs from the aforementioned studies with regard to the experimental parameters. \textcite{KahnemanTversky1972a} and \textcite{EspondaVespaYuksel2024} only use a single set of parameters---a prior of $85\%$ and a signal accuracy of $80\%$---while we use a broader set of experimental parameters. Our result is consistent with the broader belief updating literature which finds that conservatism bias and base-rate neglect are the primary biases in belief-updating experiments. We structurally estimate the \textcite{Grether1980} model parameter in \cref{sec:results-grether} and show our subjects exhibit both biases. We find that conservatism bias is stronger than base-rate neglect which explains the under-updating pattern.\footnote{\textcite{Benjamin2019} conjectures that conservatism bias is likely to be more dominant when the prior is close to 50-50. In fact, we find the most under-updating when priors are close to 50-50.} 

\Cref{fig:over-update} presents how our over-update and over-update-ratio measures vary by treatment, actual prior, and signal accuracy. Overall the data indicate that the average over-update and over-update-ratio measures are larger in the Low Confidence treatment for every actual prior value. The middle panel for the 80\% signal accuracy shows a similar pattern, but with a more pronounced treatment effect.\footnote{We also plot the CDF of the subject average for the over-update and over-update-ratio by treatment, presented in Section S1 of the Supplementary Material. We find that the distribution of the over-update in the Low Confidence treatment first-order stochastically dominates the distribution of the over-update in the High Confidence treatment. Regarding the over-update-ratio, the Low Confidence treatment is more likely to have larger values of this measure.}

\Cref{tbl:overupdate-ols} presents linear regressions for over-update and over-update-ratio, including specifications both with and without subject fixed-effects. The independent variables include an indicator for the High Confidence treatment and a constant term. On average we see that High Confidence treatment decreases the over-update measure by 3.8 percentage points ($p<0.001$) and decreases the over-update-ratio by 0.25 units ($p=0.004$).

\begin{table}
\centering
\def\sym#1{\ifmmode^{#1}\else\(^{#1}\)\fi}
\begin{threeparttable}
\caption{Ordinary least squares regressions of over-update measures \label{tbl:overupdate-ols}}
\begin{tabular}{l*{4}{d{4.6}}}
\toprule
&\multicolumn{4}{c}{Dependent variable}\\
\cmidrule{2-5}
&\multicolumn{2}{c}{over-update}&\multicolumn{2}{c}{over-update-ratio}\\
\cmidrule(r){2-3}\cmidrule(l){4-5}
                    &\multicolumn{1}{c}{(1)}&\multicolumn{1}{c}{(2)}&\multicolumn{1}{c}{(3)}&\multicolumn{1}{c}{(4)}\\
\midrule
High Confidence treatment&      -3.804&      -3.804&      -0.251&      -0.252\\
                    &     (0.585)&     (0.585)&     (0.085)&     (0.085)\\
Constant            &      -8.139&      -8.139&      -0.488&      -0.488\\
                    &     (0.876)&     (0.292)&     (0.088)&     (0.042)\\
\midrule
Subject fixed-effects&\multicolumn{1}{c}{No}&\multicolumn{1}{c}{Yes}&\multicolumn{1}{c}{No}&\multicolumn{1}{c}{Yes}\\
\(R^2\)             &       0.010&       0.013&       0.003&       0.003\\
Subjects            &\multicolumn{1}{c}{\(118\)}&\multicolumn{1}{c}{\(118\)}&\multicolumn{1}{c}{\(118\)}&\multicolumn{1}{c}{\(118\)}\\
Observations        &\multicolumn{1}{c}{\(2360\)}&\multicolumn{1}{c}{\(2360\)}&\multicolumn{1}{c}{\(2359\)}&\multicolumn{1}{c}{\(2359\)}\\
\bottomrule
\end{tabular}
\begin{tablenotes}[para,flushleft] \footnotesize \item \emph{Notes:} Standard errors (in parentheses) are clustered at the subject level. Observations with undefined log-ratio dropped.
\end{tablenotes}
\end{threeparttable}
\end{table}

We also perform a non-parametric Wilcoxon signed-rank test to determine whether the over-update and over-update-ratio differ between the High and Low Confidence treatments. We find that the distribution of each measure differs between treatments ($p<0.0001$ for each).

\begin{result}
In the High Confidence treatment subjects exhibit more under-updating relative to the Low Confidence treatment.
\end{result}

This result is consistent with \cref{hypothesis1}. Interestingly, given a noisier prior belief, subjects' beliefs are closer to the Bayesian benchmark.\footnote{This is likely due to noisier priors canceling out some of the other updating biases.} Finally, given that we find more under-updating in the High Confidence treatment, we ask if this effect is driven by confidence over multiple priors; we investigate this question next.

\subsubsection{An Instrumental Variables Regression}

We are primarily interested in how confidence over multiple priors affects the degree of over-updating. However, endogeneity is a concern between our over-updating measures and confidence measures. \Cref{fig:confidence} shows that prior confidence varies with actual priors. Accurate perception of a prior is more difficult for priors closer to 50\%, which naturally results in lower confidence. Actual priors are also correlated with the over-update and over-update-ratio measures because updated beliefs are a function of priors. We resolve this endogeneity problem by using the treatment as an instrument in a generalized two-stage least squares regression with subject fixed-effects, as presented in \cref{tbl:overupdate-iv}. The $F$-statistic for the first-stage regression is 320, indicating that our treatment is a strong instrument for our confidence measure \parencite{StockYogo2005}.

\begin{table}
\centering
\def\sym#1{\ifmmode^{#1}\else\(^{#1}\)\fi}
\begin{threeparttable}
\caption{OLS and instrumental variable regressions of over-update measures \label{tbl:overupdate-iv}}
\begin{tabular}{l*{4}{d{4.5}}}
\toprule
&\multicolumn{4}{c}{Dependent variable}\\
\cmidrule{2-5}
&\multicolumn{2}{c}{over-update}&\multicolumn{2}{c}{over-update-ratio}\\
\cmidrule(r){2-3}\cmidrule(l){4-5}
                    &\multicolumn{1}{c}{(1)}&\multicolumn{1}{c}{(2)}&\multicolumn{1}{c}{(3)}&\multicolumn{1}{c}{(4)}\\
                    &\multicolumn{1}{c}{OLS}&\multicolumn{1}{c}{FE2SLS\(^\dag\)}&\multicolumn{1}{c}{OLS}&\multicolumn{1}{c}{FE2SLS\(^\dag\)}\\
\midrule
Prior confidence, \(q^*\)\qquad\qquad&       0.020&      -0.175&       0.001&      -0.012\\
                    &     (0.021)&     (0.030)&     (0.002)&     (0.004)\\
Constant            &     -11.670&       4.537&      -0.717&       0.349\\
                    &     (1.723)&     (2.470)&     (0.174)&     (0.340)\\
\midrule
First-stage \(F\)-stat&            &\multicolumn{1}{c}{\(320.36\)}&            &\multicolumn{1}{c}{\(320.36\)}\\
Subjects            &\multicolumn{1}{c}{\(118\)}&\multicolumn{1}{c}{\(118\)}&\multicolumn{1}{c}{\(118\)}&\multicolumn{1}{c}{\(118\)}\\
Observations        &\multicolumn{1}{c}{\(2360\)}&\multicolumn{1}{c}{\(2360\)}&\multicolumn{1}{c}{\(2359\)}&\multicolumn{1}{c}{\(2359\)}\\
\bottomrule
\end{tabular}
\begin{tablenotes}[para,flushleft] \footnotesize \item \emph{Notes:} Standard errors (in parentheses) are clustered at the subject level. Subject fixed-effects included. Observations with undefined log-ratio dropped. Prior confidence \(q^*\) is measured in percentage points (between 0 and 100). \item \tnote{\dag}Fixed-effect two-stage least squares (FE2SLS) regressions use an indicator of High Confidence treatment as the instrument.
\end{tablenotes}
\end{threeparttable}
\end{table}

The results are consistent with \cref{hypothesis2}.\footnote{Columns 1 and 3 of \cref{tbl:overupdate-iv} report positive and non-significant coefficient estimates for prior confidence (the probability report) using an ordinary least squares regression.} Columns 2 and 4 show a statistically significant negative coefficient with the instrumental variable. We find that a percentage-point increase in prior confidence reduces the over-update measure by 0.175 percentage points ($p<0.001$) and the over-update-ratio by 0.012 units ($p=0.005$).

\begin{result}
Greater under-updating in the High Confidence treatment is due to greater confidence over multiple priors.
\end{result}

\subsection{Grether Model Estimation Results} \label{sec:results-grether}

A possible mechanism for our results is that subjects adjust the weights they place on their prior beliefs. This could also be interpreted as salience or attention weights \parencite{BordaloConlonGennaioliKwonShleifer2023}. We estimate the parameters of the \textcite{Grether1980} model using the IV regression specification in \cref{eq:grether-reg-spec,eq:grether-reg-recovery} to study how our subjects respond to our treatments. We employ the actual prior as the instrument for the log-prior-ratio to account for measurement error in subjects' reported prior beliefs.\footnote{A subject may misreport their latent prior belief yet use the latent belief for the updated belief.} The $F$-statistic for the first stages are all sufficiently large ($>600$), validating that the actual prior is a strong instrument for the log-prior-ratio. \Cref{tbl:grether-iv} presents results for signal accuracy of 60\% (column 1), 80\% (column 2), both these pooled (column 3), and pooled with subject fixed-effects (column 4).\footnote{Section S2.3 of the Supplementary Material presents the standard Grether regression using ordinary least squares.}

\begin{table}
\centering
\def\sym#1{\ifmmode^{#1}\else\(^{#1}\)\fi}
\begin{threeparttable}
\caption{Grether model TSLS regressions of log updated belief ratio \label{tbl:grether-iv}}
\begin{tabular}{l*{4}{d{4.6}}}
\toprule
&\multicolumn{4}{c}{Signal accuracy}\\
\cmidrule{2-5}
&\multicolumn{1}{c}{60\%}&\multicolumn{1}{c}{80\%}&\multicolumn{2}{c}{Pooled}\\
\cmidrule(r){2-2}\cmidrule(r){3-3}\cmidrule{4-5}
                    &\multicolumn{1}{c}{(1)}&\multicolumn{1}{c}{(2)}&\multicolumn{1}{c}{(3)}&\multicolumn{1}{c}{(4)}\\
\midrule
Reduced-form regression:                &            &            &            &            \\
\quad \(\alpha_L\)  &       0.634&       0.326&       0.349&       0.354\\
                    &     (0.113)&     (0.042)&     (0.040)&     (0.041)\\
\quad \(\beta_L\)   &       0.751&       0.774&       0.763&       0.767\\
                    &     (0.062)&     (0.058)&     (0.043)&     (0.042)\\
\quad \((\alpha_H-\alpha_L)\)&      -0.247&      -0.101&      -0.111&      -0.119\\
                    &     (0.108)&     (0.032)&     (0.030)&     (0.030)\\
\quad \((\beta_H-\beta_L)\)&       0.147&       0.080&       0.113&       0.108\\
                    &     (0.078)&     (0.067)&     (0.051)&     (0.051)\\
\midrule
Linear combinations:             &            &            &            &            \\
\quad \(\alpha_H\)  &       0.388&       0.224&       0.237&       0.235\\
                    &     (0.104)&     (0.041)&     (0.039)&     (0.038)\\
\quad \(\beta_H\)   &       0.898&       0.854&       0.876&       0.875\\
                    &     (0.039)&     (0.053)&     (0.033)&     (0.033)\\
\midrule
\(p\)-value of \(F\)-test:&&\\ \quad\(\alpha_L = \beta_L = 1\)&       <0.001&       <0.001&       <0.001&       <0.001\\
\quad\(\alpha_H = \beta_H = 1\)&       <0.001&       <0.001&       <0.001&       <0.001\\
\quad\(\alpha_H=\alpha_L\)&       0.023&       0.001&       <0.001&       <0.001\\
\quad\(\beta_H=\beta_L\)&       0.059&       0.231&       0.027&       0.035\\
\midrule Subject fixed-effects&\multicolumn{1}{c}{No}&\multicolumn{1}{c}{No}&\multicolumn{1}{c}{No}&\multicolumn{1}{c}{Yes}\\
First-stage \(F\)-stat&\multicolumn{1}{c}{\(2318.53\)}&\multicolumn{1}{c}{\(614.7\)}&\multicolumn{1}{c}{\(1941.18\)}&\multicolumn{1}{c}{\(1929.85\)}\\
\(R^2\)             &       0.662&       0.567&       0.606&\multicolumn{1}{c}{\(0.608\)}\\
Subjects            &\multicolumn{1}{c}{\(58\)}&\multicolumn{1}{c}{\(60\)}&\multicolumn{1}{c}{\(118\)}&\multicolumn{1}{c}{\(118\)}\\
Observations        &\multicolumn{1}{c}{\(1025\)}&\multicolumn{1}{c}{\(1068\)}&\multicolumn{1}{c}{\(2093\)}&\multicolumn{1}{c}{\(2093\)}\\
\bottomrule
\end{tabular}
\begin{tablenotes}[para,flushleft] \footnotesize \item \emph{Notes:} Standard errors (in parentheses) are clustered at the subject level. Observations with undefined log-ratio dropped. The null hypothesis of Bayesian updating requires \(\alpha = 1\) and \(\beta = 1\). Fixed-effect two-stage least squares (FE2SLS) regressions use an indicator of High Confidence treatment as the instrument.
\end{tablenotes}
\end{threeparttable}
\end{table}

In the pooled regressions we find that the weight on prior beliefs, $\beta$, is larger by about 0.113 units in the High Confidence treatment compared to the Low treatment (Column 3, $p=0.027$), indicating that subjects place more weight on the prior in the High Confidence treatment. The signal accuracy remains fixed in our experiment, thus we would expect the weights that subjects place on the signals would remain unchanged across the treatments. However, this is not the case. We estimate that the weight on the signals, $\alpha$, is smaller by about $0.111$ units in the High Confidence treatment compared to the Low treatment (Column 3, $p<0.001$), and the adjustment of the weights on the signals is larger than the adjustment of the weight on the priors. Our subjects respond to our treatments by placing more weight on the signals.

We estimate the \textcite{Grether1980} parameters by signal accuracy in columns 1 and 2 of \cref{tbl:grether-iv}. We see that the overall direction of the results is qualitatively similar. The difference in the estimated \textcite{Grether1980} parameters is greater when the signal accuracy is lower. A possible explanation is that given a weaker signal, subjects increasingly rely on the signals more to form their updated beliefs when they have less confidence in their prior beliefs.

We also find that the weight placed on the signal ($\alpha$) is larger when the signal is weaker. Although we do not find over-reaction to the 60\%-accurate signal, our result is similar to those of \textcite{AugenblickLazarusThaler2024}, who find that subjects tend to under-infer from stronger signals. Our design differs from the balls-and-urns framing of \textcite{AugenblickLazarusThaler2024}; we thus validate these findings across different styles of belief-updating tasks.

\subsubsection{Individual-level Treatment Effects}

\begin{figure}[h!]
    \centering
    \includegraphics{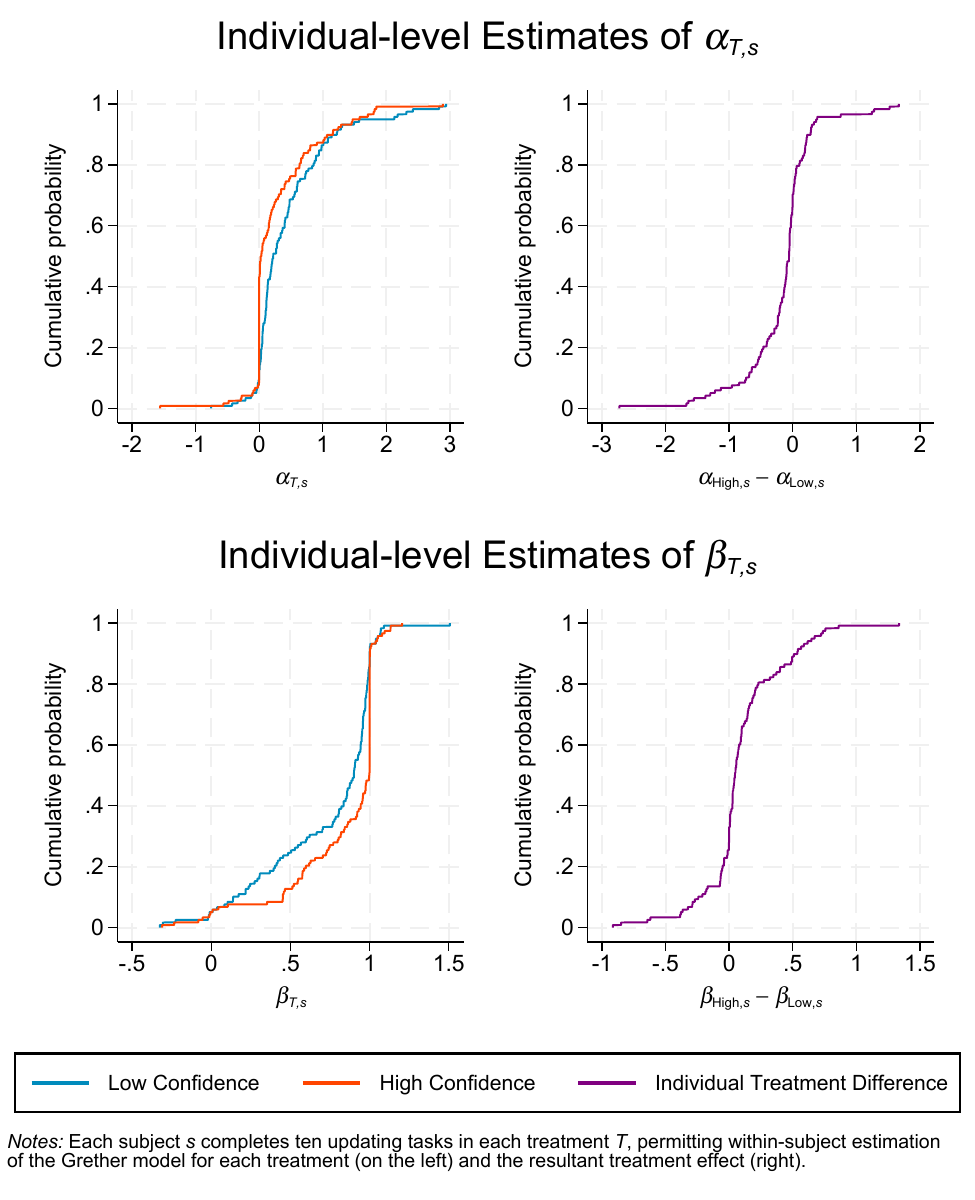}
    \caption{Subject level estimates of $\alpha$ (weight on the signal) and $\beta$ (weight on the prior) in the \textcite{Grether1980} model}
    \label{fig:grether-beta-dist}
\end{figure}

While the regression results present estimates of the \textcite{Grether1980} parameters at the aggregate level, we also offer individual-level results. The plots on the left side of \cref{fig:grether-beta-dist} depict the distribution of $\alpha_{T,s}$ and $\beta_{T,s}$ for each subject $s$ and treatment $T \in \{H,L\}$.

The CDF of $\alpha_{T,s}$ for the High Confidence treatment generally lies to the left of the corresponding CDF for the Low Confidence treatment. The CDF of $\beta_{T,s}$ for the High Confidence treatment generally lies to the right of the corresponding CDF for the Low Confidence treatment. Taken together, at the individual level, we generally observe  larger values of $\beta$ and smaller values of $\alpha$ with High Confidence treatment. Further, these graphs depict a mass at $\alpha=0$ and $\beta=1$ in the High Confidence treatment, which corresponds to subject who do not update their beliefs in any task in the High Confidence treatment. In fact about 35\% (41 out of 118) of our subjects do not update their beliefs in the High Confidence treatment.\footnote{No updating is a well-documented modal updating pattern. The survey of \textcite{Benjamin2019} notes that about one-third to one-half of individuals do not update whatsoever. Our results echo this empirical regularity.}

We also compute the within-subject difference of $\alpha_{T,s}$ and $\beta_{T,s}$ between treatments, as shown in the right-hand side plots of \cref{fig:grether-beta-dist}. Regarding $\alpha_{T,s}$, most subjects have a negative difference; these subjects place more weight on the signal in the Low Confidence treatment than in the High Confidence treatment. Regarding $\beta_{T,s}$, most subjects have a positive difference; these subjects place more weight on the prior in the High Confidence treatment than the Low Confidence treatment.

\begin{result}
When updating, subjects place more weight on their prior and less weight on the signals in the High Confidence treatment than in the Low Confidence treatment.
\end{result}

Overall, our estimation of the \textcite{Grether1980} parameters shows that our subjects place more weight on their prior belief in the High Confidence treatment. This result is consistent with predictions of noisy cognition models \parencite{Woodford2020, EnkeGraeber2023} \footnote{\textcite{AugenblickLazarusThaler2024} \textcite{BaBohrenImas2023} also use a noisy cognition model to explain the over- and under-updating observed in their experiment} and model of salience \parencite{BordaloConlonGennaioliKwonShleifer2023}. In these models, the agent places more weight on parameters given more accurate perception or greater salience. These models would predict that our subjects place more weight on the prior and less weight on the signal in the High Confidence treatment relative to the Low Confidence treatment, which is consistent with our \textcite{Grether1980} model results.

Another non-Bayesian updating rule that could be consistent with the over-update is the maximum likelihood updating \parencite{Dempster1967,gilboaschmeidler1993}, where the agent uses the most likely prior for updating after observing the signal. In our experiment, this will be prior with the highest probability of success This model assumes Bayesian updating, we should observe over-updating in the Low confidence treatment, which we did not. However, the idea that the agent ``switches'' to the most likely prior could potentially explain the smaller magnitude of under-updating in the low confidence treatment.

\subsection{Magnitude of Update}

In our presentation of results thus far, we have only considered confidence across multiple priors. Our experiment also allows us to study our other notion of confidence, which is related to the dispersion of a belief distribution. The actual proportion of successful projects represents this notion of confidence. Uncertainty regarding the true state (a success or failure) of the selected project is greater when the proportion of successful projects is closer to 50\%.

\begin{figure}
    \centering
\includegraphics{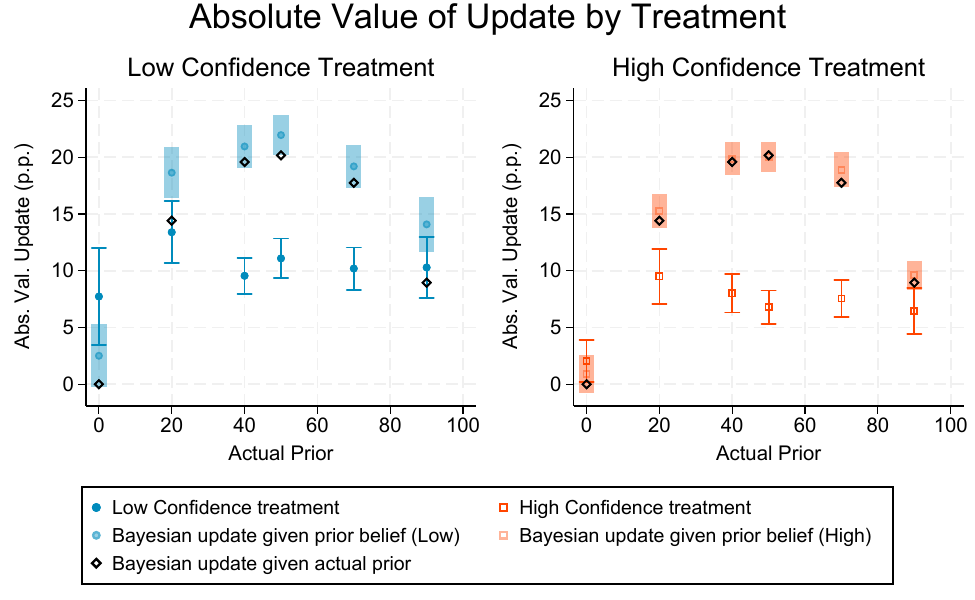}
    \caption{Mean Magnitude of Update by Treatment \\ \footnotesize \textit{Notes:} The average is the pooled data from both signal reliability.}
    \label{fig:abs-update}
\end{figure}

\Cref{fig:abs-update} depicts an inverse U-shape pattern in the Bayesian beliefs. A Bayesian agent would have the largest magnitude of update when the prior belief is close to 50\%. This pattern of Bayesian updating is similar when the Bayesian update is computed using the reported prior belief or the actual prior. Further we see that the magnitude of our subjects' updates remains similar regardless of the actual prior.\footnote{This explains the U-shaped pattern in our over-update measure in \Cref{fig:over-update}.} This pattern of updating behavior is there even when we split our sample by signal accuracy as shown in Section S1 of the Supplementary Material.

\begin{result}
    Inconsistent with Bayesian updating, subjects update by the same magnitude (in p.p.) regardless of the actual prior. 
\end{result}

This result is consistent with base-rate neglect, which has been widely documented in the literature, and of which our experiment also finds evidence. In the case of perfect base-rate neglect, we should expect a U-shape pattern, with the least updating at the prior of 50\%. However, in our experiment we see that the magnitude of the update is insensitive to the prior belief. Our subjects' behavior is consistent with a heuristic in which one updates beliefs by a fixed amount regardless of the prior beliefs. Our data show that people respond to confidence over multiple priors and that they do not respond to the confidence in a single belief distribution. Both of these results are at odds with extant theory.

\begin{figure}
    \centering
    \includegraphics{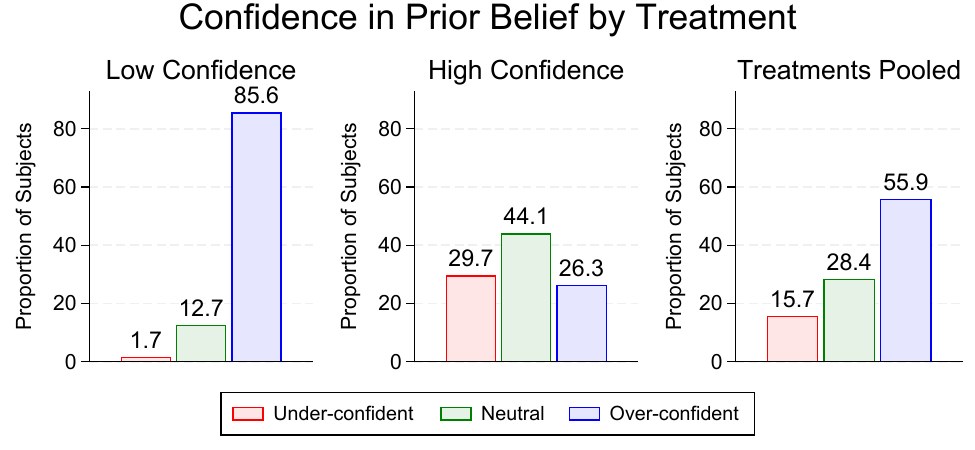}
    \caption{Subject proportions by confidence in prior belief}
    \label{fig:confidence-prior}
\end{figure}

\begin{figure}
    \centering
    \includegraphics{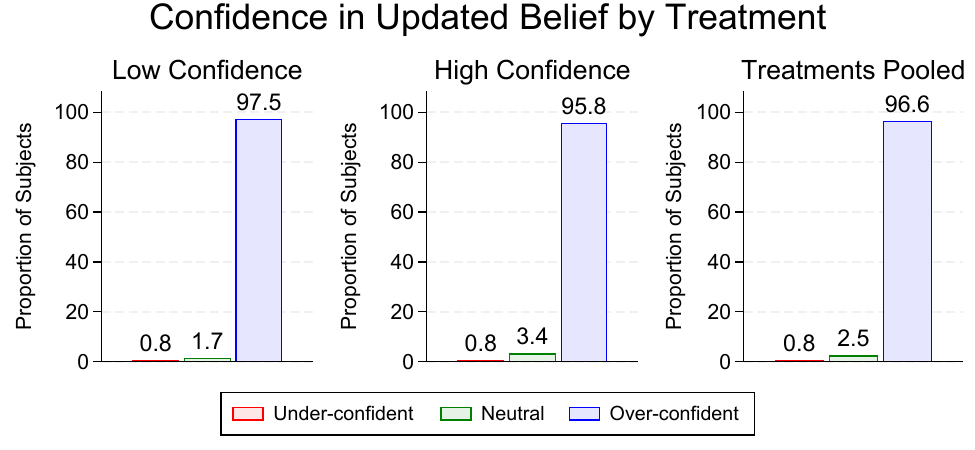}
    \caption{Subject proportions by confidence in updated belief}
    \label{fig:confidence-update}
\end{figure}


\subsection{Overconfidence in Stating Beliefs}

Our final set of results categorizes subjects based on their over-confidence in their stated prior and updated beliefs. If subjects state that they are 70\% confident that their reported belief is within three percentage points of the true value, then we should expect that 70\% of the time their guess is within three percentage points of the true value. Given that we have multiple confidence elicitation for each subject, we can test if our subjects are overconfident at the subject level. 

For each subject we compute the average reported confidence for both prior and updated beliefs and the associated 95\% confidence interval. We then compute the proportion of the subject's elicited beliefs for both prior and updated beliefs that are within three percentage points of the actual prior or the true Bayesian posterior. We define a subject as over-confident if the proportion falls above the 95\% confidence interval of the subject's average confidence, under-confident if below, and neutral if within the interval.

\Cref{fig:confidence-prior,fig:confidence-update} demonstrate that subjects are largely over-confident when it comes to reporting both their priors and their updated beliefs. Regarding prior beliefs (\cref{fig:confidence-prior}), however, the modal subject is neutral in the High Confidence treatment, with the proportions of over- and under-confident types being relatively balanced. The vast majority of subjects are over-confident regarding prior beliefs in the Low Confidence treatment. With respect to updated beliefs, \cref{fig:confidence-update} shows that subjects are overwhelmingly over-confident in both treatments relative to the Bayesian posterior. In general, our subjects could have increased their expected earnings by reporting lower confidence values.

We also define a continuous measure of confidence for each subject: we take the subject's mean stated confidence-in-beliefs and subtract the proportion of elicited beliefs (both prior and posterior) that are actually within three percentage points of the true value. A positive value indicates an over-confident subject and a negative value an under-confident subject.

\begin{figure}
    \centering
    \includegraphics[trim=0.2in 0.2in 0.2in 0]{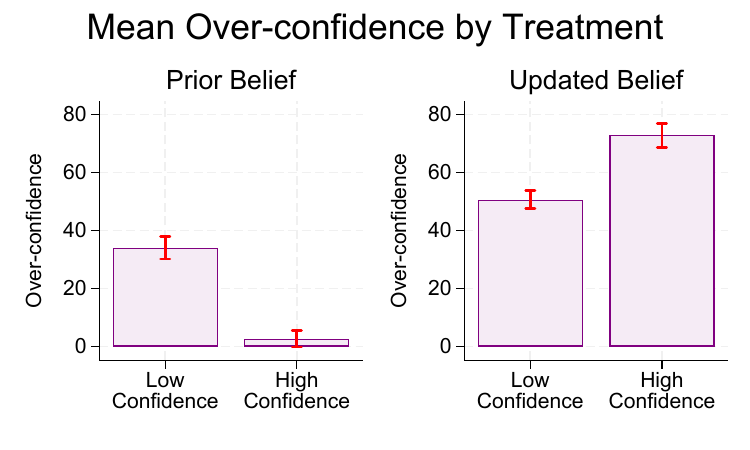}
    \caption{Mean (continuous) over-confidence measure with 95\% confidence intervals}
    \label{fig:mean-overconfidence}
\end{figure}

\begin{result}
    Subjects are overly confident in their prior beliefs in the Low Confidence treatment. Subjects are over-confident in their updated beliefs in both treatments.
\end{result}

\Cref{fig:mean-overconfidence} shows the mean continuous over-confidence measure.\footnote{We also plot the distribution of the measure in Section S1 of the Supplementary Material. The vast majority of our subjects are over-confident, echoing our results.} Results with the continuous measure are similar to those with the discrete measure. With respect to the prior, in the Low Confidence treatment, we see that the average confidence stated by our subjects is about 37 percentage points more than the proportion in which their stated prior is within three percentage points of the actual value. For the High Confidence treatment, we see that our over-confidence measure is about three percentage points and this is not statistically different from zero ($p=0.0697$).

With respect to updated beliefs, we observe more over-confidence than in the case of prior beliefs. This suggests that people are making mistakes in updating their beliefs and are over-confident in their ability to update. In the High Confidence treatment, subjects state confidence that is on average $73$ percentage points higher than the actual proportion of the times their stated beliefs are within three percentage points of the actual Bayesian belief, while in the Low Confidence treatment, subjects state confidence that is on average 51 percentage points more than the proportion of the times their stated beliefs are within three percentage of the actual Bayesian belief.

A surprising result is that the degree of over-confidence is significantly larger in the High Confidence treatment compared to the Low Confidence treatment ($p<0.001$). In the High Confidence treatment subjects have more accurate priors \textit{and} are more confident in their prior, yet they are more over-confident in their updated beliefs.

\begin{result}
    Subjects are more over-confident in stating their updated beliefs in the High Confidence treatment.
\end{result}

\begin{figure}
    \centering
    \includegraphics{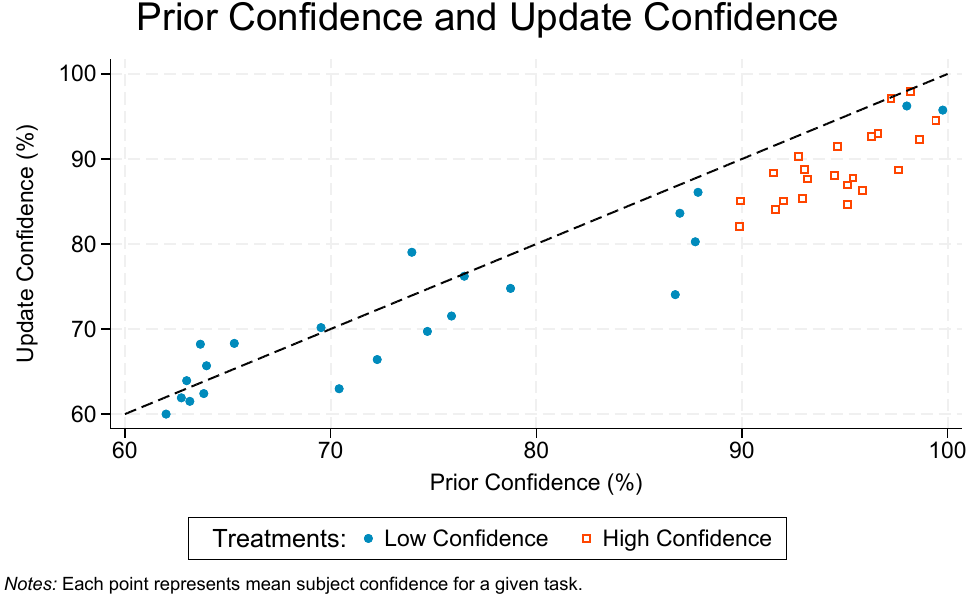}
    \caption{Relationship between confidence in prior and confidence in updated beliefs}
    \label{fig:confidence-correlation}
\end{figure}

We also plot the relationship between confidence in prior beliefs and confidence in updated beliefs. \Cref{fig:confidence-correlation} depicts a positive correlation for the mean value across subjects for each task. This suggests that, on average, their confidence-in-prior-belief directly translates into confidence-in-updated-belief, implying that subjects may think that they are Bayesian or accurate in their belief updating process and that their margin of error is within three percentage points.\footnote{Recall that we informed subjects that they would earn a \$3 bonus if their reported updated belief is within three percentage points of a value computed from a statistical process, Bayes' theorem. This suggests that the subjects believe that they update their beliefs consistent with Bayes' theorem.} This provides suggestive evidence that the majority of the subjects believe that they are Bayesian updaters when in fact they are not. We do see a drop in confidence from the prior to the updated belief (most of the scatterplot falls below the 45 degree line).

\section{Conclusion} \label{sec:discussion}

Our study examines a critical relationship between confidence over multiple prior beliefs and belief updating. We find that confidence over multiple priors matters when it shouldn't, and confidence in a belief distribution doesn't matter when it should. We document a systematic deviation from Bayes' rule, where increasing confidence over multiple prior leads to less updating. Our subjects' behavior deviates significantly from the Bayesian prediction.

Our novel design varies subjects' confidence over multiple priors and elicits incentive-compatible confidence reports regarding beliefs. We find that greater confidence over multiple prior beliefs leads to more under-updating when a Bayesian agent should not respond to this notion of confidence. We find that subjects are insensitive to an alternative notion of confidence that corresponds to dispersion within a single belief distribution. Our results contribute to a broader understanding of belief updating, suggesting that confidence over multiple prior beliefs plays a role in how individuals update their beliefs through the weight placed on the priors and signals in the updating process.

Our results also present striking evidence of over-confidence when subjects report both their prior and updated beliefs. For updated beliefs, we observe a greater degree of over-confidence in the High Confidence treatment, in which subjects report prior beliefs with greater accuracy and are more confident in those prior beliefs. This suggests that having more accurate prior beliefs does not necessarily lead to increasingly accurate updating behavior. Our finding challenges the notion that having increasingly accurate priors and confidence in those priors inherently results in better-calibrated updated beliefs. We conclude by noting first our contribution to the elicitation of beliefs literature and second the stylized facts we present about the relationship between confidence and belief-updating. These facts should motivate further models of belief-formation processes.

\clearpage

\appendix

\renewcommand{\thesection}{\Alph{section}}

\titleformat{\section}[block]{\normalfont\Large\bfseries}
  {Appendix \thesection.}{1em}{}

\linespread{1.5}\selectfont
\crefalias{section}{Appendix}

\section{Proofs}

\subsection{Incentive Compatibility of Belief and Confidence Elicitation}\label{app:incentive}

We want to show that appending the confidence elicitation does not distort the belief reported in the first part of the elicitation. 

Our belief and confidence elicitation has two stages. First subjects report a belief---a prior and then eventually a posterior as described in \cref{sec:design}. Subjects earn a fixed amount for a guess within three percentage points of the actual prior or the true Bayesian posterior. After reporting a belief, subjects report their \emph{confidence} that their stated belief is within three percentage points of the actual prior or the true Bayesian posterior, in the following elicitation protocol. Subjects are given the option to choose between sticking to their belief report, which obtains a fixed payment if the reported belief is within three percentage points of the corresponding true value, or a lottery with the same fixed payment with $x$\% chance and nothing otherwise ($x$ is elicited using a BDM). If $x \leq q$, the subject is paid for an accurate belief. If $x > q$, the subject is paid via lottery.

To model the subject's decision problem, we let a reported belief be denoted as $p$ about the binary state, and their confidence $q$. The subject's action then is to report $(p,q)$.

First we show that for any reported belief $p$, the optimal choice in the confidence elicitation stage is to report $q = q^*$, where $q^*$ is the confidence associated with the belief reported in the first stage. Let $x$ be the chance of winning the lottery, and $y>0$ be the prize for winning the lottery and the guess being correct. For simplicity, we normalize the payoff for guessing incorrectly and losing the lottery to zero. The expected payoff is

\begin{align*}
   \overbrace{P(x \leq q)}^{\substack{\text{Prob.~that BDM} \\ \text{selects guess}}}
   \times 
   \underbrace{y q^*}_{\substack{\text{Expected payoff} \\ \text{of guess}}}
   +
   \overbrace{P(x > q)}^{\substack{\text{Prob.~that BDM} \\ \text{selects lottery}}}
   \times 
   \underbrace{y \int_q^1 x \, f_{x \mid x > q}(x) \, dx}_{\substack{\text{Expected payoff} \\ \text{of lottery}}}
   ,
\end{align*}
where $f_{x \mid x > q}$ is the conditional probability function density that the random number drawn from the BDM mechanism is larger than the confidence reported by the agent. Recall that the agent has an underlying subjective belief that their guess will obtain the prize with probability $q^*$. Thus the subjective expected value of the guess is $y q^*$. Because $x$ is drawn from a continuous uniform distribution over $[0,1]$, the expected payoff becomes
\begin{align*}
    q \times yq^* + (1-q) \, y \int^1_q \frac{x}{1-q} \, \textrm{d}x = y \left( q \times q^* +  \frac{1}{2} - \frac{q^2}{2} \right).
\end{align*}
The first order condition with respect to $q$ is then
\begin{align*}
    && yq^* -yq &= 0 & \Rightarrow && q &= q^*. &&
\end{align*}

We now consider how a subject chooses a belief $p$ to report. The expected payoff of a belief and confidence pair $(p,q)$ is
\begin{align*}
    q \times yq + (1-q) \, y \int^1_{q} \frac{x}{1-q} \, \textrm{d}x &= y \left( q^{2} +  \frac{1}{2} - \frac{q^{2}}{2}\right) \\
     &= y \left( \frac{q^{2}}{2} + \frac{1}{2} \right)
\end{align*}

Notice that the function is increasing in $q$ when $q\in [0,1]$. The optimal action for the agent is to report a belief $p$ that has the largest confidence $q$. In this case, it is the point that has the largest probability mass within three percentage points.

If only the belief about the binary state were elicited, the optimal strategy would be to report the belief in which the subject is most confident.  Adding the confidence elicitation to the belief elicitation does not distort the belief reported in the first stage.

\subsection{Second-order Beliefs and Belief Updating} \label{proofs}

Consider a finite state space $\Omega$ and signal space $\Sigma$. Let us assume that the agent considers $N$ possible priors $\pi_{0,i}$, where $\pi_{0,i} \in \Delta(\Omega)$ for all $i \in \{1,\hdots,N\}$. The agent has a belief distribution over the priors, assigning subjective probability mass $k_{0,i}$ to prior $\pi_{0,i}$ for all $i \in \{1,\hdots,N\}$. For any $\omega\in\Omega$ and $\sigma\in\Sigma$, we denote $\pi_0(\omega)$ as the prior probability that state $\omega$ is realized and $\pi_1(\omega|\sigma)$ as the posterior belief that state $\omega$ is realized after observing signal $\sigma$. Let $P(\sigma|\omega)$ be the conditional probability of observing $\sigma$ when $\omega$ is the realized state. Finally let $ k_{1,i}(\sigma)$ denote the subjective probability mass assigned to prior $\pi_{0,i}$ after observing $\sigma$.

\begin{prop} \label{prop:equivalence}
The average Bayesian posterior after observing signal $\sigma$ is
\begin{equation*}
   \pi^{\mathrm{Bayes}}_1(\omega|\sigma) \coloneqq P(\omega|\sigma) = \sum^N_{i=1} k^{\mathrm{Bayes}}_{1,i}(\sigma) \, \pi^{\mathrm{Bayes}}_{1,i}(\omega|\sigma),
\end{equation*}
which is the same as updating with the average prior belief
\begin{equation*}
   \pi^{\mathrm{Bayes}}_1(\omega|\sigma)  = \frac{\pi_0(\omega) \, P(\sigma|\omega)}{\sum_{\omega'\in\Omega}\pi_0(\omega') \, P(\sigma|\omega')}.
\end{equation*}
\end{prop}

\begin{proof}
Let us update each prior $\pi_{0,i}$ and its respective weight $k_{0,i}$ individually using Bayes' rule to obtain
\begin{align*}
    && \pi^\mathrm{Bayes}_{1,i}(\omega|\sigma) &= \frac{\pi_{0,i}(\omega) \, P(\sigma | \omega)}{P_i(\sigma)} & \text{ and } && k^\mathrm{Bayes}_{1,i}(\omega|\sigma) &\coloneqq \frac{k_{0,i} \, P_i(\sigma) }{P(\sigma)}, &&
\end{align*}
where $P(\sigma)= \sum_{\omega'\in\Omega} \sum^N_{i=1}k_{0,i}\pi_{0,i}(\omega') P(\sigma | \omega')$\footnote{The summation converges absolutely because we are working with a probability measure, allowing us to switch the order of the summation.} is the probability of observing signal $\sigma$ given the mixture prior belief distribution or the average prior belief $\pi_0$, while $P_i(\sigma)=\sum_{\omega'\in\Omega}\pi_{0,i}(\omega') P(\sigma | \omega')$ is the probability of observing signal $\sigma$ given prior $\pi_{0,i}$. We now show that updating the belief of each prior and the mixture distribution over these is equivalent to simply updating the average prior belief.
\begin{align*}
  \pi^\mathrm{Bayes}_1(\omega|\sigma) &= \frac{\pi_0(\omega) \, P(\sigma|\omega)}{P(\sigma)} \\
    &= \frac{\sum^N_{i=1} k_{0,i} \, \pi_{0,i}(\omega) \, P(\sigma|\omega)}{P(\sigma)} \\
    &= \frac{\sum^N_{i=1} k_{0,i} \, P_i(\sigma) \, \pi_{0,i}(\omega) \, P(\sigma|\omega) / P_i(\sigma)  }{P(\sigma)} \\ 
    &= \frac{\sum^N_{i=1} k_{0,i} \, P_i(\sigma) \, \pi_{1,i}^\mathrm{Bayes} (\omega|\sigma) }{P(\sigma)} \\
    &= \sum^N_{i=1} k_{1,i}^\mathrm{Bayes}(\omega|\sigma) \, \pi_{1,i}^\mathrm{Bayes}(\omega|\sigma) .
\end{align*}
\end{proof}

This property is not exclusive to Bayesian updating. Assume the updating rule for the first-order belief has the functional form
\begin{equation}\label{eq:non-bayes}
    \pi_1(\omega|\sigma) = \frac{\pi_0(\omega) \, T[P(\sigma|\omega)]}{\sum_{\omega'\in\Omega} \pi_0(\omega') \, T[P(\sigma|\omega')]},
\end{equation}
where $T:\mathbb{R}_\geq\rightarrow\mathbb{R}_\geq$, and restrict $\sum_{\omega'\in\Omega} \pi_0(\omega') \, T[P(\sigma|\omega')]>0$ for any prior and signal realization to ensure the beliefs are well-defined. If we assume the updating of the second-order beliefs is Bayesian but uses the distorted probability of the realization of $\sigma$, then
\begin{equation} \label{eq:non-bayes-2nd}
     k_{1,i}(\omega|\sigma) \coloneqq \frac{k_{0,i} \sum_{\omega'\in\Omega} \pi_0(\omega') \, T[P(\sigma|\omega')] }{\sum_{\omega'\in\Omega} \sum^N_{i=1} k_{0,i} \, \pi_{0,i}(\omega') \, T[P(\sigma | \omega')]}.
\end{equation}
Updating with the average prior belief will give us the average updated belief.

\begin{prop} \label{prop:equivalence-non-Bayesian}
Given the updating rule in \cref{eq:non-bayes} and \cref{eq:non-bayes-2nd}, the average updated belief after observing signal $\sigma$ is
\begin{equation*}
   \pi_1(\omega|\sigma) \coloneqq P(\omega|\sigma) = \sum^N_{i=1} k_{1,i}(\sigma) \, \pi_{1,i}(\omega|\sigma),
\end{equation*}
which is the same as updating with the average prior belief
\begin{equation*}
   \pi_1(\omega|\sigma) = \frac{\pi_0(\omega) \, T[P(\sigma|\omega)]}{\sum_{\omega'\in\Omega}\pi_0(\omega') \, T[P(\sigma|\omega')]}.
\end{equation*}
\end{prop}

\begin{proof}
We first define the non-Bayesian updating rule for $\pi_{1,i}$ and $k_{1,i}$, 
\begin{align*}
    && \pi_{1,i}(\omega|\sigma) &= \frac{\pi_{0,i} (\omega) \, T[P(\sigma | \omega)]}{P_i^*(\sigma)} & \text{ and } && k_{1,i}(\omega|\sigma) &\coloneqq \frac{k_{0,i} \, P_i^*(\sigma) }{P^*(\sigma)} ,&&
\end{align*}
where $P^*(\sigma)= \sum_{\omega'\in\Omega} \sum^N_{i=1} k_{0,i} \, \pi_{0,i}(\omega') \, T[P(\sigma | \omega')]$ is the \emph{perceived} probability of observing signal $\sigma$ given the mixture prior belief distribution or the average prior belief $\pi_0$ by the non-Bayesian agent. $P_i^*(\sigma)=\sum_{\omega'\in\Omega}\pi_{0,i}(\omega') T[P(\sigma | \omega')]$ is the perceived probability of observing signal $\sigma$ given prior $\pi_{0,i}$ by the non-Bayesian agent. The proof is similar to the Bayesian case:
\begin{align*}
  \pi_1(\omega|\sigma) &= \frac{\pi_0(\omega) \, T[P(\sigma|\omega)]}{P^*(\sigma)} \\
    &= \frac{\sum^N_{i=1} k_{0,i} \, \pi_{0,i}(\omega) \, T[P(\sigma|\omega)]}{P^*(\sigma)} \\
    &= \frac{\sum^N_{i=1} k_{0,i} \, P^*_i(\sigma) \, \pi_{0,i}(\omega) \, T[P(\sigma|\omega)] \,/\, P^*_i(\sigma)  }{P^*(\sigma)} \\ 
    &= \frac{\sum^N_{i=1} k_{0,i} \, P^*_i(\sigma) \, \pi_{1,i} (\omega|\sigma) }{P^*(\sigma)} \\
    &= \sum^N_{i=1} k_{1,i}(\omega|\sigma) \, \pi_{1,i}(\omega|\sigma) .
\end{align*}
\end{proof}

\linespread{1}\selectfont

\printbibliography


\clearpage
\linespread{1}\selectfont

\titleformat{\section}[block]{\normalfont\Large\bfseries}{\thesection}{1em}{}

\let\addcontentsline\origaddcontentsline

\setcounter{tocdepth}{1}

\renewcommand{\thesection}{S\arabic{section}}
\setcounter{section}{0}
\setcounter{figure}{0}
\setcounter{table}{0}
\thispagestyle{empty}%
\begin{center}%
\phantom{.}%

\vspace{2\baselineskip}

\textbf{\Large
On Prior Confidence and Belief Updating\\
\vspace{9pt}
Supplementary Materials}

\vspace{2\baselineskip}

Kenneth Chan, Gary Charness, Chetan Dave, J.~Lucas Reddinger\\

\vspace{2\baselineskip}

May 13, 2025

\vspace{2\baselineskip}

\end{center}

\setcounter{tocdepth}{1} 
\tableofcontents

\linespread{1.5}\selectfont

\section{Supplemental Results}

\paragraph{Distributions of Over-updating and Overconfidence} \label{app:overconfidence}

\begin{figure}[p]
    \centering
    \includegraphics{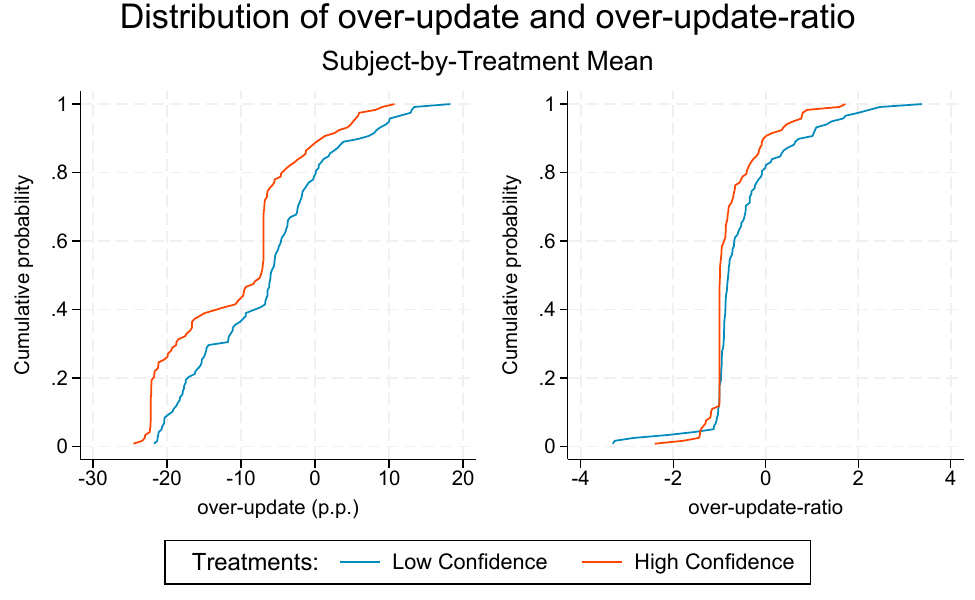}
    \caption{Distribution of the average subject over-update and over-update-ratio measures}
    \label{fig:overupdate-cdf}
\end{figure}

\begin{figure}[p]
    \centering
    \includegraphics{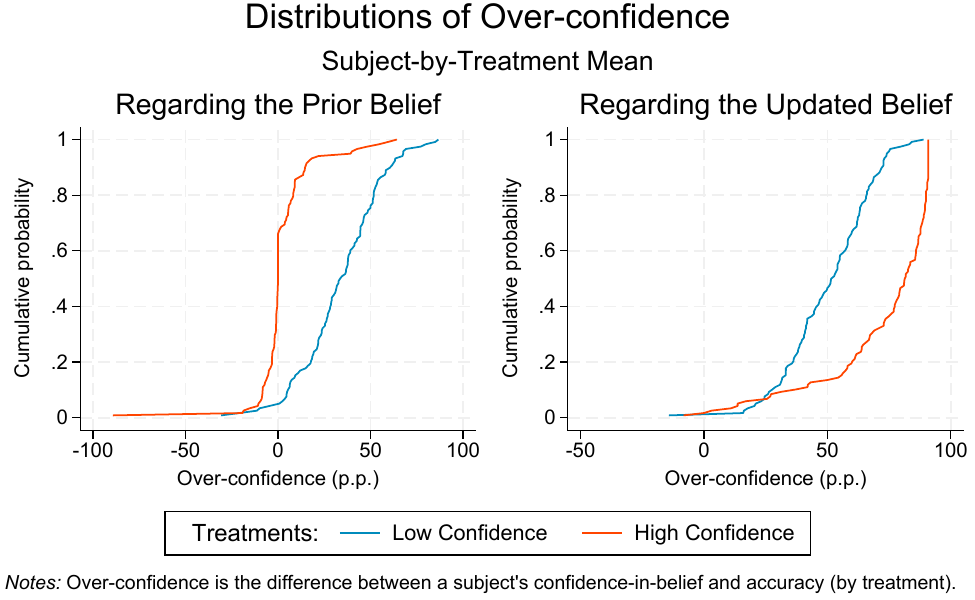}
    \caption{Distribution of the average subject continuous over-confidence measure}
    \label{fig:cdf-confidence}
\end{figure}

\Cref{fig:overupdate-cdf} depicts the distribution of the average subject over-update and over-update-ratio measures. \Cref{fig:cdf-confidence} depicts the distribution of the average subject's continuous over-confidence measure.

\paragraph{Absolute Update by Priors} \label{app:abs-update}

\begin{figure}[p]
    \centering
    \includegraphics{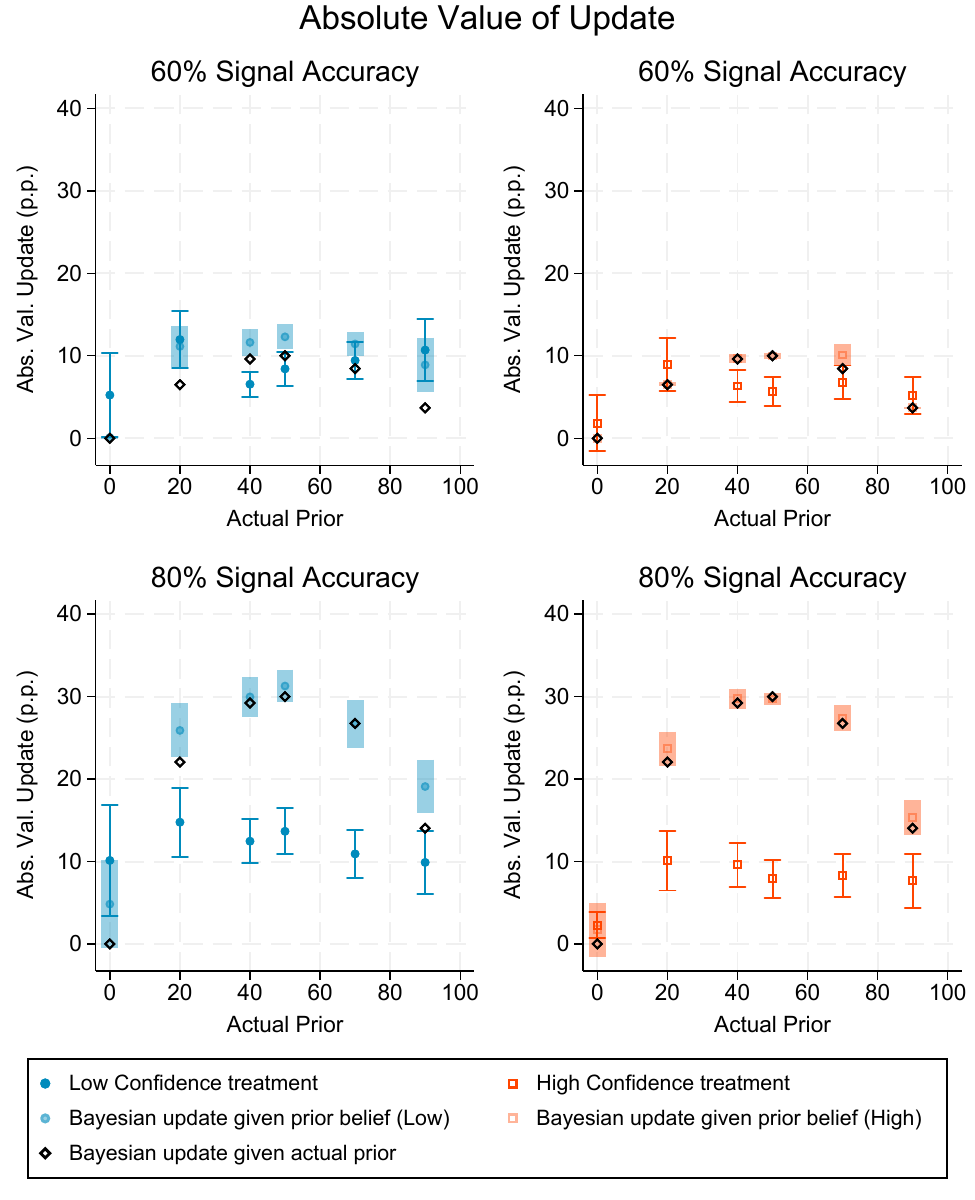}
    \caption{Magnitude of absolute value of updating by signal and treatment accuracy}
    \label{fig:abs-update-by-signal}
\end{figure}

\Cref{fig:abs-update-by-signal} plots the magnitude of the absolute value of updates for each signal accuracy and treatment pair.

\section{Robustness Checks} 

\subsection{Density Plot of Subject's Belief in Low Confidence Treatment} \label{app:density}

\begin{landscape}
    \begin{figure}[p]
    \centering
    \includegraphics[scale=0.85]{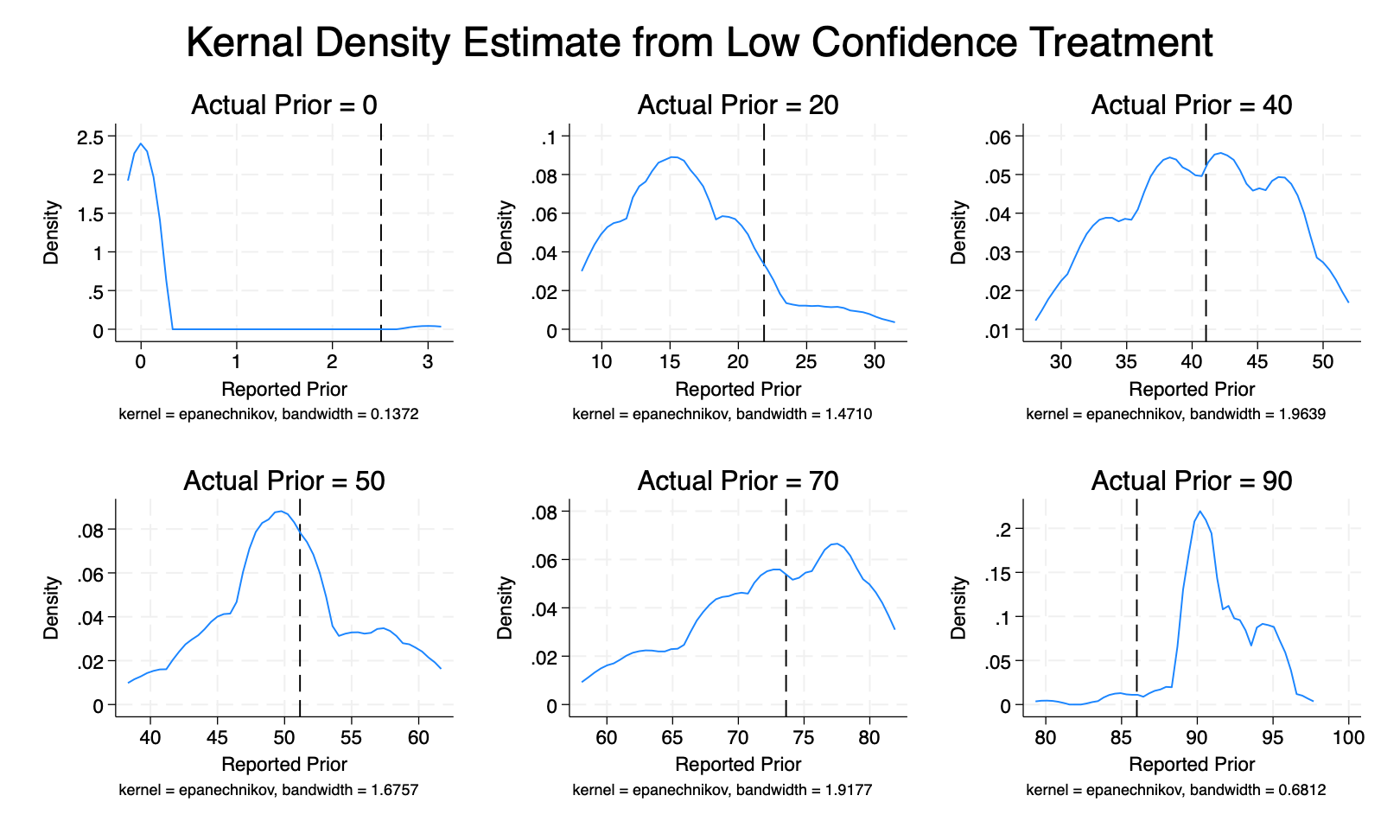}
    \caption{Density plots of subjects' reported priors in the Low Confidence treatment\\ \footnotesize \textit{Notes:} The dashed line shows the average reported prior in the low confidence treatment. We restrict the density plot to within 10 percentage points of the actual prior. }
    \label{fig:density}
\end{figure}
\end{landscape}

Our belief elicitation incentivizes subjects to report the point where subjects have the highest confidence. We plot the kernel density of the subject's belief report for the prior and the average of the subject's response in \cref{fig:density} and show that the average belief is not too different from the point with the highest confidence. This validates our assumption of using the subject's elicited belief as the average prior belief to compute the subject's average Bayesian posterior. 

\subsection{Exclusion of Subjects Who Made Mistakes}\label{app:mistakes}

As a robustness check, we exclude subjects who make a mistake. Specifically, we exclude subjects who in the updating task with a degenerate prior reported either (1) an incorrect prior or (2) not fully confident in the prior belief.  As shown in \cref{tbl:overupdate-ols-robust,tbl:overupdate-iv-robust,tbl:grether-iv-robust}, our results are qualitatively similar upon excluding subjects who made these mistakes. 

\subsection{Grether Regression with Ordinary Least Squares} \label{app:grether}

\Cref{tbl:grether-ols} presents results for the \textcite{Grether1980} model using ordinary least squares. In the pooled regressions (columns 3 and 4) we find that $\beta$ is larger by about 0.1 units in the High Confidence treatment compared to the Low treatment ($p=0.002$), indicating that subjects place more weight on the prior in the High Confidence treatment.

We estimate the parameters for each signal accuracy in columns 1 and 2 of \cref{tbl:grether-ols}. We see that the results are qualitatively similar. We find one notable exception: the weight placed on the signal ($\alpha$) is larger when the signal is weaker ($p=0.021$). Although we do not find over-reaction to the 60\%-accurate signal, our result is similar to those of \textcite{AugenblickLazarusThaler2024}, who find that people tend to under-infer from stronger signals.\footnote{Our design differs from \textcite{AugenblickLazarusThaler2024} who used ``balls and urns'' framing.} This validates findings across frames of belief-updating tasks.

\begin{table}
\centering
\def\sym#1{\ifmmode^{#1}\else\(^{#1}\)\fi}
\begin{threeparttable}
\caption{OLS regressions of over-update measures (robustness check) \label{tbl:overupdate-ols-robust}}
\begin{tabular}{l*{4}{d{4.6}}}
\toprule
&\multicolumn{4}{c}{Dependent variable}\\
\cmidrule{2-5}
&\multicolumn{2}{c}{over-update}&\multicolumn{2}{c}{over-update-ratio}\\
\cmidrule(r){2-3}\cmidrule(l){4-5}
                    &\multicolumn{1}{c}{(1)}&\multicolumn{1}{c}{(2)}&\multicolumn{1}{c}{(3)}&\multicolumn{1}{c}{(4)}\\
\midrule
High Confidence treatment&      -3.647&      -3.647&      -0.266&      -0.267\\
                    &     (0.613)&     (0.613)&     (0.096)&     (0.096)\\
Constant            &      -8.178&      -8.178&      -0.508&      -0.507\\
                    &     (0.940)&     (0.307)&     (0.096)&     (0.048)\\
\midrule
Subject fixed-effects&\multicolumn{1}{c}{No}&\multicolumn{1}{c}{Yes}&\multicolumn{1}{c}{No}&\multicolumn{1}{c}{Yes}\\
\(R^2\)             &       0.010&       0.013&       0.003&       0.003\\
Subjects            &\multicolumn{1}{c}{\(101\)}&\multicolumn{1}{c}{\(101\)}&\multicolumn{1}{c}{\(101\)}&\multicolumn{1}{c}{\(101\)}\\
Observations        &\multicolumn{1}{c}{\(2020\)}&\multicolumn{1}{c}{\(2020\)}&\multicolumn{1}{c}{\(2019\)}&\multicolumn{1}{c}{\(2019\)}\\
\bottomrule
\end{tabular}
\begin{tablenotes}[para,flushleft] \footnotesize \item \emph{Notes:} Standard errors (in parentheses) are clustered at the subject level. Observations with undefined log-ratio dropped. Excludes 17 subjects who made a mistake in the degenerate prior task.
\end{tablenotes}
\end{threeparttable}
\end{table}

\begin{table}
\centering
\def\sym#1{\ifmmode^{#1}\else\(^{#1}\)\fi}
\begin{threeparttable}
\caption{OLS and IV regressions of over-update measures (robustness check) \label{tbl:overupdate-iv-robust}}
\begin{tabular}{l*{4}{d{4.5}}}
\toprule
&\multicolumn{4}{c}{Dependent variable}\\
\cmidrule{2-5}
&\multicolumn{2}{c}{over-update}&\multicolumn{2}{c}{over-update-ratio}\\
\cmidrule(r){2-3}\cmidrule(l){4-5}
                    &\multicolumn{1}{c}{(1)}&\multicolumn{1}{c}{(2)}&\multicolumn{1}{c}{(3)}&\multicolumn{1}{c}{(4)}\\
                    &\multicolumn{1}{c}{OLS}&\multicolumn{1}{c}{FE2SLS\(^\dag\)}&\multicolumn{1}{c}{OLS}&\multicolumn{1}{c}{FE2SLS\(^\dag\)}\\
\midrule
Prior confidence, \(q^*\)\qquad\qquad&       0.015&      -0.166&       0.002&      -0.012\\
                    &     (0.022)&     (0.031)&     (0.002)&     (0.005)\\
Constant            &     -11.260&       3.939&      -0.773&       0.377\\
                    &     (1.860)&     (2.630)&     (0.199)&     (0.388)\\
\midrule
First-stage \(F\)-stat&            &\multicolumn{1}{c}{\(259.83\)}&            &\multicolumn{1}{c}{\(259.83\)}\\
Subjects            &\multicolumn{1}{c}{\(101\)}&\multicolumn{1}{c}{\(101\)}&\multicolumn{1}{c}{\(101\)}&\multicolumn{1}{c}{\(101\)}\\
Observations        &\multicolumn{1}{c}{\(2020\)}&\multicolumn{1}{c}{\(2020\)}&\multicolumn{1}{c}{\(2019\)}&\multicolumn{1}{c}{\(2019\)}\\
\bottomrule
\end{tabular}
\begin{tablenotes}[para,flushleft] \footnotesize \item \emph{Notes:} Standard errors (in parentheses) are clustered at the subject level. Subject fixed-effects included. Observations with undefined log-ratio dropped. Prior confidence \(q^*\) is measured in percentage points (between 0 and 100). Excludes 17 subjects who made a mistake in the degenerate prior task. \item \tnote{\dag}Fixed-effect two-stage least squares (FE2SLS) regressions use an indicator of High Confidence treatment as the instrument.
\end{tablenotes}
\end{threeparttable}
\end{table}

\begin{table}
\centering
\def\sym#1{\ifmmode^{#1}\else\(^{#1}\)\fi}
\begin{threeparttable}
\caption{Grether model TSLS regressions (robustness check) \label{tbl:grether-iv-robust}}
\begin{tabular}{l*{4}{d{4.6}}}
\toprule
&\multicolumn{4}{c}{Signal accuracy}\\
\cmidrule{2-5}
&\multicolumn{1}{c}{60\%}&\multicolumn{1}{c}{80\%}&\multicolumn{2}{c}{Pooled}\\
\cmidrule(r){2-2}\cmidrule(r){3-3}\cmidrule{4-5}
                    &\multicolumn{1}{c}{(1)}&\multicolumn{1}{c}{(2)}&\multicolumn{1}{c}{(3)}&\multicolumn{1}{c}{(4)}\\
\midrule
Reduced-form regression:                &            &            &            &            \\
\quad \(\alpha_L\)  &       0.586&       0.320&       0.341&       0.348\\
                    &     (0.111)&     (0.048)&     (0.045)&     (0.046)\\
\quad \(\beta_L\)   &       0.747&       0.772&       0.759&       0.763\\
                    &     (0.066)&     (0.056)&     (0.043)&     (0.043)\\
\quad \((\alpha_H-\alpha_L)\)&      -0.283&      -0.093&      -0.107&      -0.118\\
                    &     (0.116)&     (0.034)&     (0.032)&     (0.033)\\
\quad \((\beta_H-\beta_L)\)&       0.156&       0.040&       0.099&       0.097\\
                    &     (0.082)&     (0.056)&     (0.050)&     (0.050)\\
\midrule
Linear combinations:             &            &            &            &            \\
\quad \(\alpha_H\)  &       0.303&       0.227&       0.234&       0.230\\
                    &     (0.101)&     (0.044)&     (0.041)&     (0.041)\\
\quad \(\beta_H\)   &       0.903&       0.812&       0.859&       0.860\\
                    &     (0.040)&     (0.054)&     (0.034)&     (0.034)\\
\midrule
\(p\)-value of \(F\)-test:&&\\ \quad\(\alpha_L = \beta_L = 1\)&       <0.001&       <0.001&       <0.001&       <0.001\\
\quad\(\alpha_H = \beta_H = 1\)&       <0.001&       <0.001&       <0.001&       <0.001\\
\quad\(\alpha_H=\alpha_L\)&       0.014&       0.006&       0.001&       <0.001\\
\quad\(\beta_H=\beta_L\)&       0.059&       0.470&       0.048&       0.052\\
\midrule Subject fixed-effects&\multicolumn{1}{c}{No}&\multicolumn{1}{c}{No}&\multicolumn{1}{c}{No}&\multicolumn{1}{c}{Yes}\\
First-stage \(F\)-stat&\multicolumn{1}{c}{\(2437.11\)}&\multicolumn{1}{c}{\(1835.54\)}&\multicolumn{1}{c}{\(4230.23\)}&\multicolumn{1}{c}{\(4244.52\)}\\
\(R^2\)             &       0.669&       0.596&       0.627&\multicolumn{1}{c}{\(0.628\)}\\
Subjects            &\multicolumn{1}{c}{\(52\)}&\multicolumn{1}{c}{\(49\)}&\multicolumn{1}{c}{\(101\)}&\multicolumn{1}{c}{\(101\)}\\
Observations        &\multicolumn{1}{c}{\(919\)}&\multicolumn{1}{c}{\(876\)}&\multicolumn{1}{c}{\(1795\)}&\multicolumn{1}{c}{\(1795\)}\\
\bottomrule
\end{tabular}
\begin{tablenotes}[para,flushleft] \footnotesize \item \emph{Notes:} Standard errors (in parentheses) are clustered at the subject level. Observations with undefined log-ratio dropped. The null hypothesis of Bayesian updating requires \(\alpha = 1\) and \(\beta = 1\). Fixed-effect two-stage least squares (FE2SLS) regressions use an indicator of High Confidence treatment as the instrument. Excludes 17 subjects who made a mistake in the degenerate prior task.
\end{tablenotes}
\end{threeparttable}
\end{table}

\begin{table}
\centering
\def\sym#1{\ifmmode^{#1}\else\(^{#1}\)\fi}
\begin{threeparttable}
\caption{Grether model OLS regressions of log updated belief ratio \label{tbl:grether-ols}}
\begin{tabular}{l*{4}{d{4.6}}}
\toprule
&\multicolumn{4}{c}{Signal accuracy}\\
\cmidrule{2-5}
&\multicolumn{1}{c}{60\%}&\multicolumn{1}{c}{80\%}&\multicolumn{2}{c}{Pooled}\\
\cmidrule(r){2-2}\cmidrule(r){3-3}\cmidrule{4-5}
                    &\multicolumn{1}{c}{(1)}&\multicolumn{1}{c}{(2)}&\multicolumn{1}{c}{(3)}&\multicolumn{1}{c}{(4)}\\
\midrule
Reduced-form regression:                &            &            &            &            \\
\quad \(\alpha_L\)  &       0.630&       0.324&       0.348&       0.353\\
                    &     (0.114)&     (0.041)&     (0.040)&     (0.040)\\
\quad \(\beta_L\)   &       0.739&       0.723&       0.731&       0.729\\
                    &     (0.045)&     (0.047)&     (0.033)&     (0.034)\\
\quad \((\alpha_H-\alpha_L)\)&      -0.245&      -0.102&      -0.112&      -0.119\\
                    &     (0.110)&     (0.031)&     (0.030)&     (0.030)\\
\quad \((\beta_H-\beta_L)\)&       0.133&       0.082&       0.107&       0.095\\
                    &     (0.041)&     (0.040)&     (0.029)&     (0.029)\\
\midrule
Linear combinations:             &            &            &            &            \\
\quad \(\alpha_H\)  &       0.386&       0.222&       0.236&       0.234\\
                    &     (0.105)&     (0.041)&     (0.039)&     (0.038)\\
\quad \(\beta_H\)   &       0.872&       0.806&       0.838&       0.824\\
                    &     (0.032)&     (0.043)&     (0.027)&     (0.028)\\
\midrule
\(p\)-value of F-test:&&\\ \quad\(\alpha_L = \beta_L = 1\)&       <0.001&       <0.001&       <0.001&       <0.001\\
\quad\(\alpha_H = \beta_H = 1\)&       <0.001&       <0.001&       <0.001&       <0.001\\
\quad\(\alpha_H=\alpha_L\)&       0.030&       0.002&       <0.001&       <0.001\\
\quad\(\beta_H=\beta_L\)&       0.002&       0.047&       <0.001&       0.002\\
\midrule Subject fixed-effects&\multicolumn{1}{c}{No}&\multicolumn{1}{c}{No}&\multicolumn{1}{c}{No}&\multicolumn{1}{c}{Yes}\\
\(R^2\)             &       0.674&       0.584&       0.622&       0.638\\
Subjects            &\multicolumn{1}{c}{\(58\)}&\multicolumn{1}{c}{\(60\)}&\multicolumn{1}{c}{\(118\)}&\multicolumn{1}{c}{\(118\)}\\
Observations        &\multicolumn{1}{c}{\(1025\)}&\multicolumn{1}{c}{\(1068\)}&\multicolumn{1}{c}{\(2093\)}&\multicolumn{1}{c}{\(2093\)}\\
\bottomrule
\end{tabular}
\begin{tablenotes}[para,flushleft] \footnotesize \item \emph{Notes:} Standard errors (in parentheses) are clustered at the subject level. Observations with undefined log-ratio dropped. The null hypothesis of Bayesian updating requires \(\alpha = 1\) and \(\beta = 1\).
\end{tablenotes}
\end{threeparttable}
\end{table}

\clearpage

\section{Interface}

\begin{figure}[p]
\fbox{\includegraphics[width=\textwidth,trim=1cm 2cm 1cm 1cm,clip,page=1]{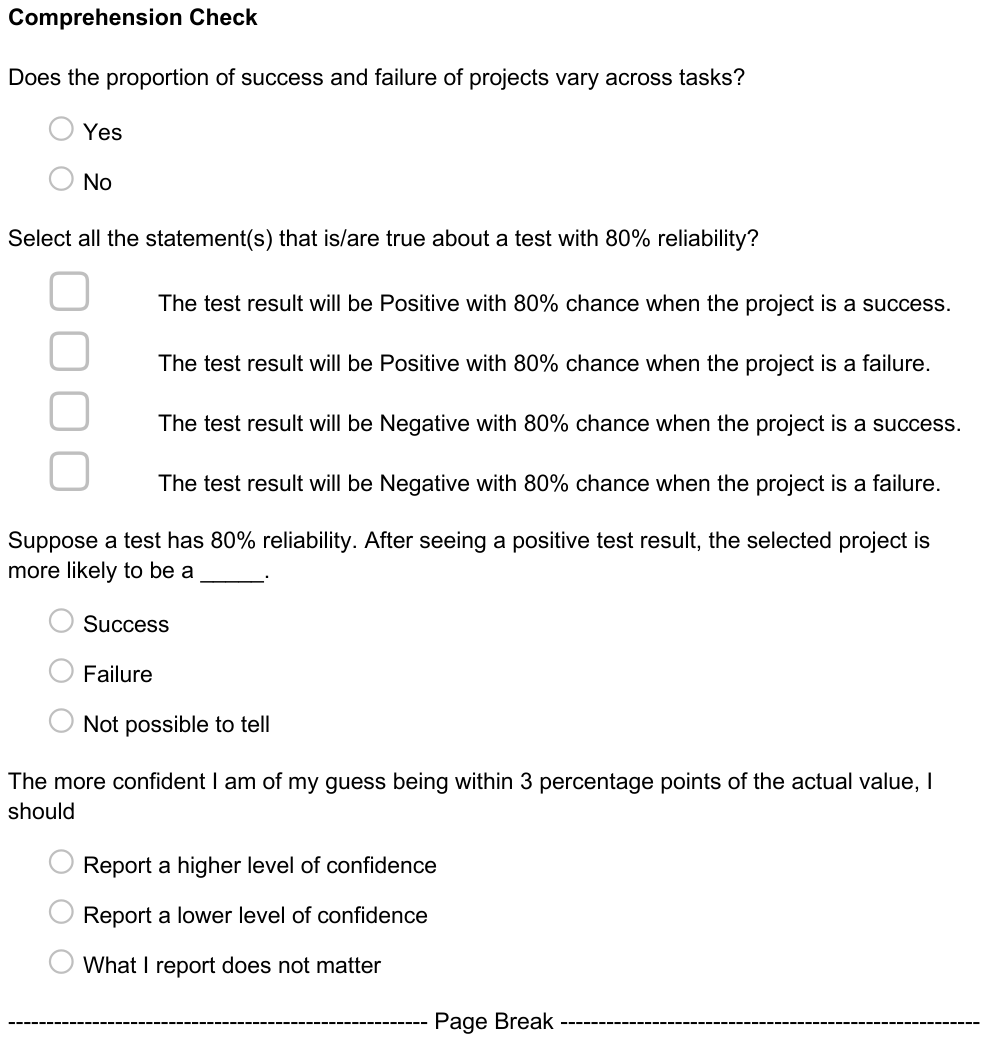}}
\caption{Interface, part 1 of 3}\label{sup-fig-interface-01}
\end{figure}

\begin{figure}[p]
\fbox{\includegraphics[width=\textwidth,trim=1cm 2cm 1cm 1cm,clip,page=2]{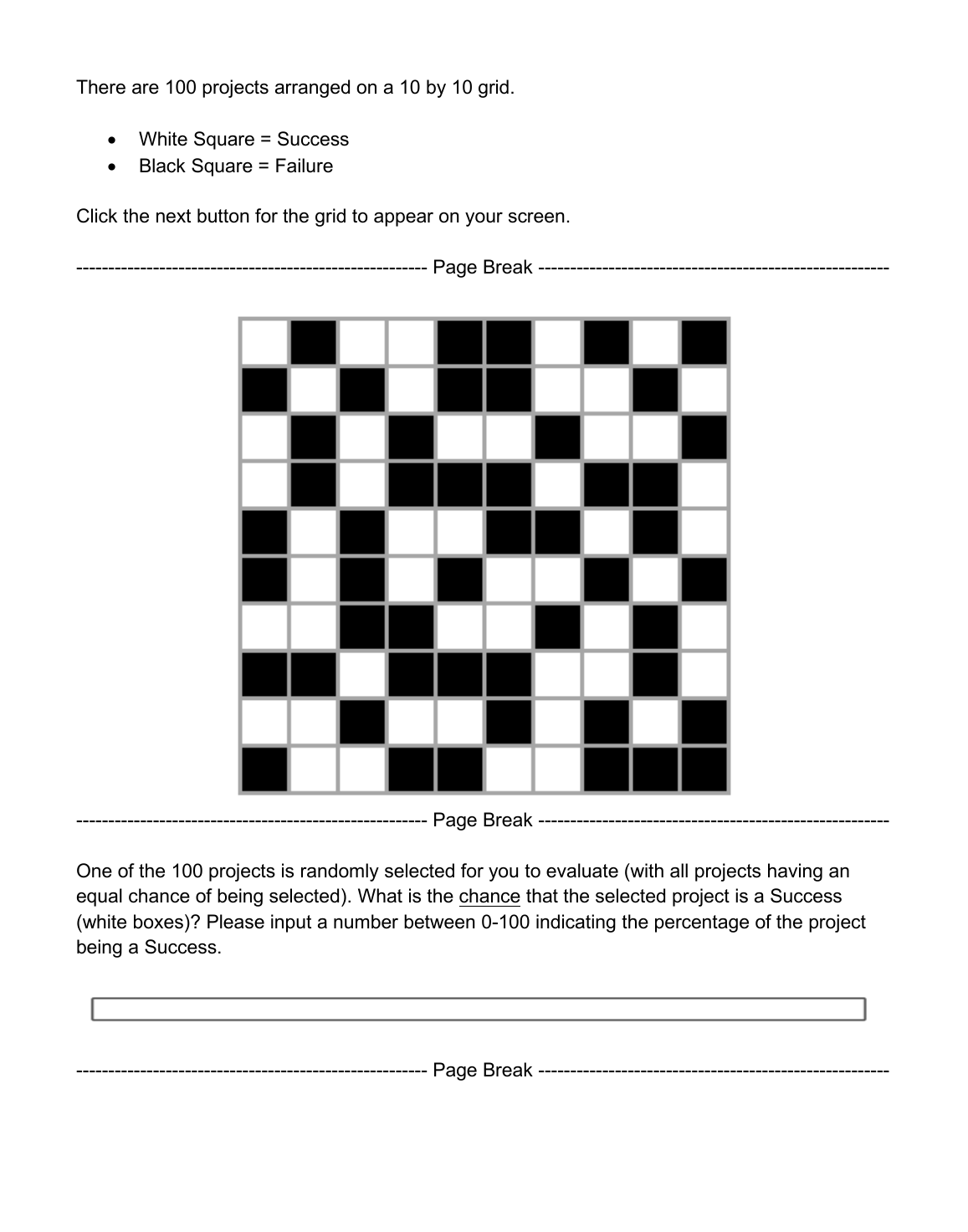}}
\caption{Interface, part 2 of 3}\label{sup-fig-interface-02}
\end{figure}

\begin{figure}[p]
\fbox{\includegraphics[width=\textwidth,trim=1cm 2cm 1cm 1cm,clip,page=3]{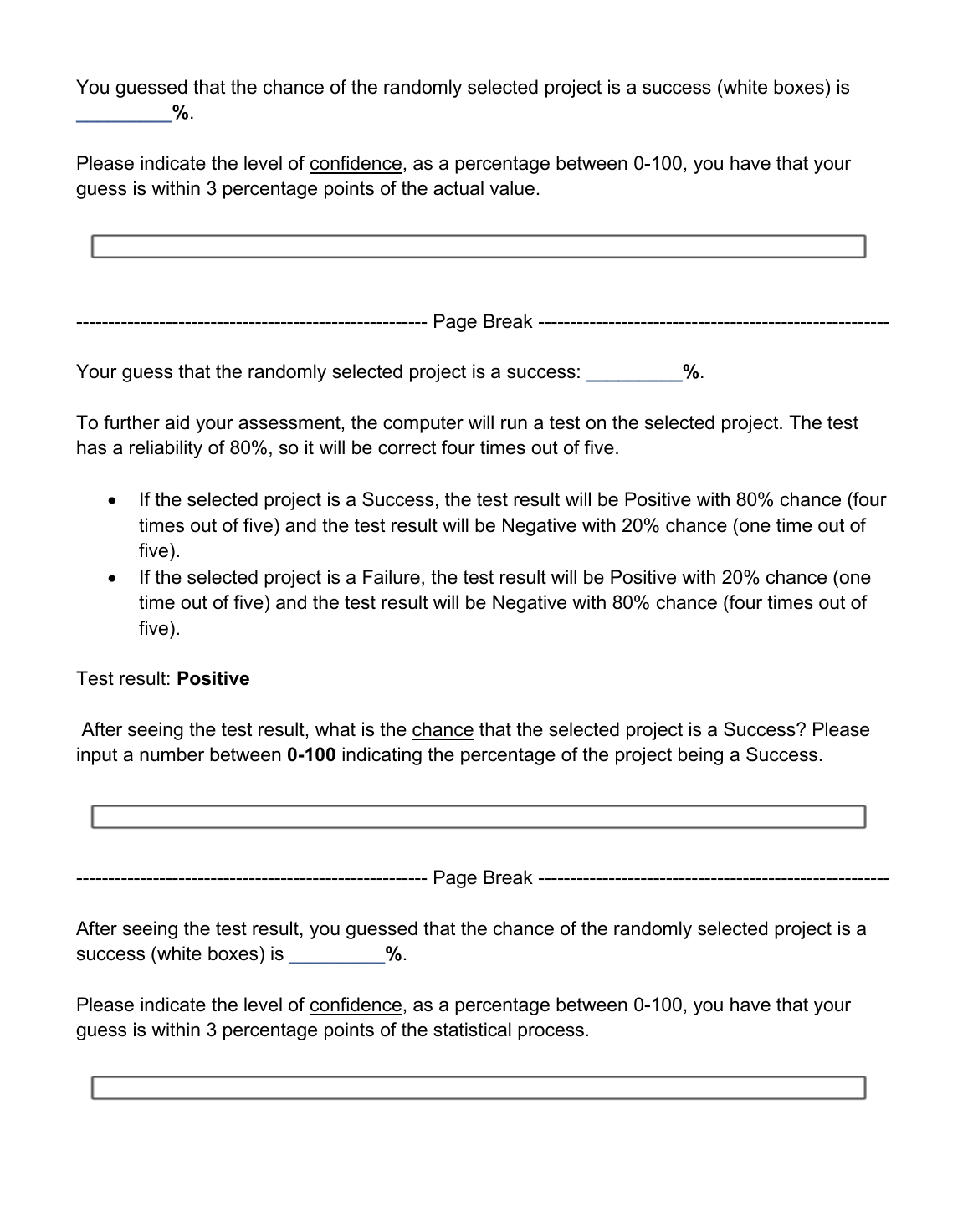}}
\caption{Interface, part 3 of 3}\label{sup-fig-interface-03}
\end{figure}

\Cref{sup-fig-interface-01,sup-fig-interface-02,sup-fig-interface-03} show the experimental interface and comprehension checks.

\section{Instruments}

\Cref{sup-fig-slide01,sup-fig-slide02,sup-fig-slide03,sup-fig-slide04,sup-fig-slide05,sup-fig-slide06,sup-fig-slide07,sup-fig-slide08,sup-fig-slide09,sup-fig-slide10,sup-fig-slide11,sup-fig-slide12,sup-fig-slide13,sup-fig-slide14,sup-fig-slide15,sup-fig-slide16,sup-fig-slide17,sup-fig-slide18,sup-fig-slide19,sup-fig-slide20,sup-fig-slide21,sup-fig-slide22,sup-fig-slide23} show the slides that we personally presented to subjects using a projector. Subjects also retained a printed copy (two slides per page) for reference during the experiment.

\begin{figure}[p]
\fbox{\includegraphics[width=\textwidth,trim=0 0 0 0,clip,page=1]{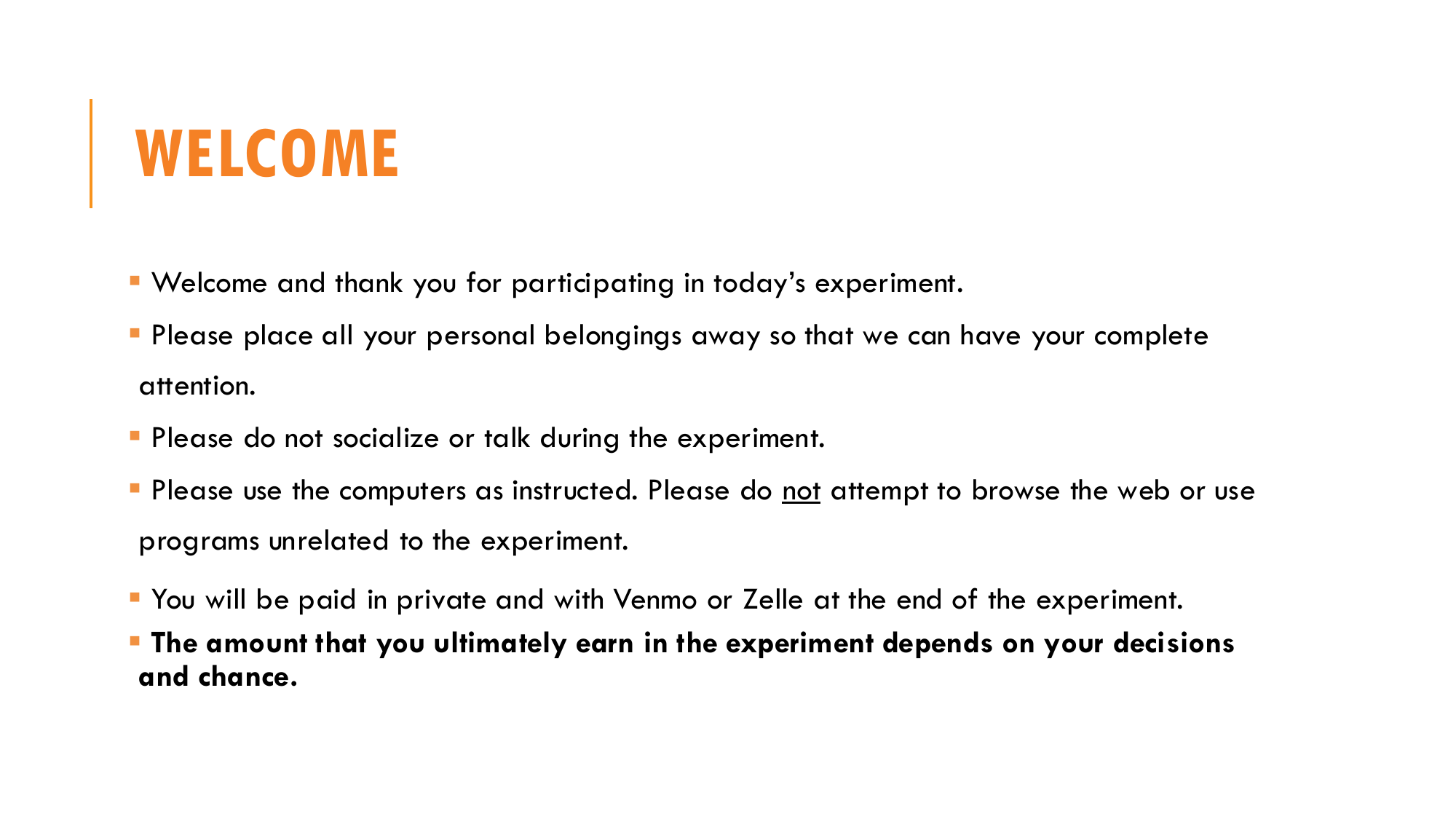}}
\caption{Slide 1 of 23}\label{sup-fig-slide01}
\end{figure}

\begin{figure}[p]
\fbox{\includegraphics[width=\textwidth,trim=0 0 0 0,clip,page=2]{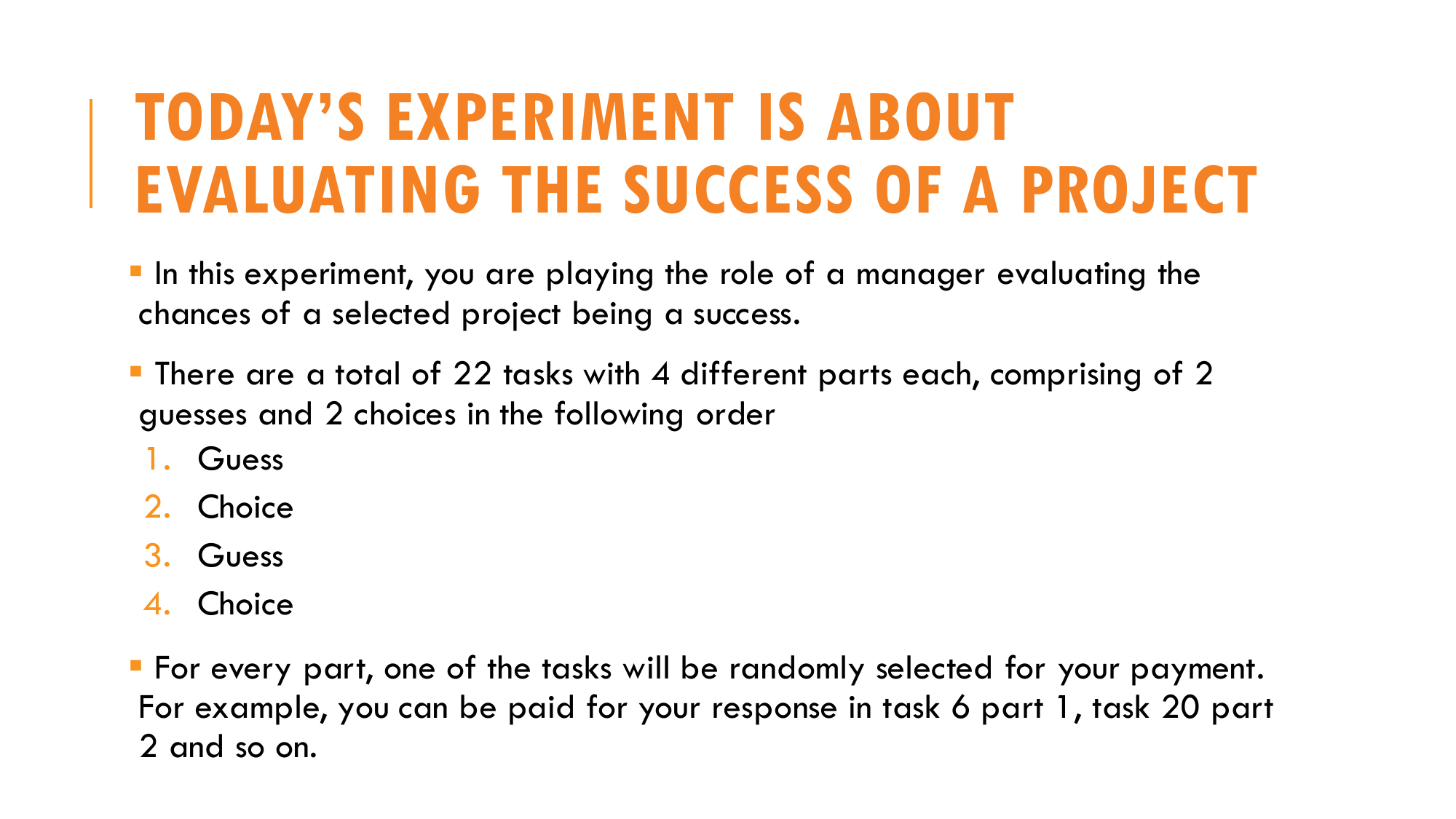}}
\caption{Slide 2 of 23}\label{sup-fig-slide02}
\end{figure}

\begin{figure}[p]
\fbox{\includegraphics[width=\textwidth,trim=0 0 0 0,clip,page=3]{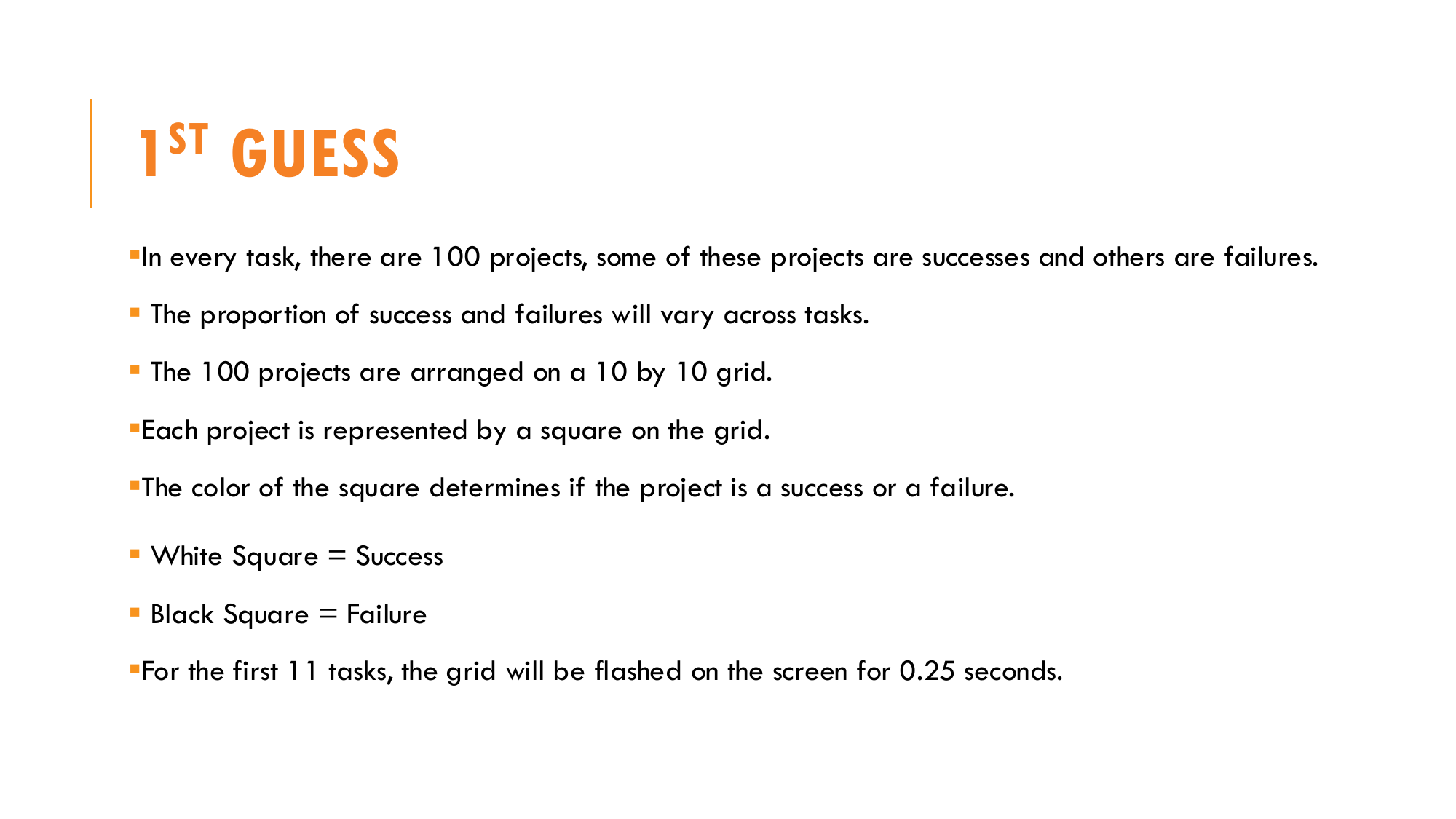}}
\caption{Slide 3 of 23}\label{sup-fig-slide03}
\end{figure}

\begin{figure}[p]
\fbox{\includegraphics[width=\textwidth,trim=0 0 0 0,clip,page=4]{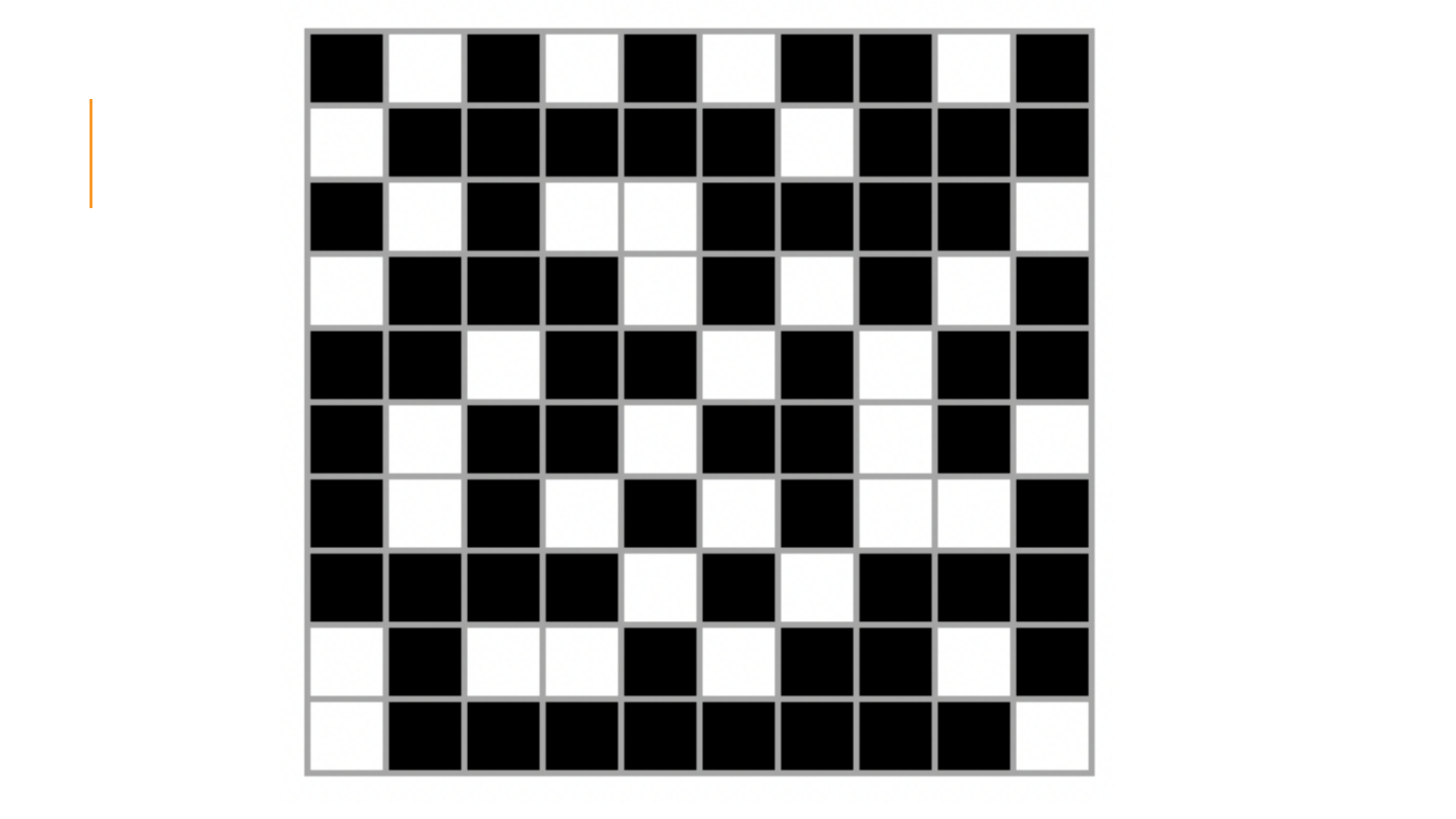}}
\caption{Slide 4 of 23}\label{sup-fig-slide04}
\end{figure}

\begin{figure}[p]
\fbox{\includegraphics[width=\textwidth,trim=0 0 0 0,clip,page=5]{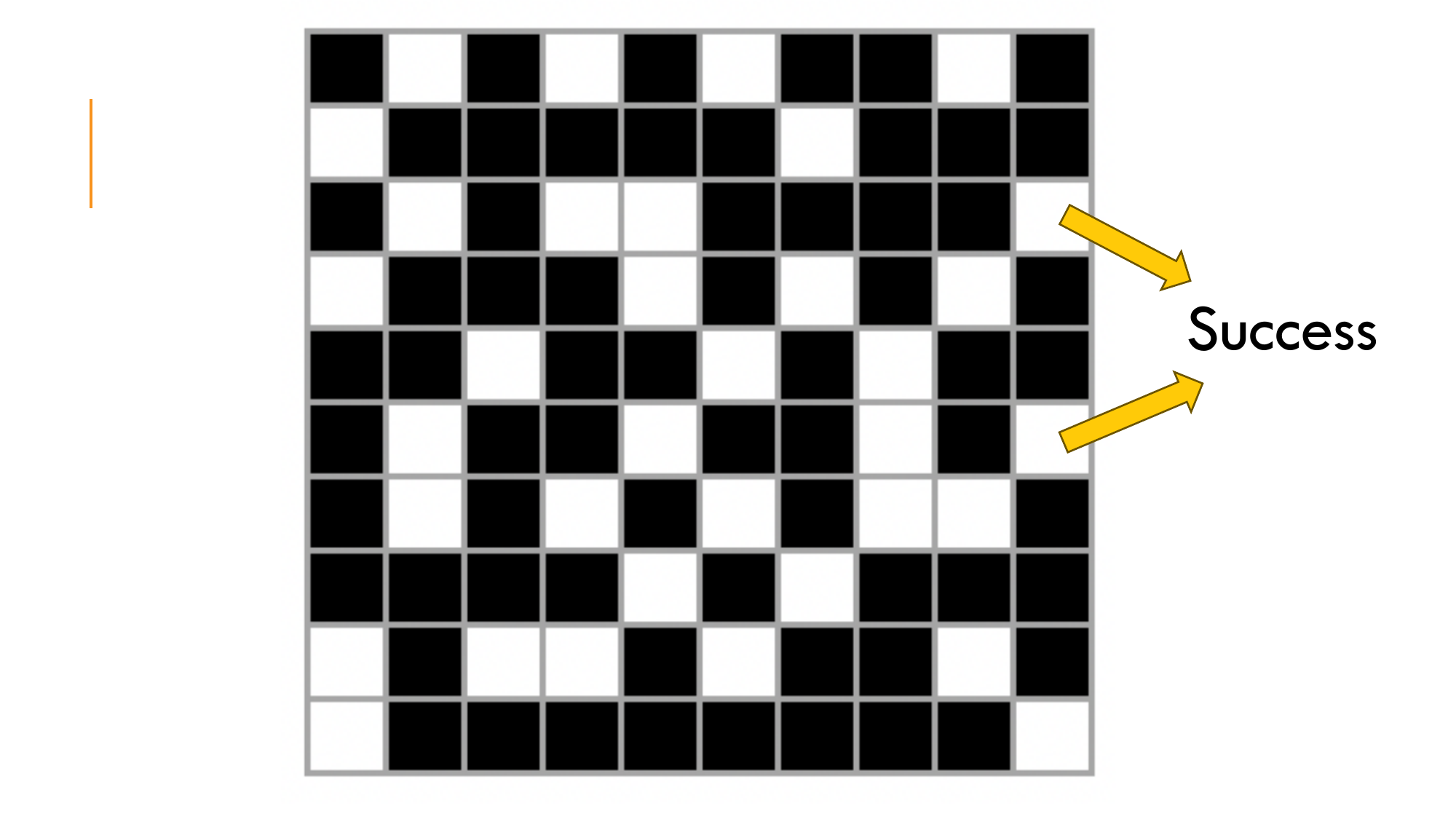}}
\caption{Slide 5 of 23}\label{sup-fig-slide05}
\end{figure}

\begin{figure}[p]
\fbox{\includegraphics[width=\textwidth,trim=0 0 0 0,clip,page=6]{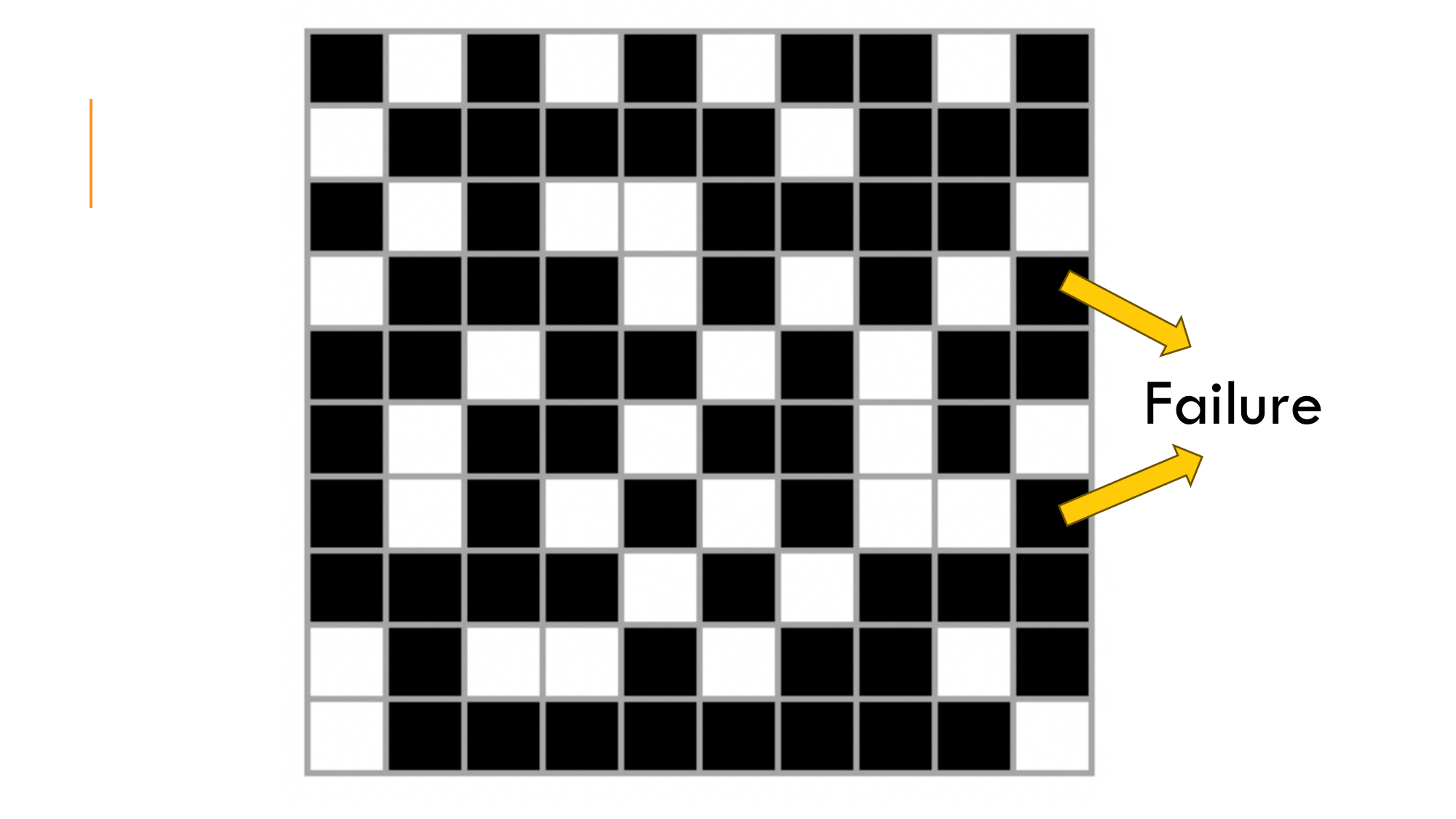}}
\caption{Slide 6 of 23}\label{sup-fig-slide06}
\end{figure}

\begin{figure}[p]
\fbox{\includegraphics[width=\textwidth,trim=0 0 0 0,clip,page=7]{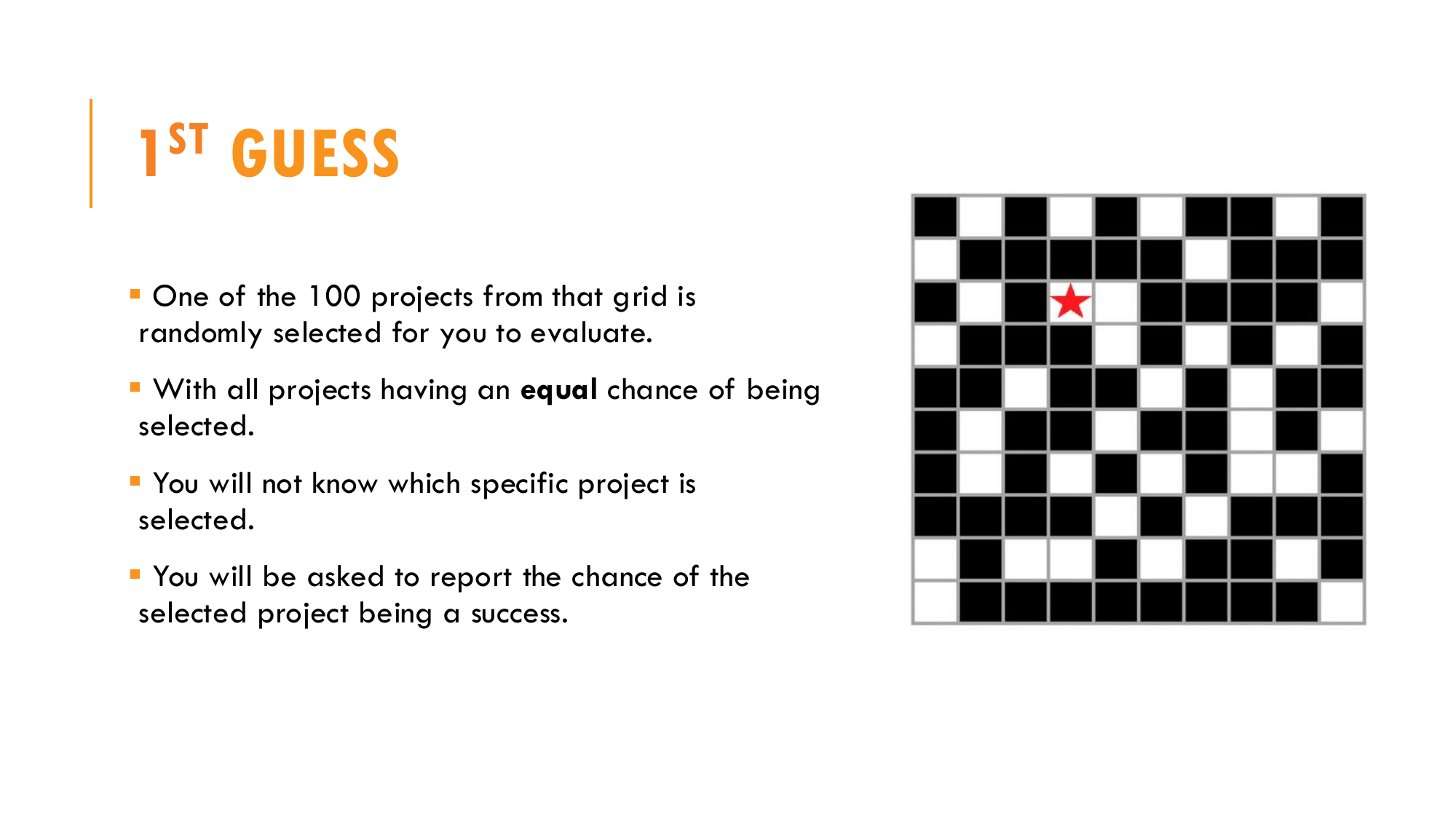}}
\caption{Slide 7 of 23}\label{sup-fig-slide07}
\end{figure}

\begin{figure}[p]
\fbox{\includegraphics[width=\textwidth,trim=0 0 0 0,clip,page=8]{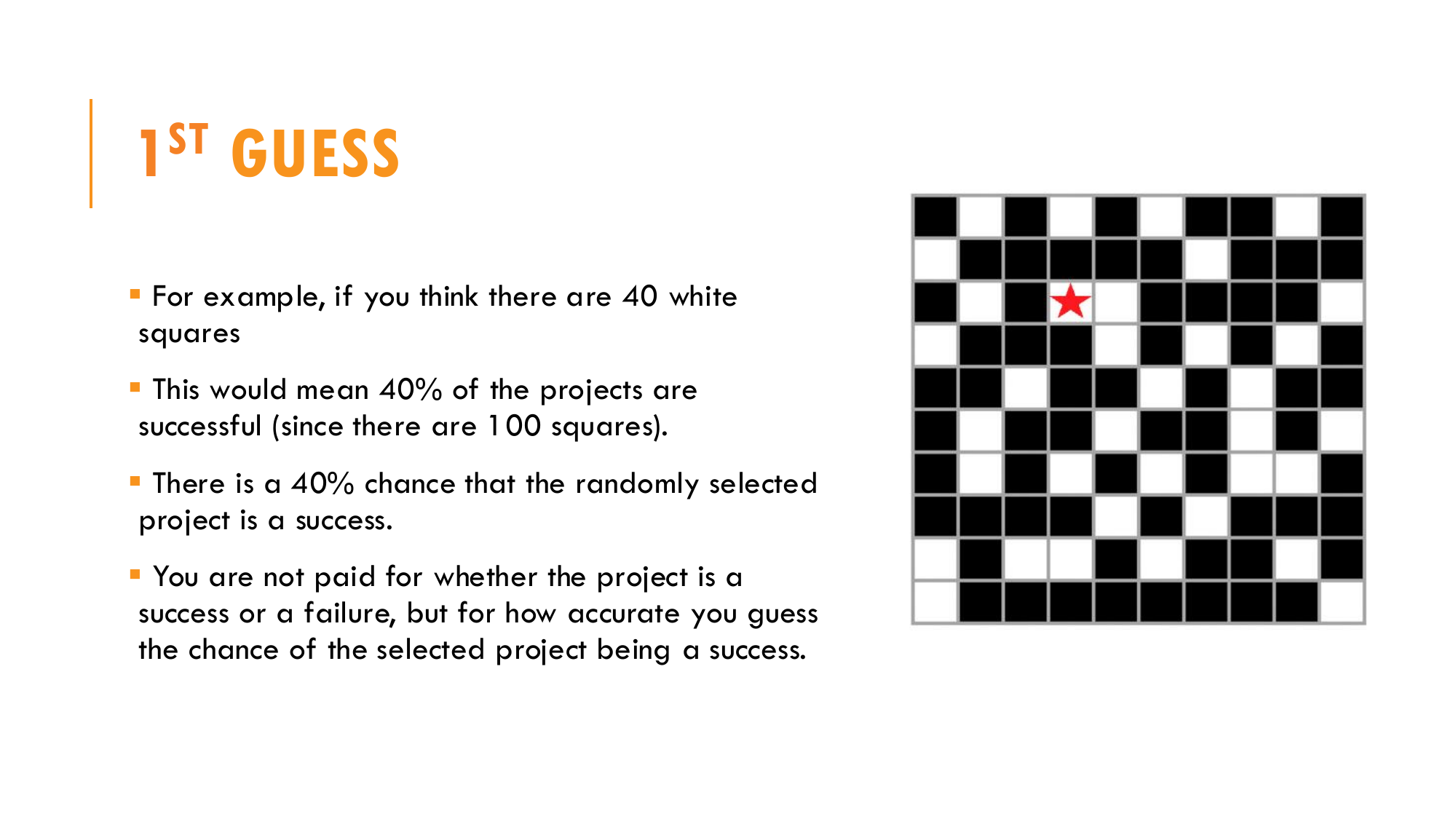}}
\caption{Slide 8 of 23}\label{sup-fig-slide08}
\end{figure}

\begin{figure}[p]
\fbox{\includegraphics[width=\textwidth,trim=0 0 0 0,clip,page=9]{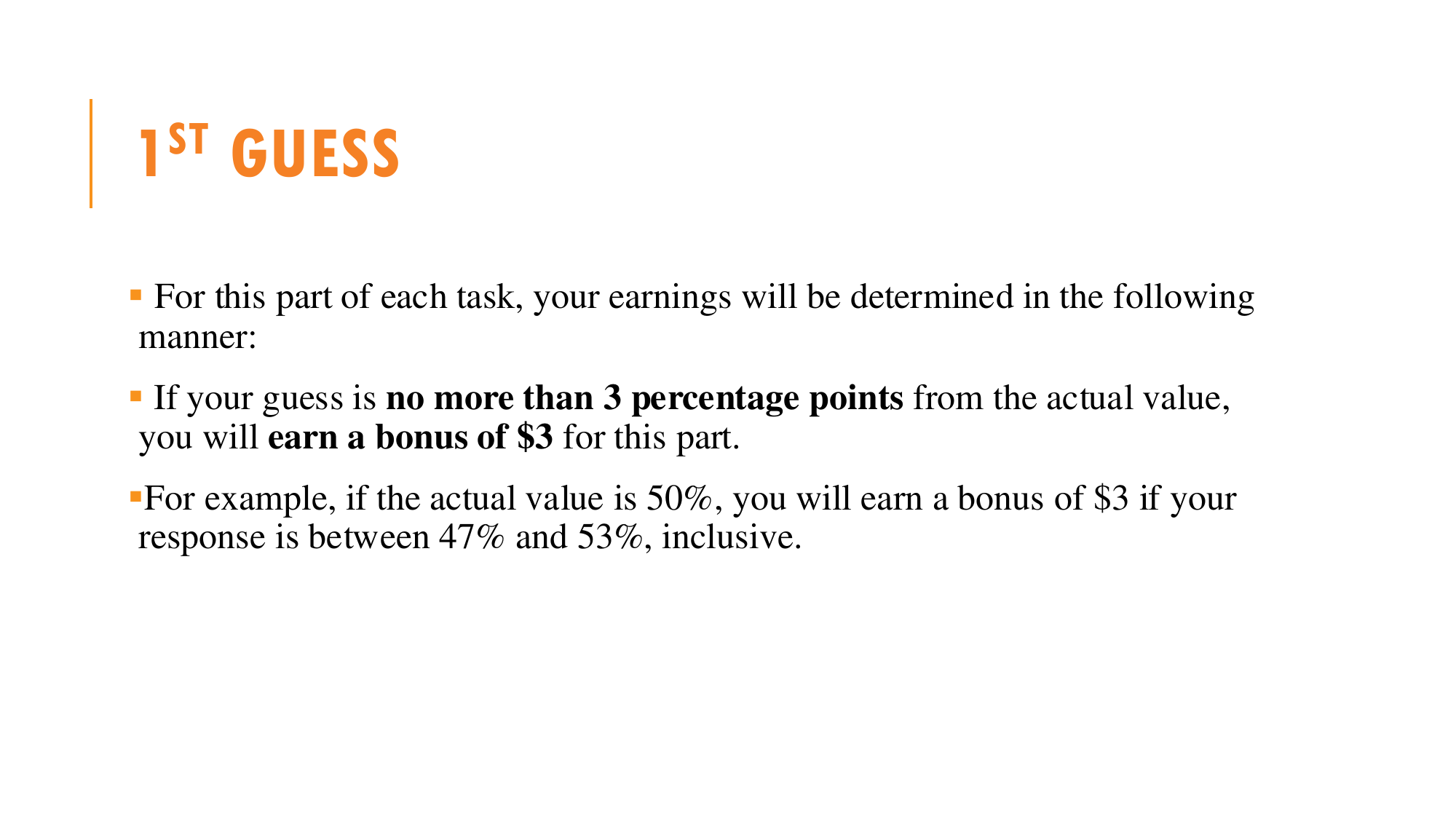}}
\caption{Slide 9 of 23}\label{sup-fig-slide09}
\end{figure}

\begin{figure}[p]
\fbox{\includegraphics[width=\textwidth,trim=0 0 0 0,clip,page=10]{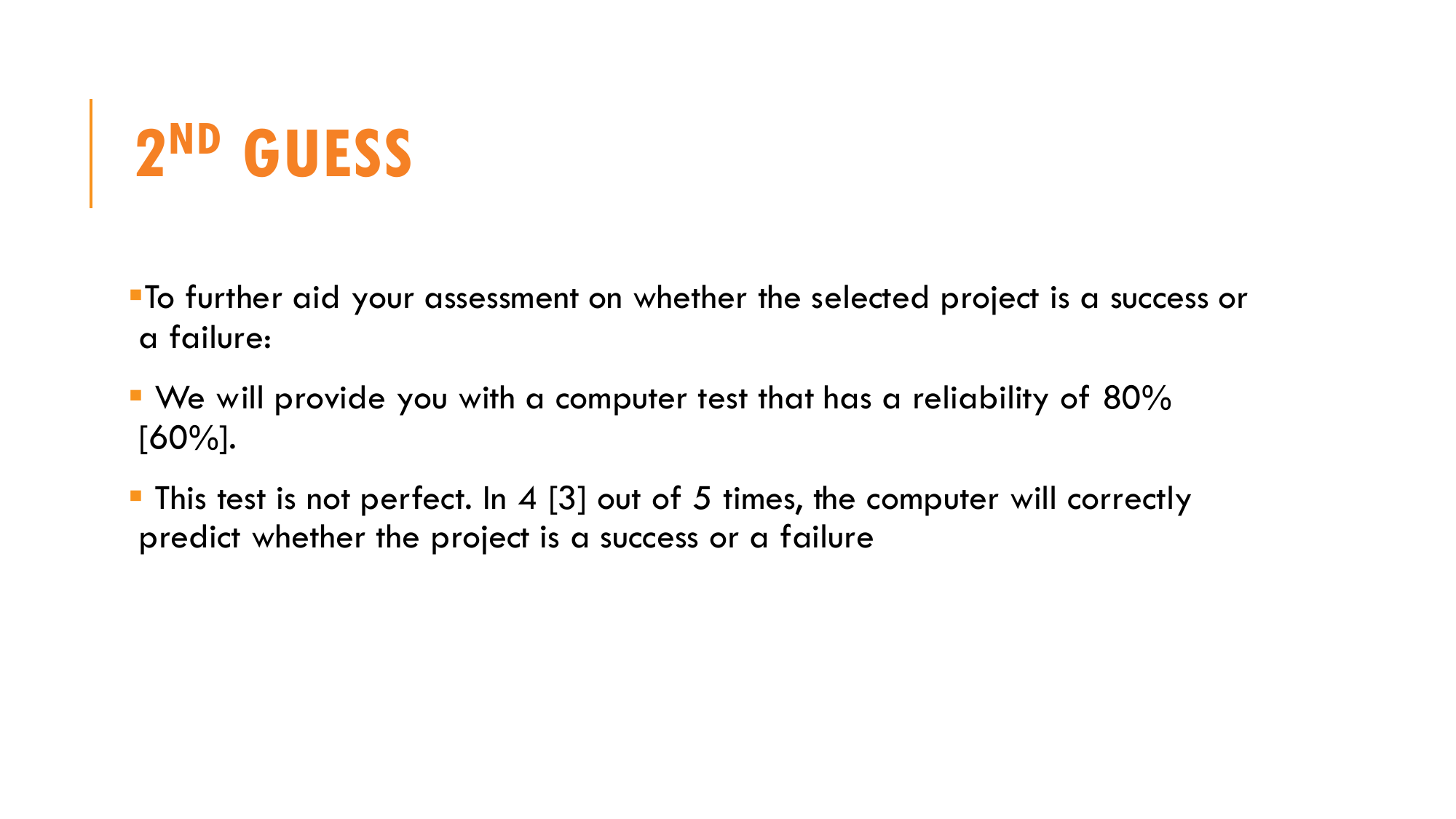}}
\caption{Slide 10 of 23}\label{sup-fig-slide10}
\end{figure}

\begin{figure}[p]
\fbox{\includegraphics[width=\textwidth,trim=0 0 0 0,clip,page=11]{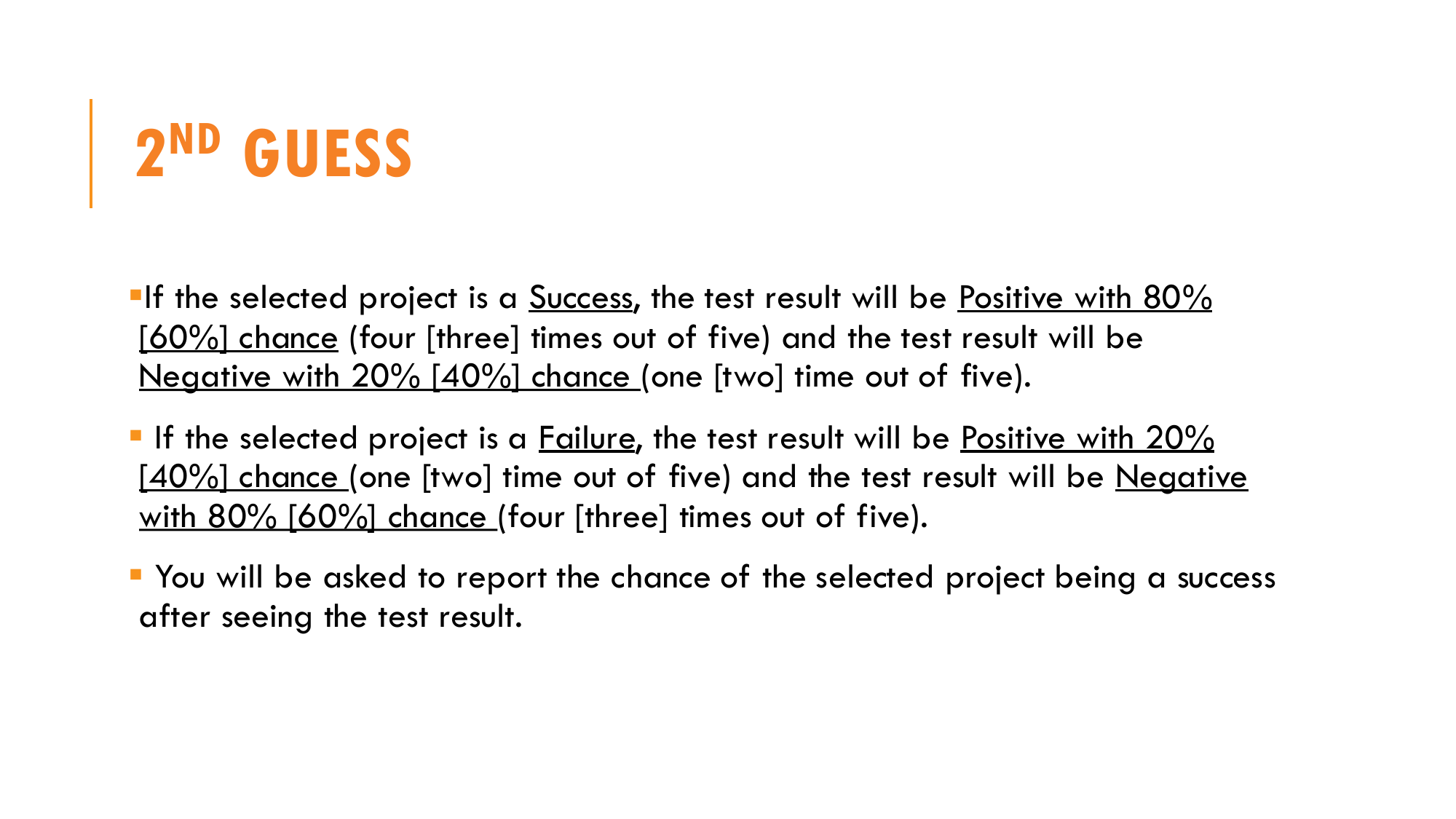}}
\caption{Slide 11 of 23}\label{sup-fig-slide11}
\end{figure}

\begin{figure}[p]
\fbox{\includegraphics[width=\textwidth,trim=0 0 0 0,clip,page=12]{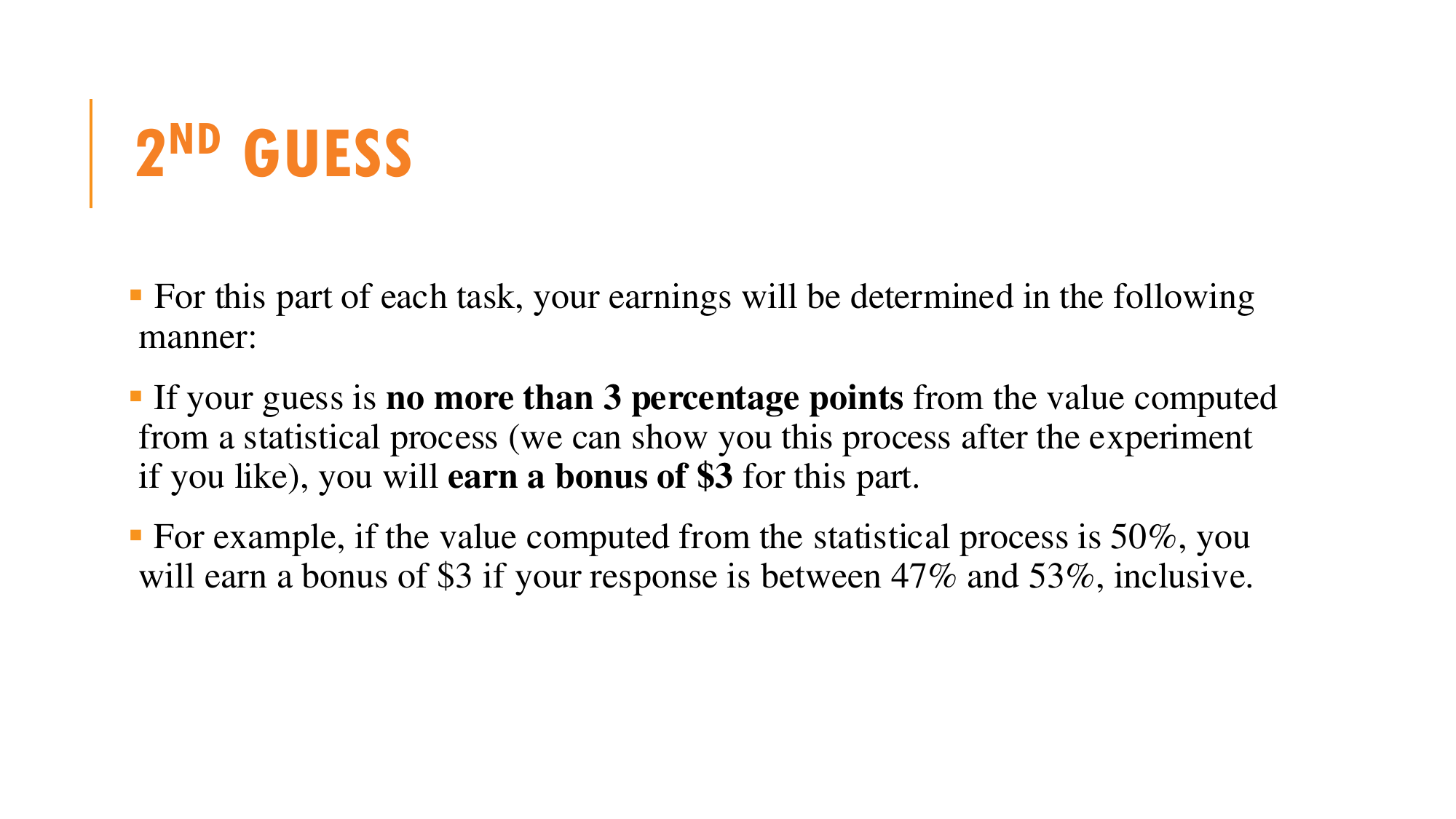}}
\caption{Slide 12 of 23}\label{sup-fig-slide12}
\end{figure}

\begin{figure}[p]
\fbox{\includegraphics[width=\textwidth,trim=0 0 0 0,clip,page=13]{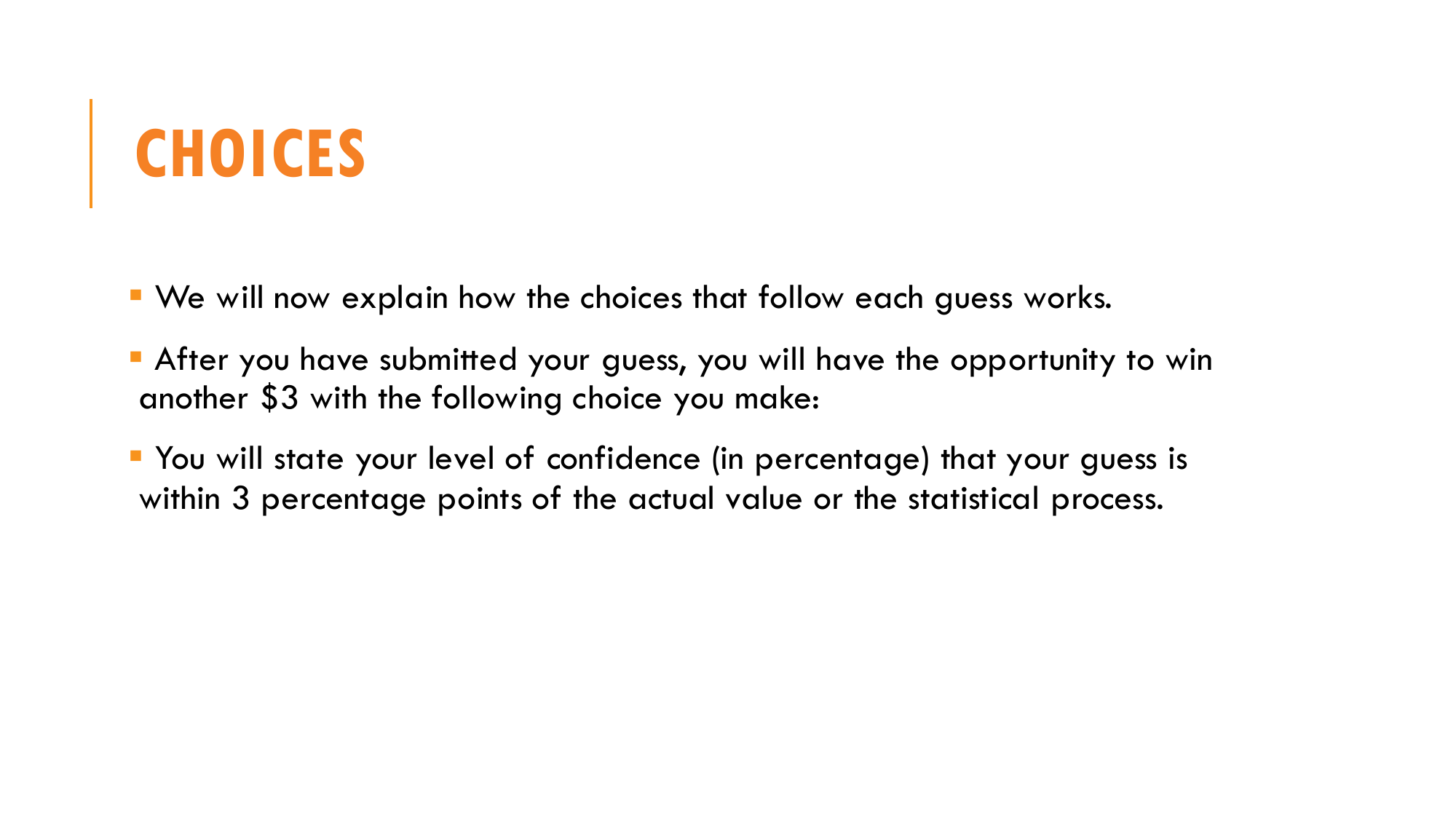}}
\caption{Slide 13 of 23}\label{sup-fig-slide13}
\end{figure}

\begin{figure}[p]
\fbox{\includegraphics[width=\textwidth,trim=0 0 0 0,clip,page=14]{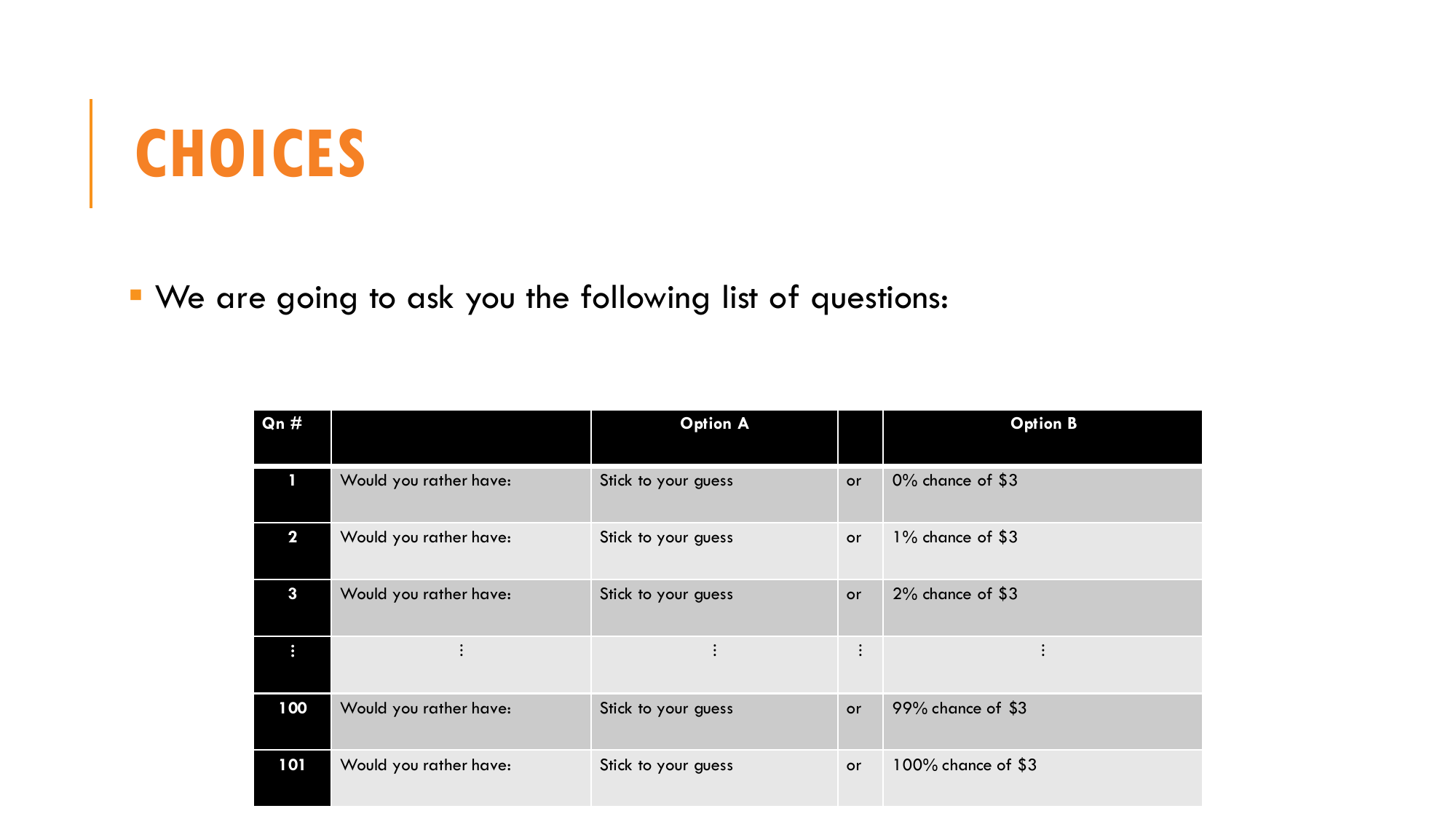}}
\caption{Slide 14 of 23}\label{sup-fig-slide14}
\end{figure}

\begin{figure}[p]
\fbox{\includegraphics[width=\textwidth,trim=0 0 0 0,clip,page=15]{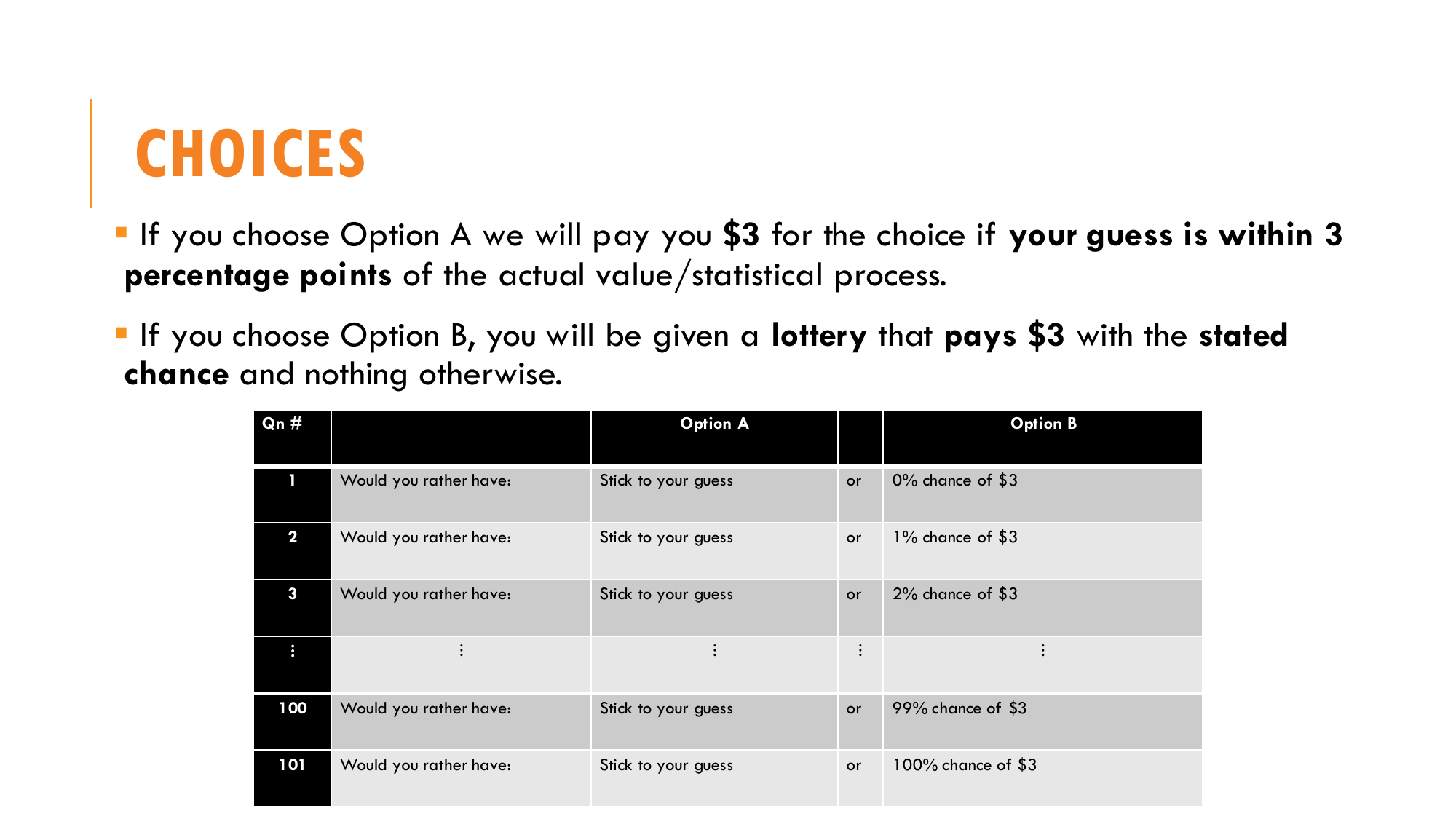}}
\caption{Slide 15 of 23}\label{sup-fig-slide15}
\end{figure}

\begin{figure}[p]
\fbox{\includegraphics[width=\textwidth,trim=0 0 0 0,clip,page=16]{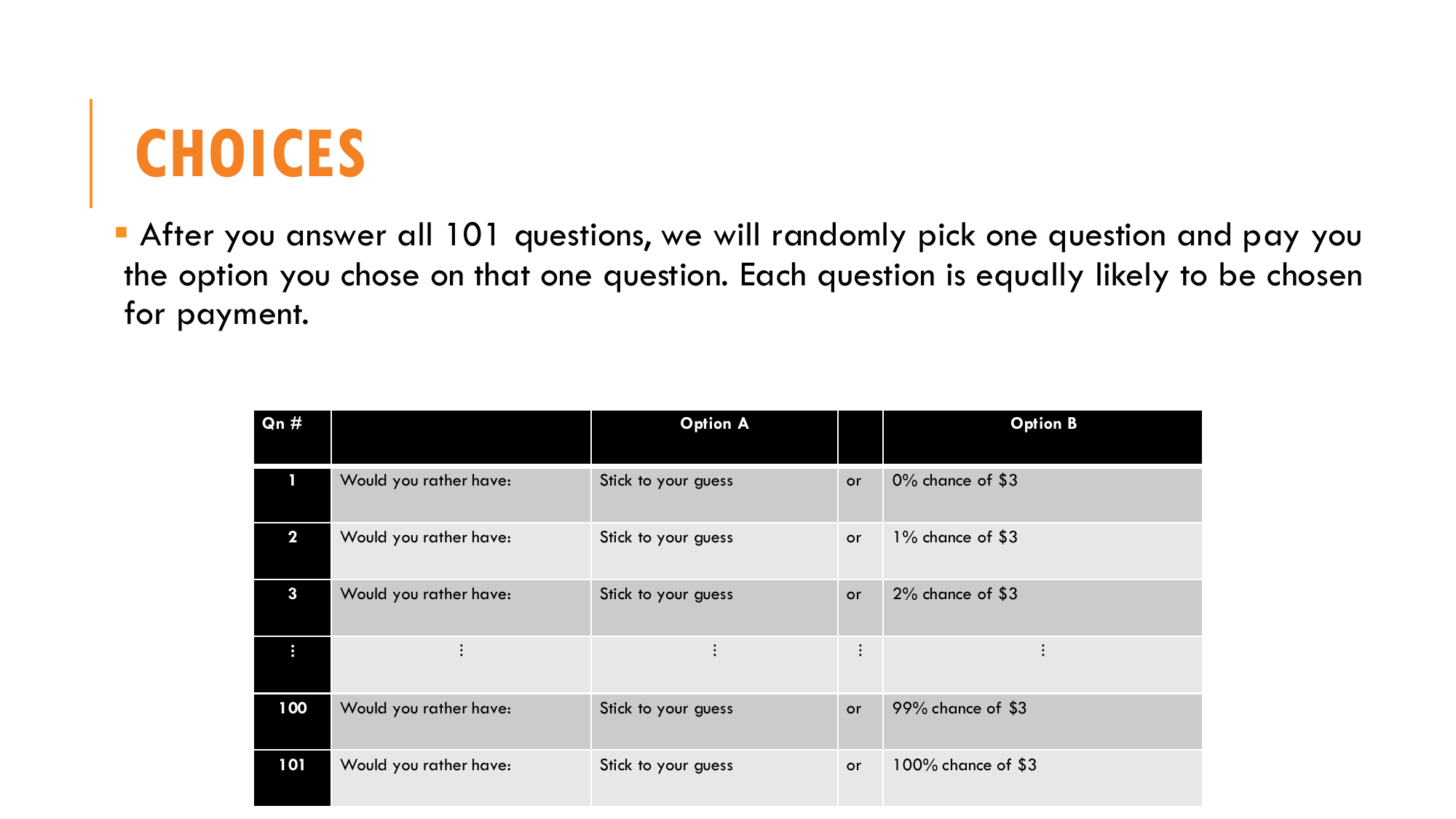}}
\caption{Slide 16 of 23}\label{sup-fig-slide16}
\end{figure}

\begin{figure}[p]
\fbox{\includegraphics[width=\textwidth,trim=0 0 0 0,clip,page=17]{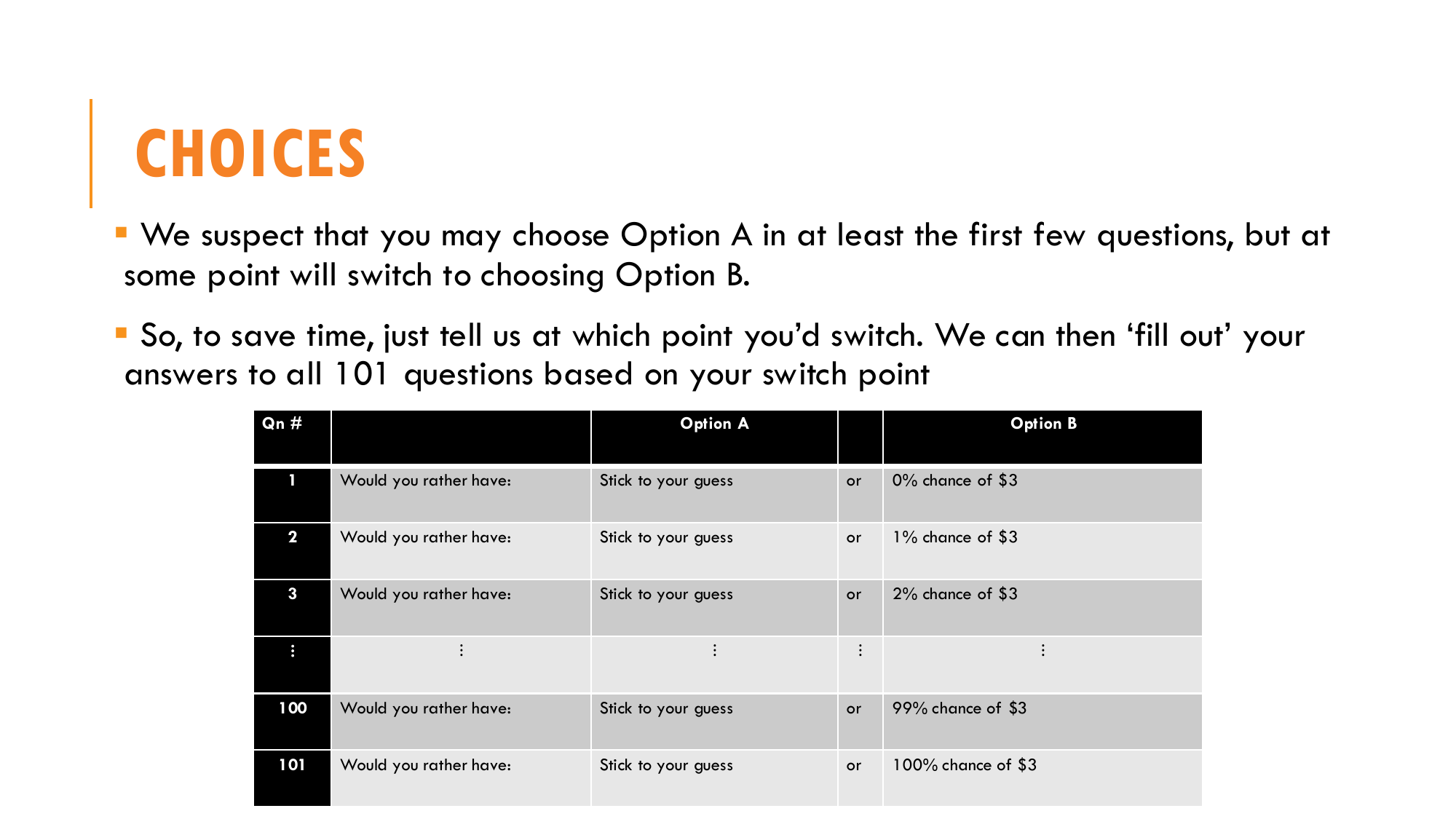}}
\caption{Slide 17 of 23}\label{sup-fig-slide17}
\end{figure}

\begin{figure}[p]
\fbox{\includegraphics[width=\textwidth,trim=0 0 0 0,clip,page=18]{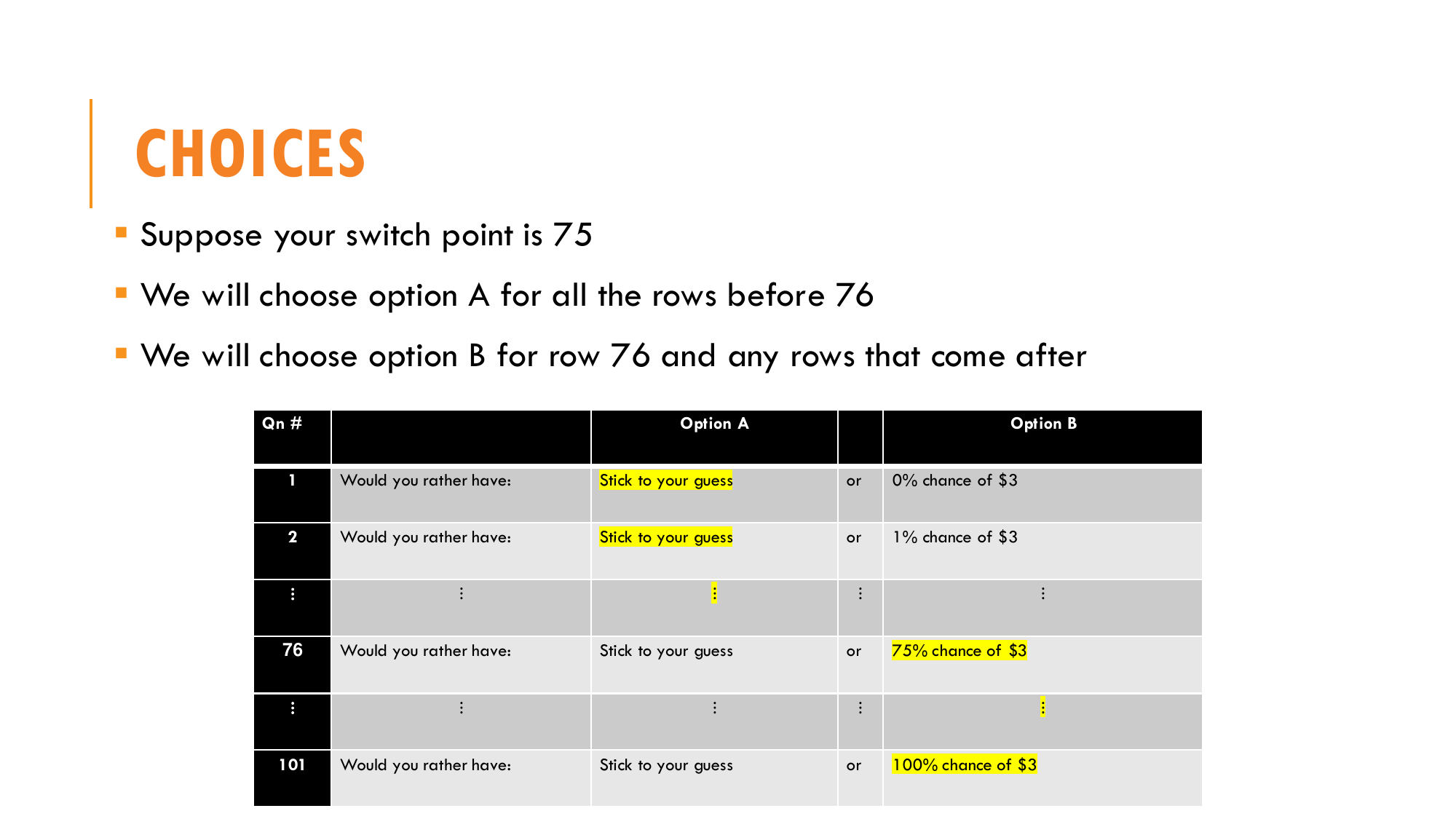}}
\caption{Slide 18 of 23}\label{sup-fig-slide18}
\end{figure}

\begin{figure}[p]
\fbox{\includegraphics[width=\textwidth,trim=0 0 0 0,clip,page=19]{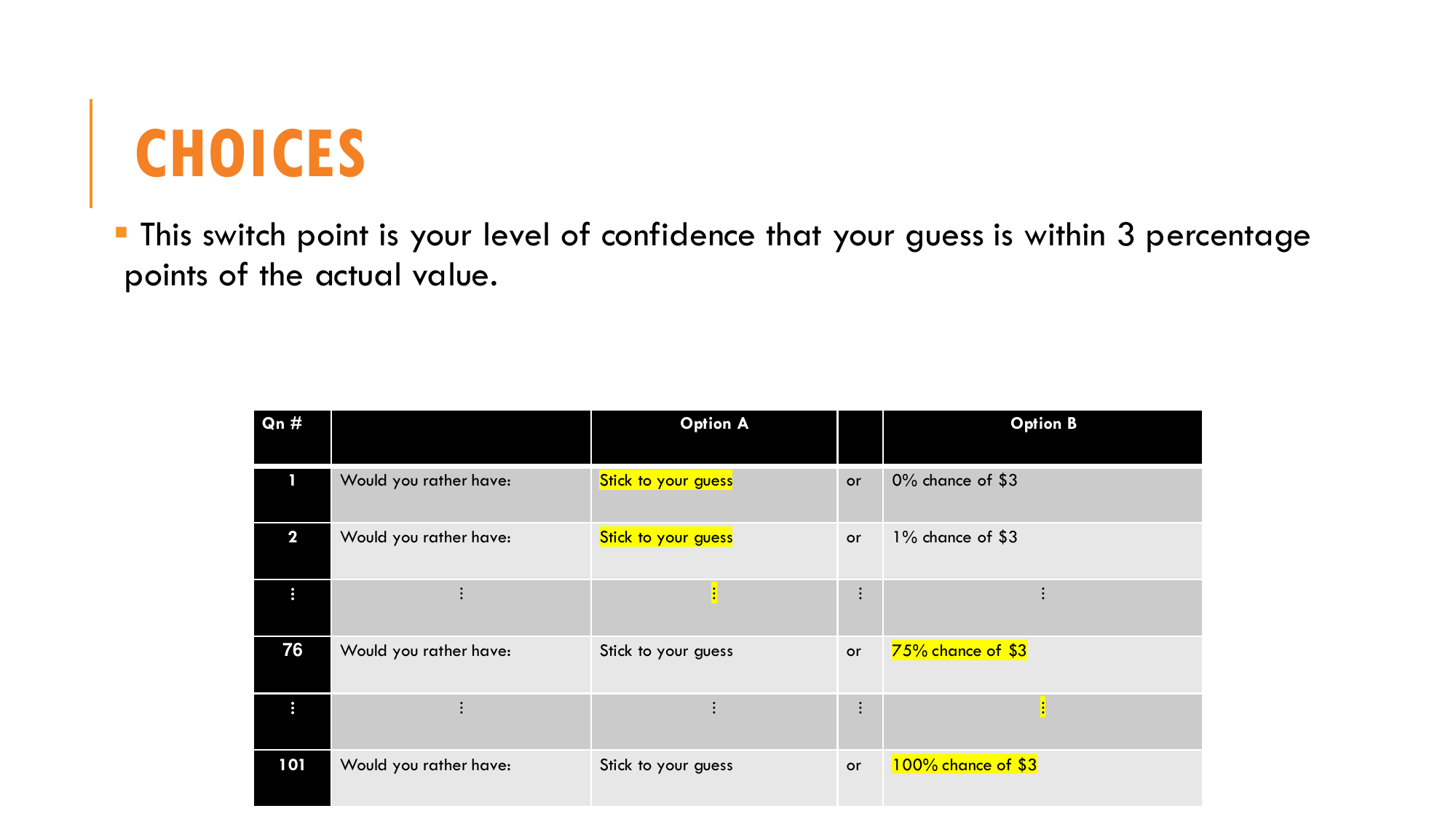}}
\caption{Slide 19 of 23}\label{sup-fig-slide19}
\end{figure}

\begin{figure}[p]
\fbox{\includegraphics[width=\textwidth,trim=0 0 0 0,clip,page=20]{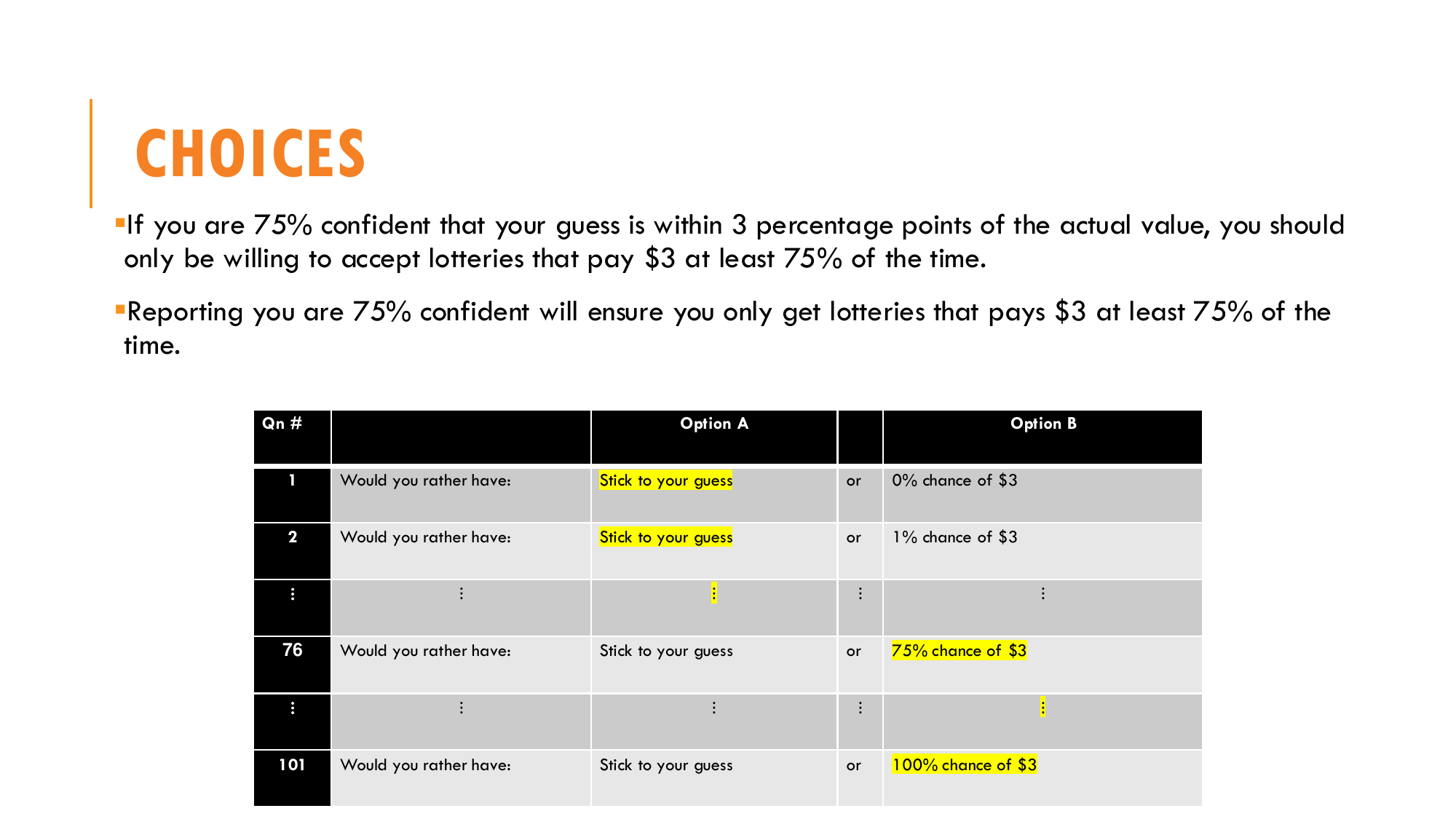}}
\caption{Slide 20 of 23}\label{sup-fig-slide20}
\end{figure}

\begin{figure}[p]
\fbox{\includegraphics[width=\textwidth,trim=0 0 0 0,clip,page=21]{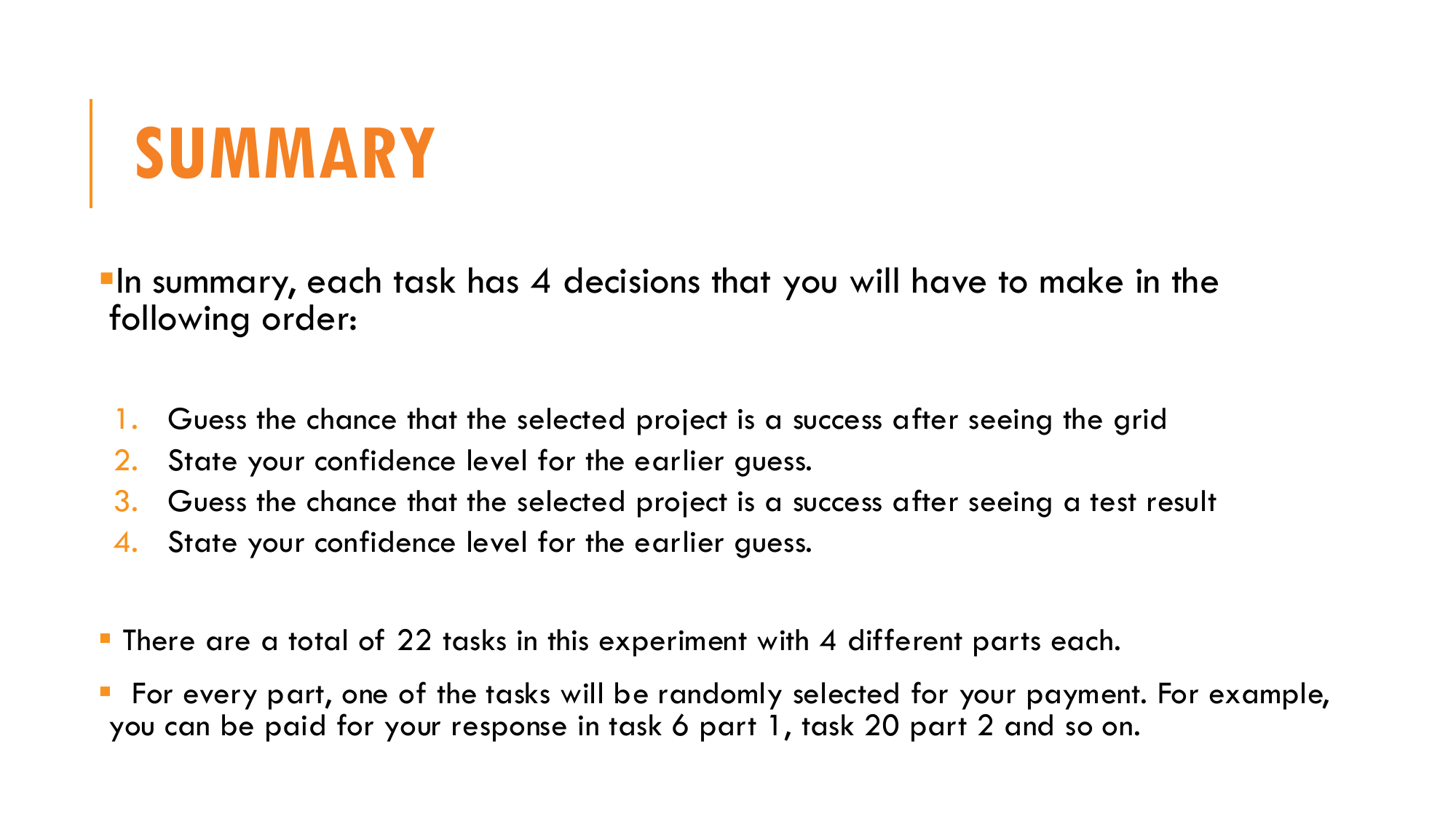}}
\caption{Slide 21 of 23}\label{sup-fig-slide21}
\end{figure}

\begin{figure}[p]
\fbox{\includegraphics[width=\textwidth,trim=0 0 0 0,clip,page=22]{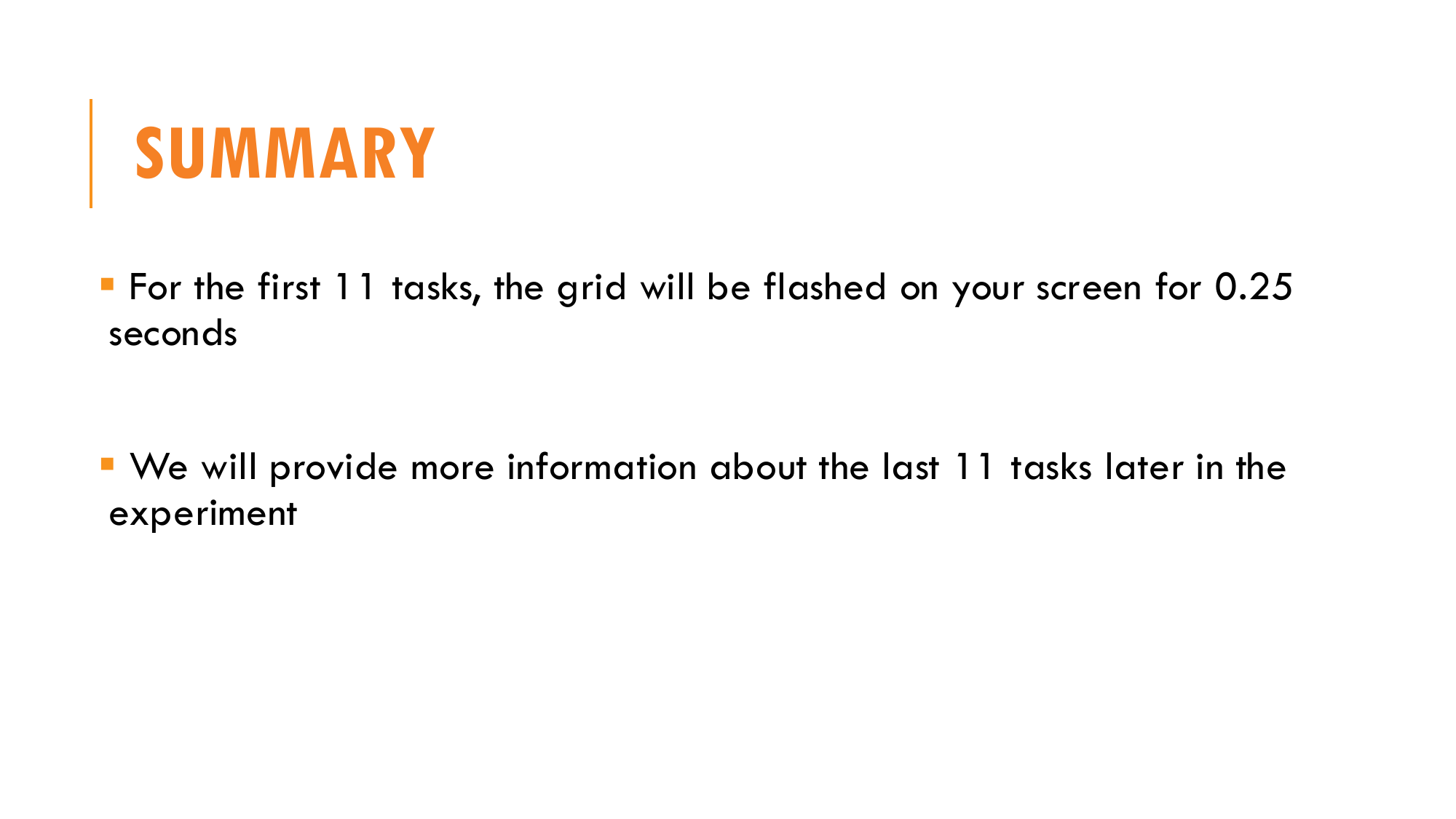}}
\caption{Slide 22 of 23}\label{sup-fig-slide22}
\end{figure}

\begin{figure}[p]
\fbox{\includegraphics[width=\textwidth,trim=0 0 0 0,clip,page=23]{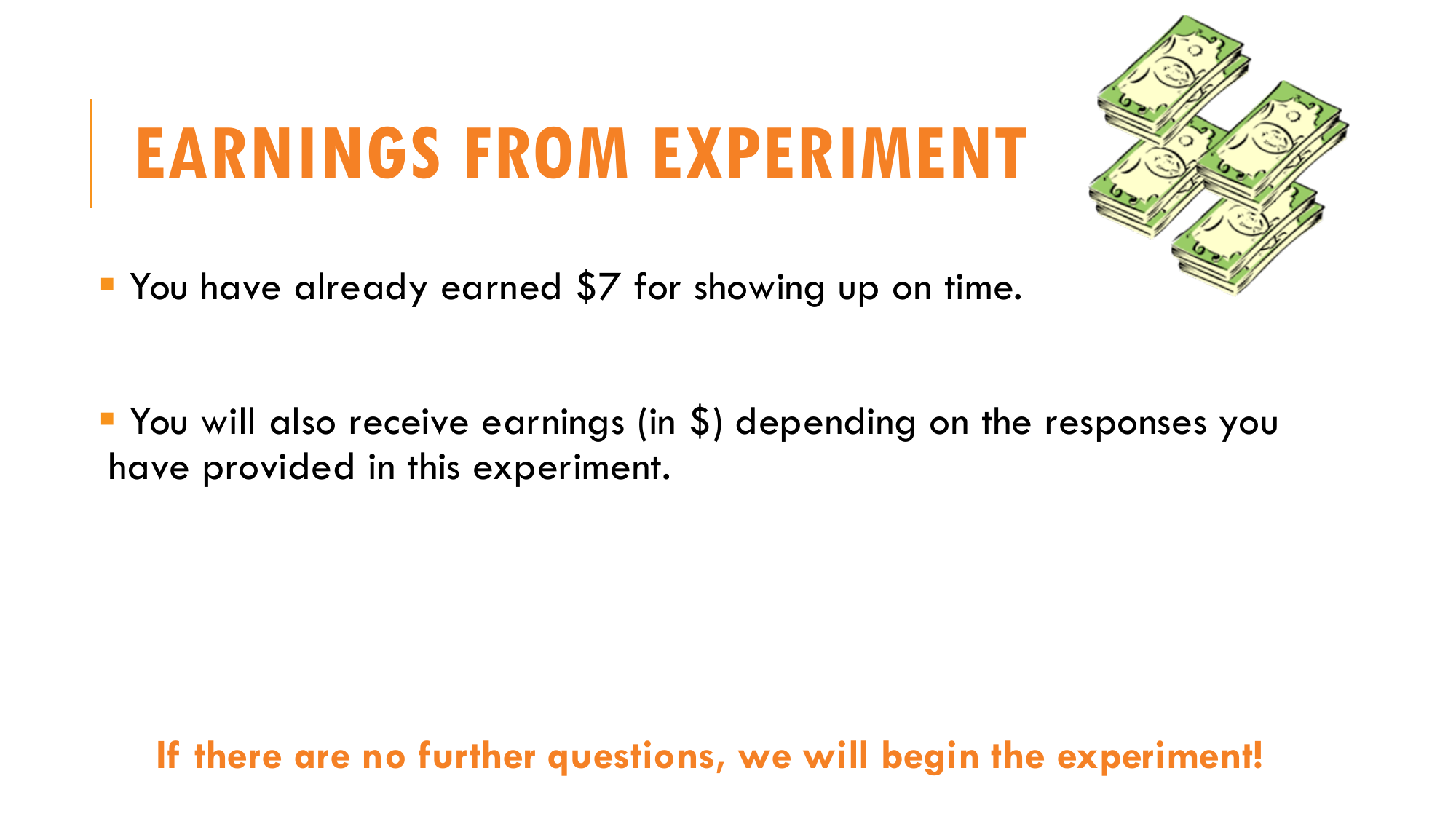}}
\caption{Slide 23 of 23}\label{sup-fig-slide23}
\end{figure}

\end{document}